\documentclass[12pt,preprint,showpacs,,nofootinbib]{revtex4}
\usepackage{subfigure}
\usepackage{color}
\usepackage{graphicx}

\makeatletter

\providecommand{\tabularnewline}{\\}


\begin{document}

\begin{flushright}
hep-ph/0504230
\end{flushright}

\newcommand{\oalphas}{O(\alpha_{s})}
\newcommand{\met}{\not\!\! E_{T}}

\title{Next-to-Leading Order Corrections to Single Top Quark Production
and Decay at the Tevatron: 2. $t$-channel Process}

\author{Qing-Hong Cao}
\email{cao@pa.msu.edu}
\affiliation{Department of Physics $\&$ Astronomy, Michigan State University,
East Lansing, MI 48824, USA}
\author{Reinhard Schwienhorst}
\email{schwier@pa.msu.edu}
\affiliation{Department of Physics $\&$ Astronomy, Michigan State University,
East Lansing, MI 48824, USA}
\author{Jorge A. Benitez}
\email{benitez@pa.msu.edu}
\affiliation{Department of Physics $\&$ Astronomy, Michigan State University,
East Lansing, MI 48824, USA}
\author{Raymond Brock}
\email{brock@pa.msu.edu}
\affiliation{Department of Physics $\&$ Astronomy, Michigan State University,
East Lansing, MI 48824, USA}
\author{C.-P. Yuan}
\email{yuan@pa.msu.edu}
\affiliation{Department of Physics $\&$ Astronomy, Michigan State University,
East Lansing, MI 48824, USA}

\begin{abstract}
We present a study of the $t$-channel mode of single top quark production
at the upgraded Tevatron $p\bar{p}$ collider, including the next-to-leading
order (NLO) QCD corrections to the production and the decay of a single
top quark. The narrow width approximation was adopted in order to
preserve the spin of the top quark in its production and decay. We
discuss the effects of different $\oalphas$ contributions on the
inclusive cross section as well as various kinematic distributions
after imposing the relevant cuts to select $t$-channel single top
signal events. 
\end{abstract}
\maketitle

\section{Introduction\label{sec:Introduction}}

In the Standard Model (SM), the charged-current weak interaction links
the top quark with a down-type quark with an amplitude proportional
to the Cabibbo-Kobayashi-Maskawa (CKM) matrix $V_{tq}$ ($q=d,\, s,\, b$).
At hadron colliders, this interaction leads to a single top quark
final state through three possible processes: the $s$-channel decay
of a virtual $W$ ($q\bar{q}'\rightarrow W^{*}\rightarrow t\bar{b}$),
the $t$-channel exchange of a virtual $W$ ($bq\rightarrow tq'$
and $b\bar{q}'\rightarrow t\bar{q}$, also referred to as $W$-gluon
fusion), and the associated production of a top quark and a $W$ boson
($bg\rightarrow tW^{-}$). Since the cross section for each of these
processes is proportional to the CKM matrix element $|V_{tb}|^{2}$,
a measurement of each single top quark production cross section determines
$|V_{tb}|$. A study of spin correlations in single top quark production
can be used to test the $V-A$ structure of the top quark charged-current
weak interaction. Such structure implies that the top quark should
be polarized.

Among the three processes, the largest cross section at both the Tevatron
proton-antiproton collider at Fermilab and the proton-proton Large
Hadron Collider (LHC) at CERN is due to the $t$-channel. The next
largest cross section is from the $s$-channel at the Tevatron, and
from associated production at the LHC. The $s$-channel production
cross section is small at the LHC because it involves a quark-antiquark
collision. Similarly, associated production cross section is relatively
large at the LHC because the gluon parton distribution functions (PDF)
grow more rapidly with decreasing $x$ than the light quark PDF. The
$s$-channel and $t$-channel processes should be observed for the
first time in Run II at the Tevatron; whereas the observation of associated
production will likely have to await the LHC. 

Single top quark production is also a very important background to
many searches for new physics. For example, the $s$-channel process
is a significant background to Higgs searches at the Tevatron in the
production process $q\bar{q}'\rightarrow WH$ with decay $H\rightarrow b\bar{b}$~\cite{Stange:1993ya,Stange:1994bb,Belyaev:1995gb}
and other new physics searches~\cite{Cao:2003tr}. At the LHC, the
$Wt$ associated production is an important background to Higgs searches
in the decay channel $H\to WW$~\cite{Moretti:1997ng} and the primary
charged Higgs boson production channel $bg\to tH^{\pm}$ with $H^{\pm}\to\tau\nu$~\cite{Odagiri:1999fs}.
As the largest single top process, the $t$-channel production is
an important background to many new physics searches.

In Tevatron Run~I, searches for single top quark production were
performed by both the D\O~\cite{Abbott:2000pa} and CDF collaborations~\cite{Acosta:2004er}.
At the 95\% confidence level, the D\O~limit on the $s$-channel
production is 17~picobarn (pb) and the CDF limit is 18~pb. At the
same confidence level, the D\O~limit on the $t$-channel production
cross section is 22~pb and the CDF limit is 13~pb. Searches for
single top quark production in Run~II have begun, and the limits
from CDF are 13.6 pb~in the $s$-channel and 10.1~pb in the $t$-channel~\cite{Acosta:2004bs}.
The $s$-channel and $t$-channel single top quark processes can be
probed separately at the Tevatron by taking advantage of $b$-quark
tagging using displaced vertices and differences in the kinematic
distributions of the $b$-tagged and non-$b$-tagged jets. Usually,
only one $b$-tagged jet can be expected in the $t$-channel case
while two $b$-tagged jets can be expected in the $s$-channel case.
This is because the $\bar{b}$ quark produced with the top quark tends
to be collinear with the initial state gluon in the $t$-channel,
giving it a large pseudo-rapidity ($\eta$) and low transverse momentum
($p_{T}$) and thus making it challenging to $b$-tag this jet experimentally.
It is important to separate the two processes since they have different
systematic uncertainties and different sensitivities to new physics
\cite{Tait:2000sh}. 

The extraction of a signal is more challenging for single top quark
production than for top pair production since there are fewer objects
in the final state and the overall event properties are less distinct
from the large $W$+jets background. Therefore, an accurate calculation
including higher order quantum chromodynamics (QCD) corrections is
needed. The next-to-leading order (NLO) $\oalphas$ corrections to
single top quark production have already been carried out in Refs.~\cite{Smith:1996ij,Bordes:1994ki},
which shows an uncertainty on the total cross section of about $5\%$
by varying the factorization and renormalization scales. For the $s$-channel,
multiple soft gluon resummation effects have been calculated~\cite{Mrenna:1997wp}.
In order to confront theory with experimental data, where kinematical
cuts are necessary in order to detect a signal, it is crucial to accurately
model event topologies of single top quark events. For this, Refs.~\cite{Harris:2002md,Sullivan:2004ie}
have calculated the differential cross sections for on-shell single
top quark production. The complete NLO calculations including both
the single top quark production and decay have been done independently
by two groups recently~\cite{Campbell:2004ch,Cao:2004ky,Cao:2004ap}.
In both calculations, the narrow width approximation was adopted in
order to link top quark production with its consequent decay~\cite{Campbell:2004ch,Cao:2004ky}.
In Ref.~\cite{Campbell:2004ch}, various kinematics distributions
are examined both with and without top quark decay at NLO. In our
previous study~\cite{Cao:2004ap}, we presented a detailed phenomenological
analysis of $s$-channel single top quark production at the Tevatron,
focusing on signal cross sections and kinematical distributions. To
complete our study of single top processes at the Tevatron, we will
focus on the single top quark production process with the largest
production cross section in this paper, namely the $t$-channel. A
more realistic study which disentangles the signal of single top quark
events from copious backgrounds and further separates the $s$-channel
from the  $t$-channel will be given in the future.

The paper is organized as follows. In Sec.~\ref{sec:Cross-Section-Inclusive},
we first present the inclusive cross section for the $t$-channel
single top process using both the fixed mass narrow width approximation
and the ``modified'' narrow width approximation. We then investigate
its dependence on the top quark mass as well as the renormalization
and factorization scales. In Sec.~\ref{sec:Single-Top-Acceptance},
we examine the acceptance of single top quark events for various sets
of kinematic cuts. In Sec.~\ref{sec:EventDistr}, we explore the
kinematical information in the final state objects. Section~\ref{sec:Conclusions}
contains our conclusions.

\section{Cross Section (Inclusive Rate)\label{sec:Cross-Section-Inclusive}}

In order to accurately measure $V_{tb}$ using the single top process,
the uncertainties in the theoretical prediction need to be minimized.
In this section, we show the inclusive production rates and discuss
their dependence on the factorization and renormalization scales and
the top quark mass.

We present numerical results for the production of single top quark
events considering the leptonic decay of the $W$-boson from the top
quark decay at the upgraded Tevatron (a 1.96 TeV $p\bar{p}$ collider).
Unless otherwise specified, we use the NLO parton distribution function
(PDF) set CTEQ6M~\cite{Pumplin:2002vw}, defined in the $\overline{MS}$
scheme, and the NLO (2-loop) running coupling $\alpha_{s}$ with $\Lambda_{\overline{MS}}$
provided by the PDFs. For the CTEQ6M PDFs, $\Lambda_{\overline{MS}}^{(4)}=0.326~{\textrm{GeV}}$
for four active quark flavors. The values of the relevant electroweak
parameters are: $\alpha=1/137.0359895$, $G_{\mu}=1.16637\times10^{-5}\,{\rm GeV}^{-2}$,
$M_{W}=80.40\,{\rm GeV}$, $M_{Z}=91.1867\,{\rm GeV}$, and $\sin^{2}\theta_{W}=0.231$~\cite{Cao:2004yy}.
The square of the weak gauge coupling is $g^{2}=4\sqrt{2}M_{W}^{2}G_{\mu}$.
As in the \textcolor{black}{$s$-channel study}~\cite{Cao:2004ap},
here we also focus our attention only on the positively charged electron
(i.e., positron), though our analysis procedure also applies to the
muon final state. Including $\oalphas$ corrections to $W\to\bar{q}q^{\prime}$,
the decay branching ratio of the $W$ boson into leptons is $Br(W\to\ell^{+}\nu)=0.108$~\cite{Cao:2004yy}.
Unless otherwise specified, we will choose the top quark mass to be
$m_{t}=178\,{\rm GeV}$~\cite{Azzi:2004rc,Abazov:2004cs} and the
renormalization scale ($\mu_{R}$) as well as the factorization scale
($\mu_{F}$) to be equal to the top quark mass. The top quark mass
and scale dependencies of the $t$-channel production cross section
are investigated in this section.

\subsection{Theoretical Cutoff Dependence\label{sub:Theoretical-Cutoff-Dependence}}

The NLO QCD differential cross sections are calculated using the one-cutoff
phase space slicing (PSS) method~\cite{Giele:1991vf,Giele:1993dj}.
This procedure introduces a theoretical cutoff parameter ($s_{min}$)
in order to isolate soft and collinear singularities associated with
real gluon emission subprocesses by partitioning the phase space into
soft, collinear and hard regions such that\begin{equation}
\left|\mathcal{M}^{{\rm r}}\right|^{2}=\left|\mathcal{M}^{{\rm r}}\right|_{{\rm soft}}^{2}+\left|\mathcal{M}^{{\rm r}}\right|_{{\rm collinear}}^{2}+\left|\mathcal{M}^{{\rm r}}\right|_{{\rm hard}}^{2}\,.\label{eq:nlomat}\end{equation}
 In the soft and collinear regions the cross section is proportional
to the Born-level cross section. Using dimensional regularization,
we can evaluate the real gluon emission diagrams in n-dimensions under
the soft gluon approximation in the soft region, or the collinear
approximation in the collinear region, and can integrate out the corresponding
phase space volume analytically. The resulting divergences are cancelled
by virtual corrections or absorbed into the perturbative parton distribution
functions in the factorization procedure. Since the cutoff is introduced
in the calculation only for this technical reason and is unrelated
to any physical quantity, the inclusive rate must not depend on it.
In other words, the sum of all contributions, virtual, resolved, and
unresolved corrections must be independent of $s_{min}$. This is
the case as long as $s_{min}$ is small enough so that the soft and
collinear approximations are valid. However, numerical cancellation
in the Monte Carlo (MC) integration becomes unstable if $s_{min}$
is too small. Furthermore, the jet-finding algorithm and other infrared-safe
experimental observables should also be defined in a way such that
they are consistent with the choice of $s_{min}$. In practice, one
wants to choose the largest $s_{min}$ possible within these constraints
in order to minimize the processing time of the MC integration program.
For our study, we found a value of $s_{min}=1\,{\rm GeV}^{2}$ to
be appropriate for studying the $t$-channel single top process. For
comparison, in Ref.~\cite{Cao:2004ap}, we found that a value of
$s_{min}=5\,{\rm GeV}^{2}$ is adequate for studying the $s$-channel
single top process. A detailed discussion of the phase space slicing
method can be found in Ref.~\cite{Cao:2004ky}.

\begin{figure}
\includegraphics[%
  scale=0.5]{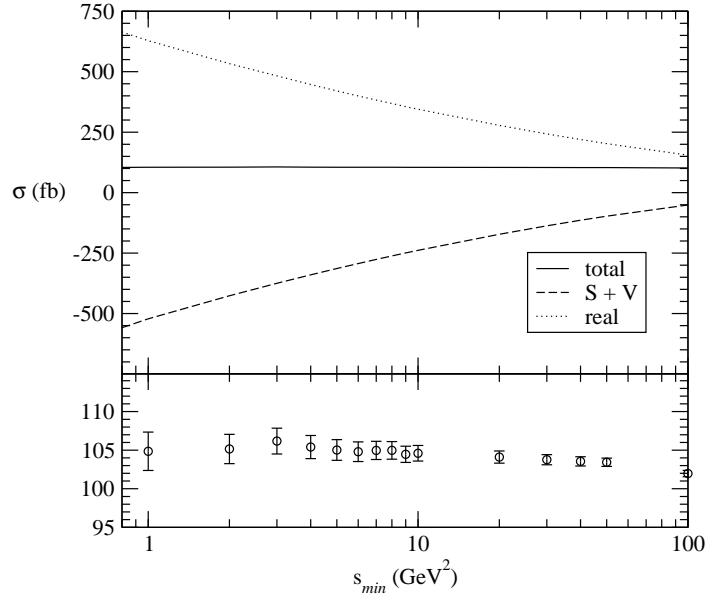}

\caption{The theoretical cutoff $s_{min}$ dependence of the inclusive $t$-channel
single top quark cross section at the Tevatron with $\mu_{R}=\mu_{F}=m_{t}$
for $m_{t}=178\,{\rm GeV}$. The decay branching ratio $t\rightarrow bW^{+}(\rightarrow e^{+}\nu)$
has been included.\label{fig:deps-smin}}
\end{figure}

Figure~\ref{fig:deps-smin} illustrates that the total $t$-channel
production cross section is indeed insensitive to $s_{min}$ for a
large range of $s_{min}$ values. The figure shows the sum of the
virtual and unresolved real corrections (${\rm s}+{\rm v}$) as well
as the resolved contribution (${\rm real}$) to the $t$-channel single
top process as a function of $s_{min}$. Although contributions from
the individual pieces vary, their sum (${\rm total}$) remains essentially
constant for a large range of $s_{min}$\emph{.} With the choices
of $\mu_{R}=\mu_{F}=m_{t}$ at the Tevatron, we obtain an inclusive
cross section for the $t$-channel single top ($t$ only) processes
(with $W$-boson decay branching ratio) of 104.8 femtobarn (fb), which
agrees with Ref.~\cite{Campbell:2004ch}. \emph{}

\subsection{Inclusive Cross Section\label{sub:Inclusive-Cross-Section}}

To facilitate the calculation, we divided the higher-order QCD corrections
into three separate gauge invariant sets: corrections from the light
quark line of top quark production (LIGHT), corrections from the heavy
quark line of top quark production (HEAVY), and corrections from the
top quark decay (TDEC). The explicit diagrams and definitions for
these three sets can be found in Ref.~\cite{Cao:2004ky}. The inclusive
cross section as well as the individual $\oalphas$ contributions
are listed in Table~\ref{tab:inclusive}. The effects of the finite
widths of the top quark and $W$-boson have been included. As one
expects, the dominant $\oalphas$ corrections come from the heavy
quark line, because the bulk part of the radiative corrections originating
from the light quark line have been absorbed into the definition of
the light quark PDFs. For the same reason pointed out in our previous
paper~\cite{Cao:2004ap}, the TDEC contribution is very small compared
to the other two.

\begin{table}
\begin{center}\begin{tabular}{c|c|c}
\hline 
&
Cross Section &
Fraction of\tabularnewline
&
(fb)&
NLO ($\%$)\tabularnewline
\hline
Born -level&
99.2&
94.6\tabularnewline
\hline
$O(\alpha_{s})$ HEAVY&
5.56&
5.31\tabularnewline
$O(\alpha_{s})$ LIGHT&
1.03&
0.98\tabularnewline
$O(\alpha_{s})$ TDEC&
-0.81&
-0.77\tabularnewline
\hline
$O(\alpha_{s})$ sum&
5.54&
5.28\tabularnewline
\hline 
NLO&
104.8&
\tabularnewline
\hline
\end{tabular}\end{center}

\caption{Inclusive $t$-channel single top production cross section for different
subprocesses, including the top quark decay branching ratio $t\rightarrow bW^{+}(\rightarrow e^{+}\nu)$.
\label{tab:inclusive}}
\end{table}

\subsection{Top Quark Mass and Renormalization/Factorization Scale Dependence\label{sub:Top-Quark-Mass}}

In order to test the SM and measure the CKM matrix element $V_{tb}$,
one needs an accurate prediction of single top quark production and
decay so as to reduce the theoretical uncertainties. Besides the top
quark mass, the choices of renormalization and factorization scales
also contribute to the uncertainty of the theoretical prediction.
The renormalization scale $\mu_{R}$ is introduced when redefining
the bare parameters in terms of the renormalized parameters, while
the factorization scale $\mu_{F}$ is introduced when absorbing the
collinear divergences into parton distribution functions. Therefore,
both $\mu_{R}$ and $\mu_{F}$ are unphysical and the final predictions
should not depend on them. However, since we work at a fixed order
in perturbation theory, we indeed see a dependence of the predicted
cross section on $\mu_{R}$ and $\mu_{F}$. The change due to varying
the scale is formally of higher order. Since the single top quark
rate is small at the Tevatron, it is important to reduce the scale
uncertainty in order to compare the theory prediction with experimental
data. Here, we examine the top quark mass and the scale dependence
of the $t$-channel production cross section.

\begin{figure}
\includegraphics[%
  scale=0.6]{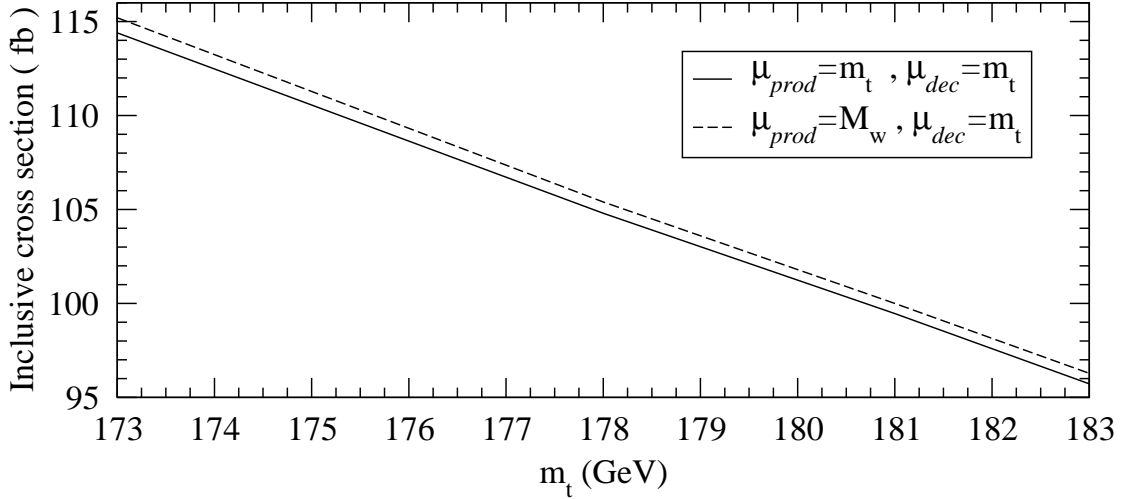}

\caption{Top quark mass dependence as well as renormalization and factorization
scale dependence of the inclusive $t$-channel single top quark cross
section at the Tevatron. The decay branching ratio of $t\to bW^{+}(\to e^{+}\nu)$
has been included. \label{fig:deps-mt}}
\end{figure}

As shown in Fig.~\ref{fig:deps-mt}, the cross section changes by
about $\pm9\%$ when the top quark mass $m_{t}$ is varied by its
current uncertainty of about $\mp5\,{\rm GeV}$ around 178 GeV. It
can also be seen that measuring the top quark mass to an uncertainty
of $1-2\,{\rm GeV}$ will reduce the theoretical uncertainty on the
single top cross section correspondingly. The reduction in uncertainty
on the $t$-channel single top production rate will improve the measurement
of the CKM matrix $V_{tb}$. 

\begin{figure}
\includegraphics[%
  scale=0.6]{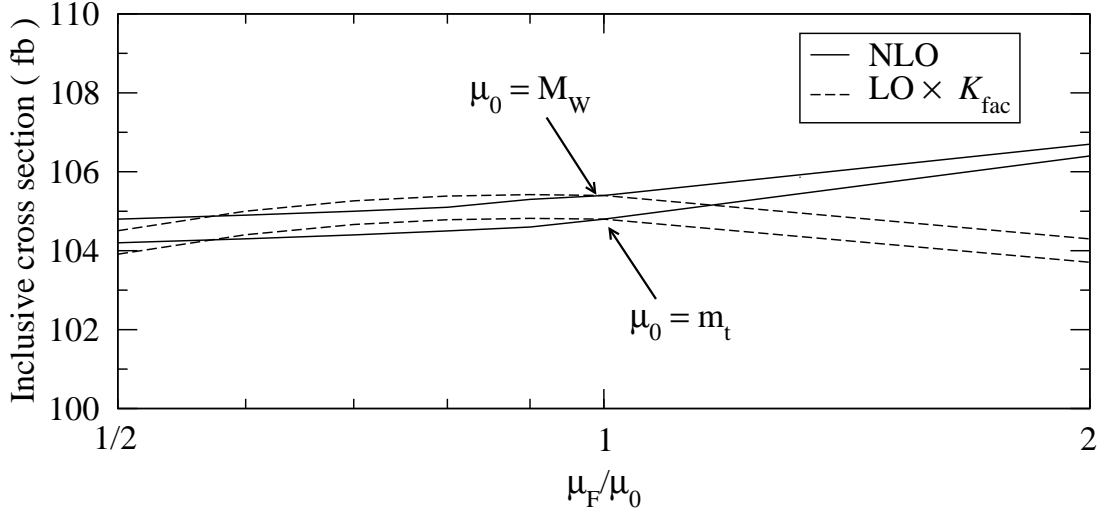}

\caption{Inclusive $t$-channel single top quark production cross section
at the Tevatron for $m_{t}=178$ GeV, versus the ratio of the factorization
scale $\mu_{F}$ to its typical value $\mu_{0}$, where $\mu_{0}=m_{t}$
(solid line) and $\mu_{0}=M_{W}$ (dashed line), respectively. The
decay branching ratio of $t\to bW^{+}(\to e^{+}\nu)$ has been included.
\label{fig:varyscale}}
\end{figure}

In order to examine the scale dependence of the $t$-channel single
top production rate, we show in Fig.~\ref{fig:deps-mt} the results
of two typical scales: one is the top quark mass ($\mu_{F}=\mu_{R}^{prod}=m_{t}$),
shown as the solid line, the other is the $W$ boson mass ($\mu_{F}=\mu_{R}^{prod}=M_{W}$),
shown as the dashed line. For the decay of top quark, we take $\mu_{R}^{dec}=m_{t}$,
which gives similar results as the choice of $\mu_{R}^{dec}=M_{W}$.
The band constrained by these two $\mu_{F}$ scales represents a range
of uncertainty due to the NLO predictions. The usual practice for
estimating the yet-to-be calculated higher order QCD correction to
a perturbative cross section is to vary around the typical scale by
a factor of 2, though in principle the ``best'' scale to be used for
estimation cannot be determined without completing the higher order
calculation. In Fig.~\ref{fig:varyscale} we show the total cross
section of the $t$-channel single top production for a range of the
scale $\mu_{F}$ which, for simplicity, is set to be equal to $\mu_{R}$.
We have multiplied the LO cross sections by a constant factor (the
``K-factor'') of 1.04 (shown as the dashed line) in order to compare
to the NLO ones (shown as the solid-line). It is clear that the NLO
calculation reduces the scale dependence. For example, when the scale
is changed from $\mu_{F}/2$ to $2\mu_{F}$ around $\mu_{F}=m_{t}$,
the LO rate varies by $-0.9\%$ to $-1.0\%$ while the NLO rate varies
only by $-0.6\%$ to $+1.3\%$. Similar results also hold for varying
the scale around $\mu_{F}=M_{W}$. 

\section{Single Top Acceptance Studies\label{sec:Single-Top-Acceptance}}

In this section we explore the final state objects of $t$-channel
single top quark events. Although the $W$-boson from the top quark
decay can decay in both leptonic and hadronic modes, we focus only
on the leptonic decay mode in this study because the all-jet production
mode of single top quark events is difficult to observe experimentally.
Therefore, the signature of $t$-channel single top quark events which
is accessible experimentally consists of one charged lepton, missing
transverse energy, together with two or three jets. Since we are studying
the effects of NLO QCD radiative corrections on the production rate
and the kinematic distributions of single top quark events at the
parton level in this paper, we do not include any detector effects,
such as jet energy resolution or $b$-tagging efficiency. Only an
approximation of kinematic acceptances of a generic detector are considered.
Since the Tevatron is $p\bar{p}$ collider, and $p\bar{p}$ is a CP-even
state, the production cross sections for $b\bar{t}(j)$ at the Tevatron
are the same as those for $t\bar{b}(j)$ (when ignoring the small
CP violation effect induced by the CKM mixing matrix in quarks). For
simplicity we therefore only consider distributions of $t$-channel
single top quark events in this work in which a $t$ quark (not including
$\bar{t}$) decays into $W^{+}(\rightarrow\ell^{+}\nu)$ and a $b$
quark. In this section, we first present the event topology of the
$t$-channel single top process, and then introduce a jet finding
algorithm and the various kinematical cuts in order to study the acceptance.

\subsection{Event Topology\label{sub:Event-Topology}}

\begin{figure}
\includegraphics[%
  width=0.60\linewidth,
  keepaspectratio]{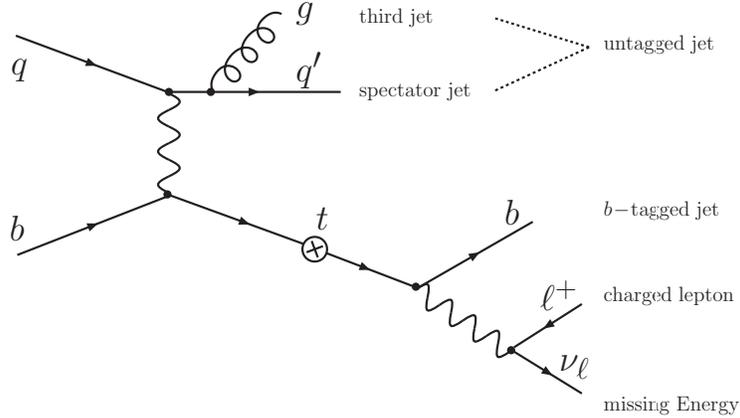}

\caption{Pictorial illustration of the notation used in this paper.\label{fig:notation}}
\end{figure}

At tree level, the collider signature of the $t$-channel single top
process includes two jets (one $b$-tagged jet from the $b$~quark
from the top quark decay, and one non-$b$-tagged jet from the light
quark), one charged lepton, and missing transverse energy ($\met$)
in the final state. This signature becomes more complicated beyond
tree level, as Fig.~\ref{fig:notation} indicates. The light quark
jet is also called {}``spectator jet'', and the label {}``untagged
jet'' refers to all jets which do not contain a $b$~or $\bar{b}$~quark.

\begin{figure}
\includegraphics[%
  width=0.60\linewidth,
  keepaspectratio]{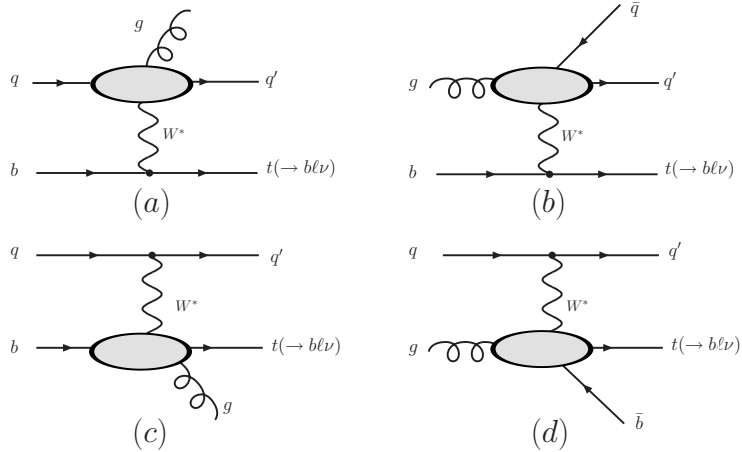}

\caption{Representative diagrams of the real emission corrections for the
$t$-channel single top process: (a) and (b) represent the real radiative
corrections to the LIGHT quark line, while (c) and (d) represent the
real radiative corrections to the HEAVY quark line. The NLO QCD corrections
are indicated by the large shaded ellipse. Detailed Feynman diagrams
can be found in Ref.~\cite{Cao:2004ky}.\label{fig:real_tchan}}
\end{figure}

At NLO, besides the charged lepton and $\met$, there may be two jets
(one $b$-tagged jet and one untagged jet) as for the Born-level,
or there may be three jets. The flavor composition of the three-jet
final state depends on the origin of the third jet. When it is a gluon
or anti-quark (cases (a-c) in Fig.~\ref{fig:real_tchan}), there
is one $b$-tagged jet and two untagged jets. When it is a $\bar{b}$
quark (case (d) in Fig.~\ref{fig:real_tchan}, also called $W$-gluon
fusion), there are two $b$-tagged jets and one untagged jet. Therefore,
prescriptions are needed to identify the $b$~jet from the top quark
decay and the light quark jet produced with the top quark.

In our study, we differentiate the following three cases:

\begin{enumerate}
\item Born-level-type exclusive two-jet events (containing the $b$~quark
and the light quark):\\
In Born-level type events, i.e. without an additional hard jet, the
final state is given by the $b$~jet from the top quark decay and
the light quark jet produced with the top quark. Experimentally, this
configuration is easily reconstructed because the $b$-tagged jet
is identified as the $b$~quark, and the untagged jet is assigned
to the spectator jet. We note that in our phenomenology study to be
presented in Sec.~\ref{sec:EventDistr} we shall concentrate on inclusive
two-jet events and exclusive three-jet events.
\item Exclusive three-jet events with one $b$~jet (containing the $b$~quark,
the light quark, and a gluon or anti-quark):\\
When the radiated gluon is reconstructed as a separate jet in the
LIGHT, TDEC, or HEAVY corrections, the final state contains one $b$~jet
and two other jets. As shown later, the transverse momentum and energy
differences can be used to separate the spectator jet from the gluon
jet (third jet). The transverse momentum of the spectator jet, which
comes from the initial quark ($q$) after emitting the effective $W$-boson,
peaks around $\sim M_{W}/2$, thus its energy is large. In comparison,
the transverse momentum of the gluon jet is small. Due to the collinear
enhancement, the resolved gluon jet prefers to move along the beam
line direction at smaller transverse momentum ($p_{T}$) for both
the LIGHT and HEAVY corrections, and along the $b$~quark moving
direction for the TDEC correction. 
\item Exclusive three-jet events with one $b$~jet and one $\bar{b}$~jet
(containing the $b$~quark, the light quark, and a $\bar{b}$~quark):\\
This final state is produced in the $W$-gluon fusion process, case
(d) of Fig.~\ref{fig:real_tchan}. In this case, the spectator jet
can be uniquely identified, but the $b$~jet from the top decay cannot
because heavy flavor tagging algorithms do not distinguish $b$~jets
from $\bar{b}$~jets experimentally. Although the likelihood of tagging
two $b$-jets is smaller than that of tagging a single $b$-jet, requiring
two $b$-tagged jets will suppress the QCD and $W$+jets backgrounds
significantly. We again can use transverse momentum differences to
separate $b$~jets from $\bar{b}$~jets. The $b$~jet from the
top quark decay has a transverse momentum peaking around $m_{t}/3$,
while the $\bar{b}$~jet from gluon splitting tends to move along
the beam line (the gluon-moving direction) due to the collinear enhancement
and is much softer. These kinematic differences also enable us to
separate the $t$-channel process from the $s$-channel process, recalling
that in the $s$-channel process, both the $b$~jet and the $\bar{b}$~jet
are preferentially produced at central rapidity and large $p_{T}$.
\end{enumerate}
The unique signature of the $t$-channel single top process is the
spectator jet in the forward direction, which can help to suppress
the copious backgrounds, such as $Wb\bar{b}$ and $t\bar{t}$ production.
Studying the kinematics of this spectator jet is important in order
to have a better prediction of the acceptance of $t$-channel single
top quark events and of the distribution of several important kinematic
variables. In this work, we study the impact of the NLO QCD corrections
on the kinematic properties of the spectator jet. As pointed out in
Ref.~\cite{Yuan:1989tc}, in the effective-$W$ approximation, a
high-energy $t$-channel single top quark event is dominated by a
longitudinal $W$~boson and the $b$~quark fusion diagram. It is
the same effective longitudinal $W$~boson that dominates the production
of a heavy Higgs boson at high energy colliders via the $W$-boson
fusion process. For a heavy SM Higgs boson, the longitudinal $W$~boson
fusion process dominates the Higgs boson production rate. Therefore,
it is also important to study the kinematics of the spectator jet
in $t$-channel single top quark events in order to have a better
prediction for the kinematics of Higgs boson events via the $WW$
fusion process, i.e. $q\bar{q}(WW)\to Hq^{\prime}\bar{q}^{\prime}$
at hadron colliders.

\subsection{Acceptance\label{sub:Acceptance}}

In order to meaningfully discuss the effects of gluon radiation in
single top quark events, we must define a jet as an infrared-safe
observable. In this study, we adopt the cone-jet algorithm~\cite{Alitti:1990aa}
as explained in our previous work~\cite{Cao:2004ap}. More specifically,
we adopt the $E$-scheme cone-jet approach (4-momenta of particles
in a cone are simply added to form a jet) with radius $R=\sqrt{\Delta\eta^{2}+\Delta\phi^{2}}$
in order to define $b$, $q$ and possibly extra $g$, $\bar{q}$,
or $\bar{b}$ jets, where $\Delta\eta$ and $\Delta\phi$ are the
separation of particles in the pseudo-rapidity $\eta$ and the azimuthal
angle $\phi$, respectively. For reference, we shall consider both
$R=0.5$ and $R=1.0$. The same $R$-separation will also be applied
to the separation between the lepton and each jet.

The kinematic cuts imposed on the final state objects are:\begin{eqnarray}
P_{T}^{\ell}\ge15\,{\rm GeV} & , & \left|\eta_{\ell}\right|\le\eta_{\ell}^{max},\nonumber \\
\met\ge15\,{\rm GeV} & ,\nonumber \\
E_{T}^{j}\ge15\,{\rm GeV} & , & \left|\eta_{j}\right|\le\eta_{j}^{max},\nonumber \\
\Delta R_{\ell j}\ge R_{cut} & , & \Delta R_{jj}\ge R_{cut},\label{eq:cuts}\end{eqnarray}
where the jet cuts are applied to both the $b$- and light quark jets
as well as any gluon or antiquark jet in the final state. Two lepton
pseudo-rapidity cuts are considered here: a loose version with $\eta_{\ell}^{max}=2.5$
and a tight version with $\eta_{\ell}^{max}=1.0$. Similarly, loose
and tight cuts are also considered for the jet pseudo-rapidity, $\eta_{j}^{max}=3.0$
and $\eta_{j}^{max}=2.0$, respectively. The minimum transverse energy
cuts on the lepton as well as the jets is 15~GeV. Each event is furthermore
required to have at least one lepton and two jets passing all selection
criteria. The cut on the separation in $R$ between the lepton and
the jets as well as between different jets is given by $R_{cut}$
which is is chosen to be $0.5$ (or $1.0$).

\begin{table}
\begin{center}\begin{tabular}{c|c|c|c|c|c|c}
\hline 
\multicolumn{2}{c|}{$\sigma$ {[}fb{]} }&
 LO &
 NLO &
Heavy&
Light&
Decay\tabularnewline
\hline
Tevatron (a) &
 $tq+tqj$ &
65.6&
64.0&
4.9&
-3.4&
-0.59\tabularnewline
&
$tq+tqj$, $\, E_{Tj}>30\,{\textrm{GeV}}$&
46.3&
44.0&
4.2&
-3.3&
-1.7\tabularnewline
&
 $tqj$&
&
25.3&
15.5&
5.1&
4.8\tabularnewline
&
$tqj,\, E_{Tj}>30\,{\textrm{GeV}}$&
&
7.4&
5.4&
1.3&
0.75\tabularnewline
\hline
Tevatron (b) &
 $tq+tqj$&
56.8&
48.1&
-0.45&
-4.4&
-2.2\tabularnewline
&
$tq+tqj$, $\, E_{Tj}>30\,{\textrm{GeV}}$&
40.1&
33.3&
0.0&
-3.3&
-2.1\tabularnewline
&
 $tqj$&
&
14.8&
10.8&
2.2&
1.7\tabularnewline
&
 $tqj,\, E_{Tj}>30\,{\textrm{GeV}}$&
&
4.5&
3.7&
0.58&
0.20\tabularnewline
\hline
Tevatron (c)&
$tq+tqj$ &
31.1&
34.0&
5.8&
-2.8&
0.82\tabularnewline
&
$tq+tqj$, $\, E_{Tj}>30\,{\textrm{GeV}}$&
24.4&
24.2&
3.8&
-2.7&
-0.7\tabularnewline
&
$tqj$&
&
11.4&
6.7&
2.4&
2.3\tabularnewline
&
$tqj,\, E_{Tj}>30\,{\textrm{GeV}}$&
&
3.9&
2.8&
0.74&
0.41\tabularnewline
\hline
\end{tabular}\end{center}

\caption{The $t$-channel single top production cross section at the Tevatron
(in fb ) for different subprocesses under various cut scenarios: (a)
is the loose version with $\eta_{l}^{max}=2.5$, $\eta_{j}^{max}=3.0$,
and $R_{cut}=0.5$, (b) is the loose version with a larger jet clustering
cone size (and jet-lepton separation cut) of $R_{cut}=1.0$, and (c)
is the tight version of cuts with $\eta_{l}^{max}=1.0$, $\eta_{j}^{max}=2.0$,
and $R_{cut}=0.5$. The first two columns show the Born-level and
full NLO cross sections, the last three columns show the contributions
from the different $O(\alpha_{s})$ processes. The decay branching
ratio $t\rightarrow bW(\rightarrow e\nu)$ is included. \label{tab:total}}
\end{table}

In Table~\ref{tab:total}, we show the single top production cross
sections in fb for the loose and tight set of cuts listed in Eq.~(\ref{eq:cuts})
for the different subprocesses. A larger value for $R_{cut}$ reduces
the acceptance significantly mainly because more events fail the lepton-jet
separation cut. While this is only a 13\% reduction at the Born-level,
the difference grows to 25\% at NLO. Hence, a smaller cone size is
preferred in order to keep the acceptance at a high level. For events
with at least three jets, imposing a harder cut on the transverse
momentum of each jet ($E_{Tj}>30\,{\rm GeV}$) decreases the contribution
from the LIGHT and HEAVY corrections by a factor of 3 to 4, but the
contribution from the TDEC correction by more than a factor 6. This
large reduction occurs because the top quark mass sets the scale for
the top quark decay contribution rather than the invariant mass of
$tq^{\prime}$ system, resulting in a softer jet $E_{T}$ spectrum
for the TDEC correction (cf. Fig.~\ref{fig:ptetaJet3}, to be discussed
in Sec.~\ref{sub:Distributions-for-Three-jet}).

\begin{figure}
\subfigure[]{\includegraphics[%
  width=0.40\linewidth,
  keepaspectratio]{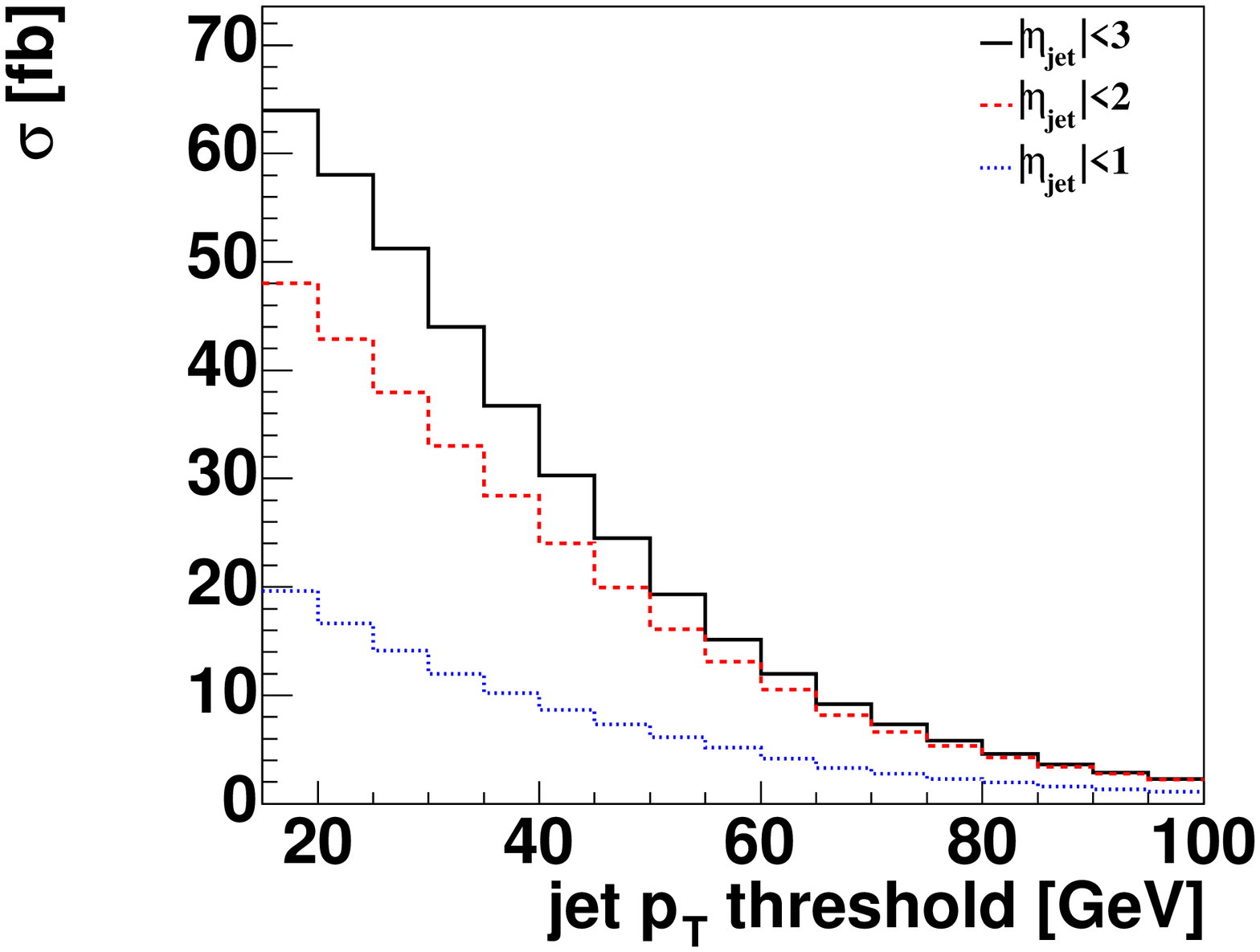}}\subfigure[]{\includegraphics[%
  width=0.40\linewidth,
  keepaspectratio]{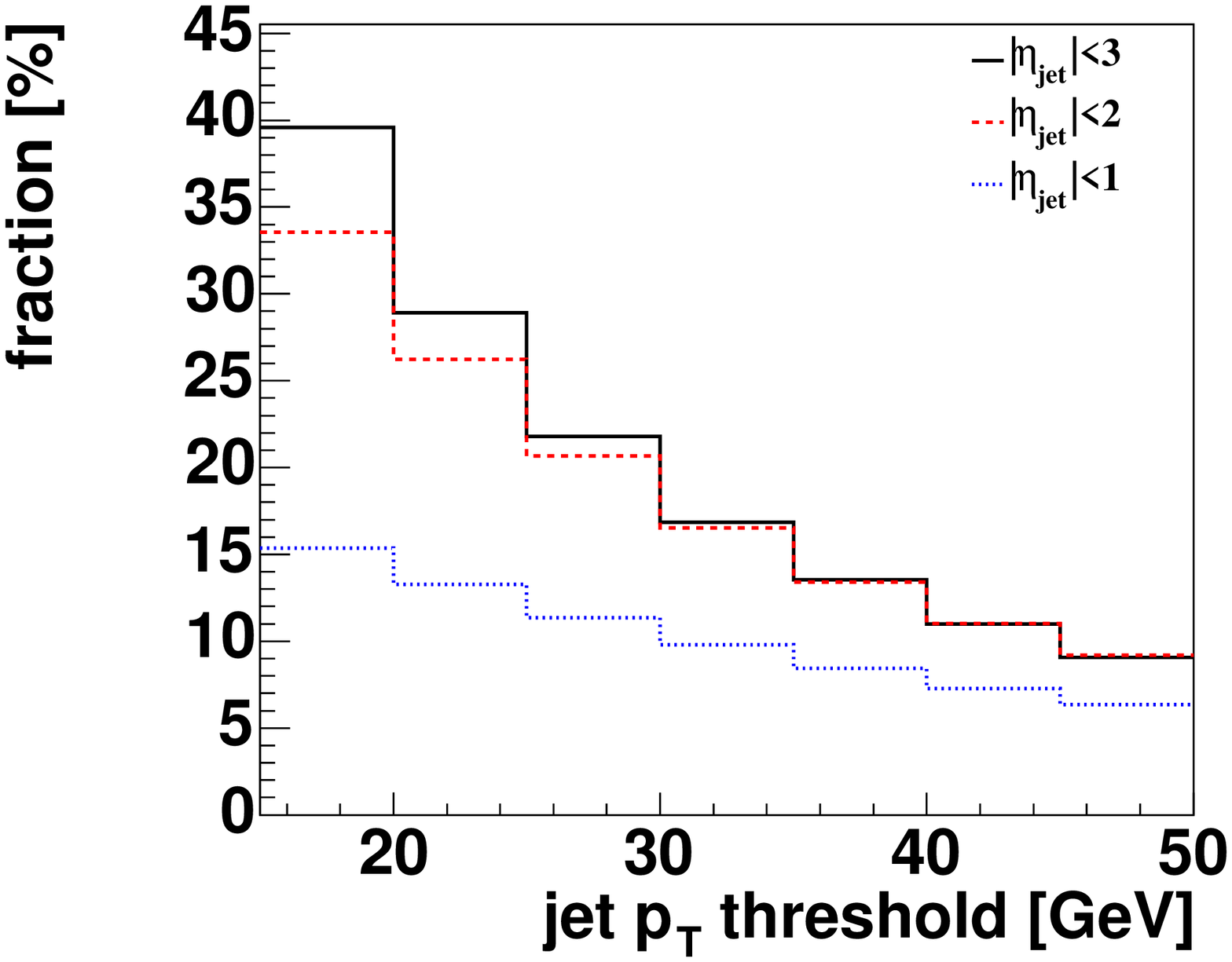}}

\caption{Cross section and fraction of three-jet events at NLO for varying
jet $p_{T}$ cuts, requiring only that $n_{jets}\geq2$, and not making
any cuts on the electron or neutrino. Shown is the total cross section
for events with two or three jets as a function of the jet $E_{T}$
cut for three different jet pseudo-rapidity cuts (a) and the fraction
of three-jet events as a function of the jet $p_{T}$ for different
jet pseudo-rapidity cuts (b). The lowest threshold considered is 15
GeV. \label{fig:njets_jet_pt}}
\end{figure}

Figure~\ref{fig:njets_jet_pt} shows how the observed cross section
changes as a function of the jet $E_{T}$ cut when applying the loose
set of cuts, including a requirement of there being at least two jets
in the event. The figure also shows the dependence of the fraction
of two-jet events and three-jet events on the jet $E_{T}$ cut. At
the Born-level, there are only two-jet events, whereas $O(\alpha_{s})$
corrections can produce an additional soft jet. The fraction of events
with these additional jets is low only for very high jet $E_{T}$
thresholds. For typical jet $E_{T}$ thresholds considered by experiments
of 15~GeV to 25~GeV these jets add a significant contribution. As
expected, the effect is not quite as large when only jets within a
very small $\eta$ range are considered because the extra jet typically
has higher $\eta$. In order to study $\oalphas$ effects it is thus
important to set the jet $\eta$ cut as high as possible and the jet
$E_{T}$ cut as low as possible. 

As mentioned before, the event rate for single top quark events is
small and it is important for experiments to maximize their acceptance.
We will thus use the loose set of cut values for the following discussion:
$\eta_{l}^{max}=2.5$, $\eta_{j}^{max}=3.0$, and $R_{cut}=0.5$,
$E_{Tj}^{min}=15$~GeV, cf. Eq.~(\ref{eq:cuts}).

\section{Single Top Quark Event Distributions\label{sec:EventDistr}}

In this section we examine the kinematic properties of single top
quark events. As discussed in the previous section, the signature
of the $t$-channel process includes at least one $b$-tagged jet,
one untagged jet, one charged lepton, and missing energy. At the Born-level,
it is straightforward to identify the single top final state using
the $b$-tagged jet and the reconstructed $W$~boson to reconstruct
the top quark and identifying the spectator jet as the light quark.
At NLO, however, an additional jet is radiated, which will complicate
the reconstruction of the top quark final state. First, this is because
the additional jet can be either a $b$-tagged jet or an untagged
jet. When it is the $b$-tagged jet, we need to select the correct
$b$~quark from the two possible jets in the finial state to reconstruct
the top quark. When it is the untagged jet, we need to select the
correct spectator jet. Second, the additional untagged jet can come
from either the production or the decay of the top quark. Production-stage
emission occurs before the top quark goes on shell and decay-stage
emission occurs only after the top quark goes on shell. In production
emission events, the $W$~boson and $b$~quark momenta will combine
to give the top quark momentum, while in the decay emission event
the gluon momentum must also be included in order to reconstruct the
top quark momentum properly. To find the best prescription for identifying
the correct $b$~jet and spectator jet, we first examine various
kinematic distributions of the final state particles. We then investigate
two top quark reconstruction prescriptions: using the leading $b$-tagged
jet and the best jet algorithm. Choosing the $b$-tagged jet prescription,
we can also improve the reconstruction efficiency for the $W$~boson.
We then study the effects of NLO corrections on distributions concerning
the reconstructed top quark, in particular spin correlations between
the final state particles. Finally, we explore the impact of the radiated
jet in exclusive three-jets events. We use only the loose set of cuts
to maximize the efficiency when examining the distributions and efficiencies
in detail.

\subsection{Final State Object Distributions\label{sub:Final-State-Object}}

\subsubsection{Charged Lepton and Missing Transverse Energy}

In this section we examine various kinematic distributions of final
state objects after event reconstruction and after applying the loose
set of cuts, cf. Table~\ref{tab:inclusive} and Eq.~(\ref{eq:cuts}).
We concentrate on inclusive two-jet events in this section because
they give more reliable infrared-safe predictions. 

\begin{figure}
\subfigure[]{\includegraphics[%
  width=0.35\linewidth,
  keepaspectratio]{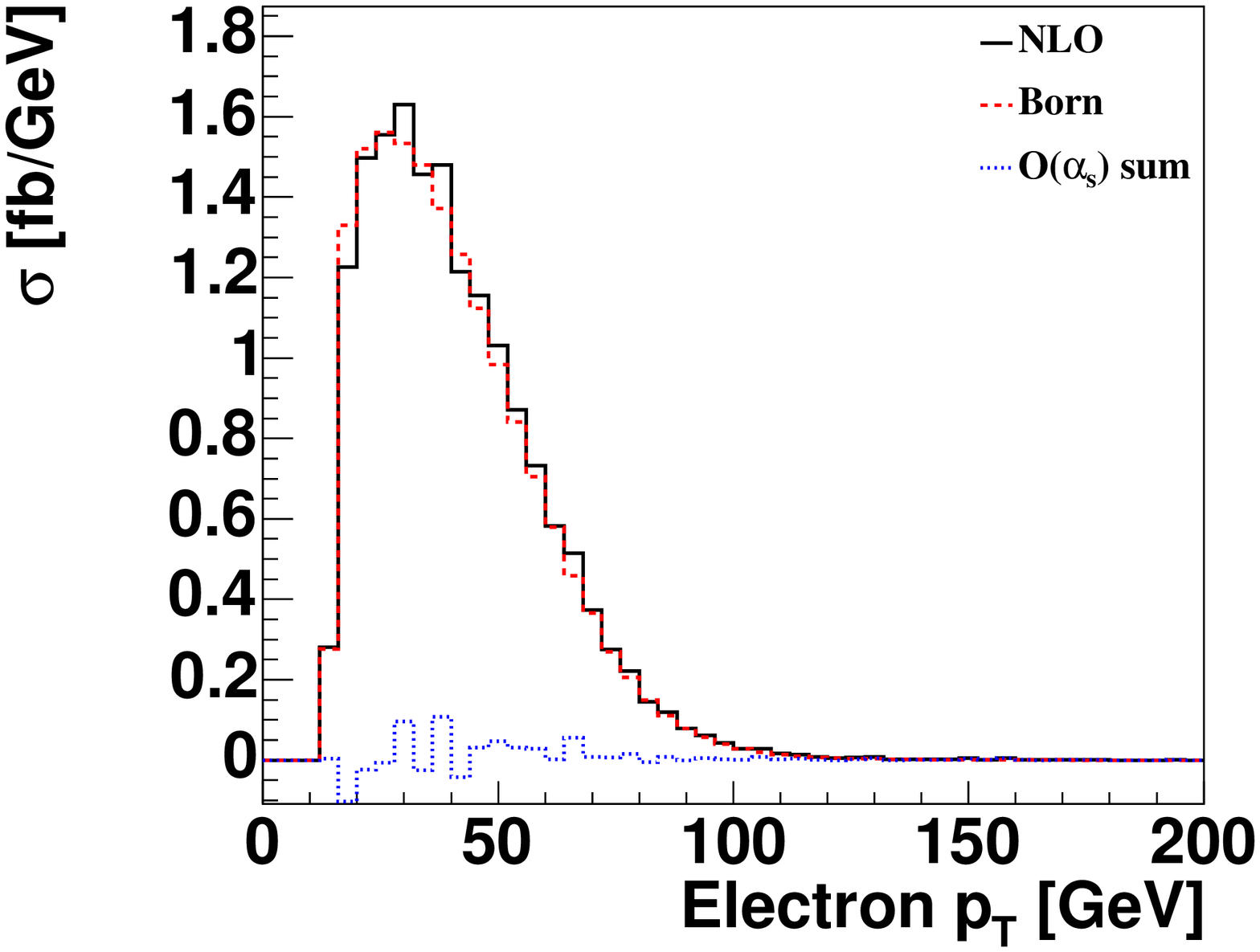}}\subfigure[]{\includegraphics[%
  width=0.35\linewidth,
  keepaspectratio]{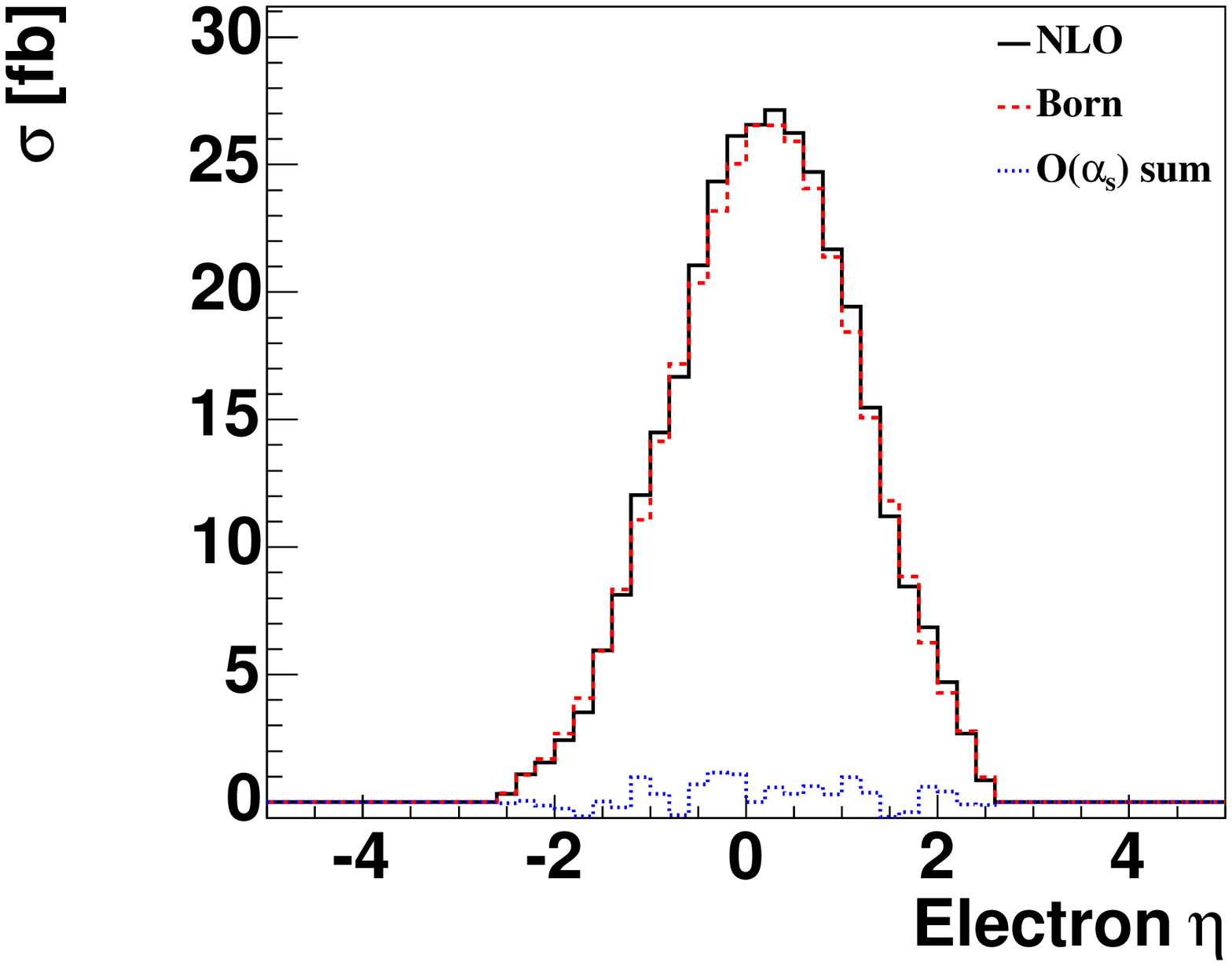}}\subfigure[]{\includegraphics[%
  width=0.35\linewidth,
  keepaspectratio]{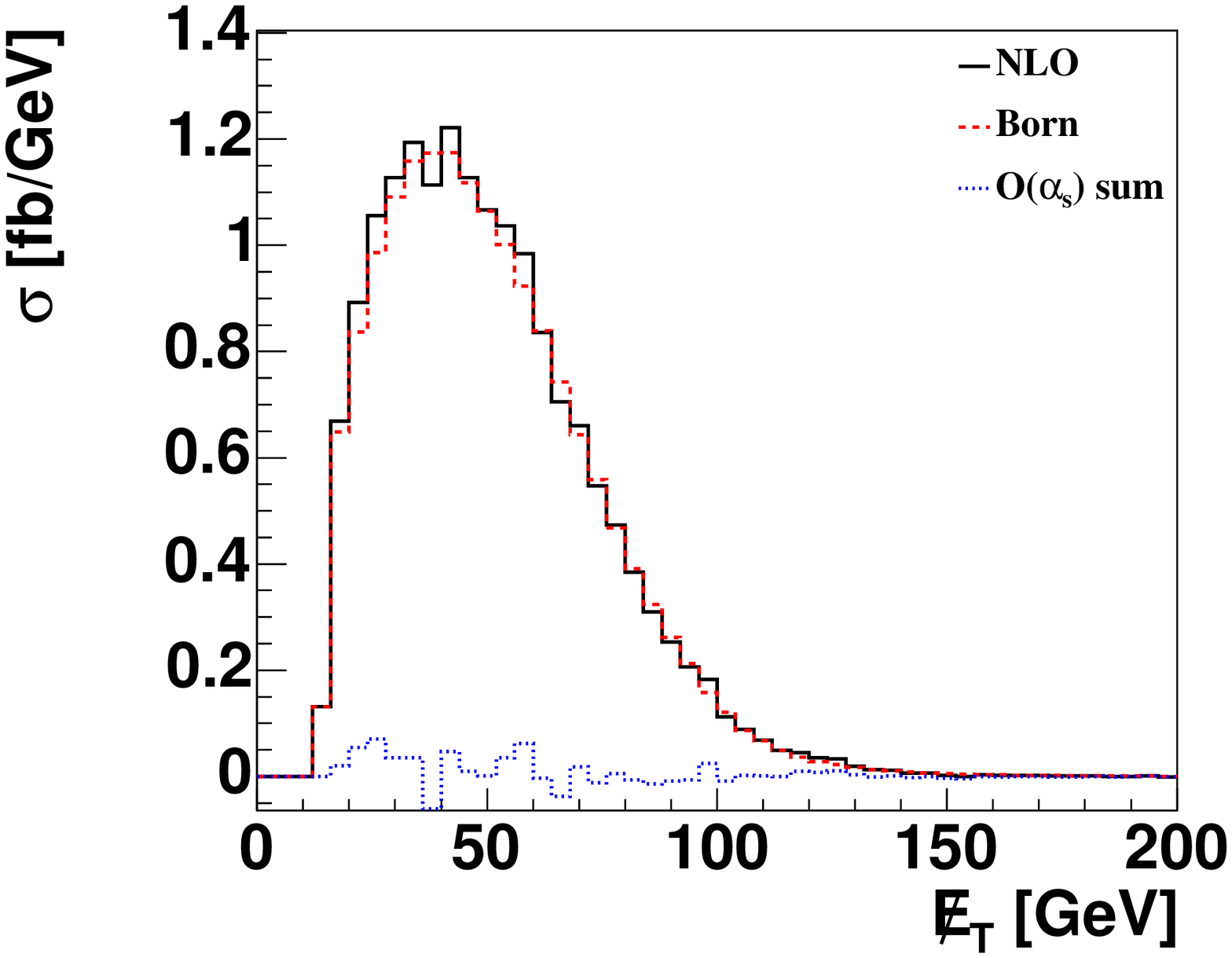}}

\caption{The transverse momentum $p_{T}$ (a) and pseudo-rapidity $\eta$
(b) distributions of the electron and the missing transverse energy
$\met$ (c) after selection cuts, comparing Born-level to $\oalphas$
corrections.\label{fig:pte-etae}}
\end{figure}

Figures~\ref{fig:pte-etae} (a) and (c) show the transverse momentum
of the electron and the missing transverse energy $\met$, respectively.
As expected, because they are leptons and not quarks, the change in
shape when going from Born-level to NLO is not very large for the
electron or the missing transverse energy. The pseudo-rapidity distribution
of the electron is given in Fig.~\ref{fig:pte-etae}(b). This distribution
widens at NLO, but again the effect is small because the overall $O(\alpha_{s})$
contribution to the event rate is small, cf. Table~\ref{tab:inclusive}.
We note that the peak position of the $\met$ distribution is at a
higher value than that of the electron $p_{T}$ because the neutrino
from the $W$-boson decay moves preferentially along the direction
of the top quark. This is due to the left-handed nature of the charged
current interaction and can easily be seen when examining the spin
correlations between the charged lepton and the top quark in the top
quark rest frame. We will comment more on this subject in Sec.~\ref{sub:Object-Correlations}.

\subsubsection{Spectator Jet}

\begin{figure}
\subfigure[]{\includegraphics[%
  width=0.35\linewidth,
  keepaspectratio]{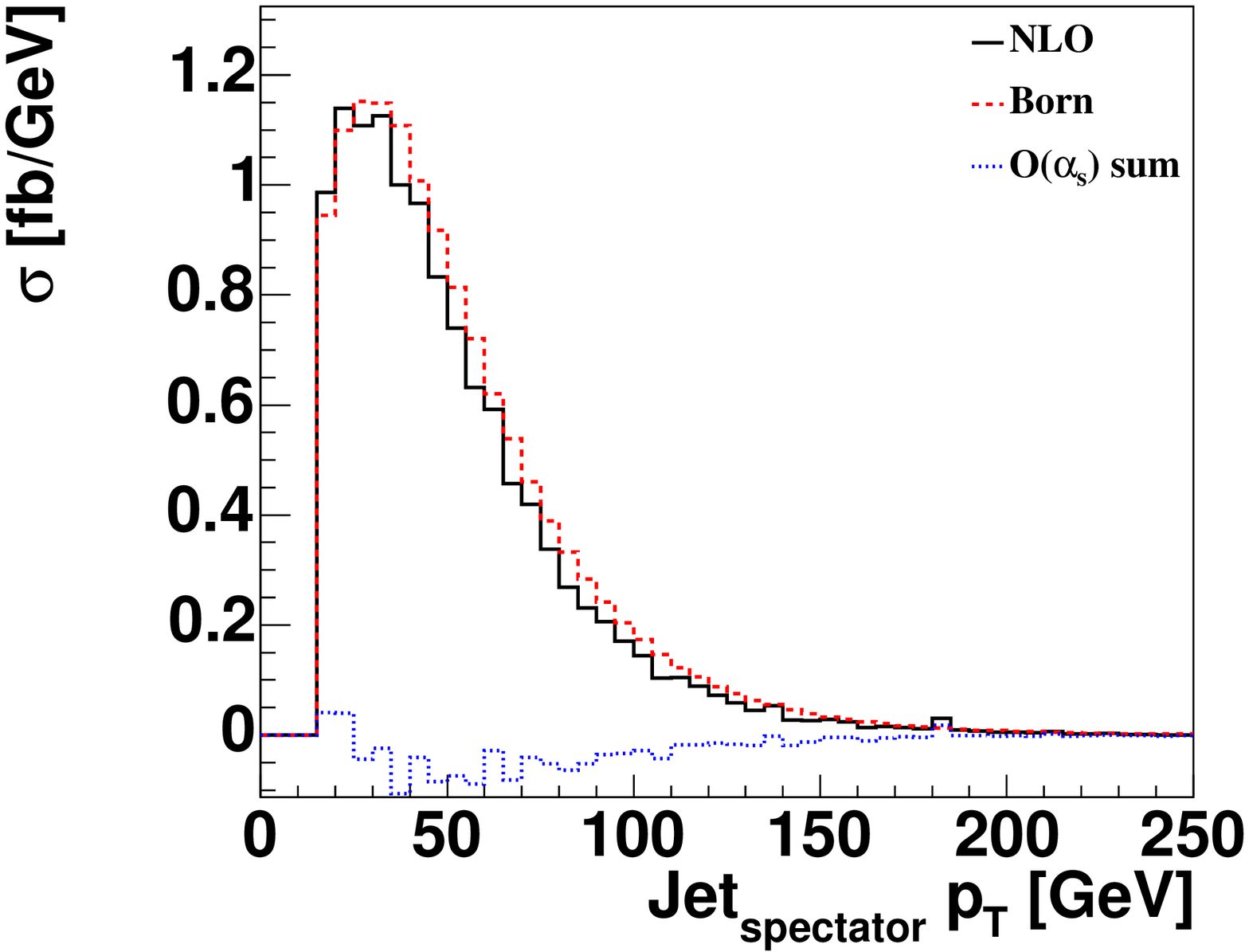}}\subfigure[]{\includegraphics[%
  width=0.35\linewidth,
  keepaspectratio]{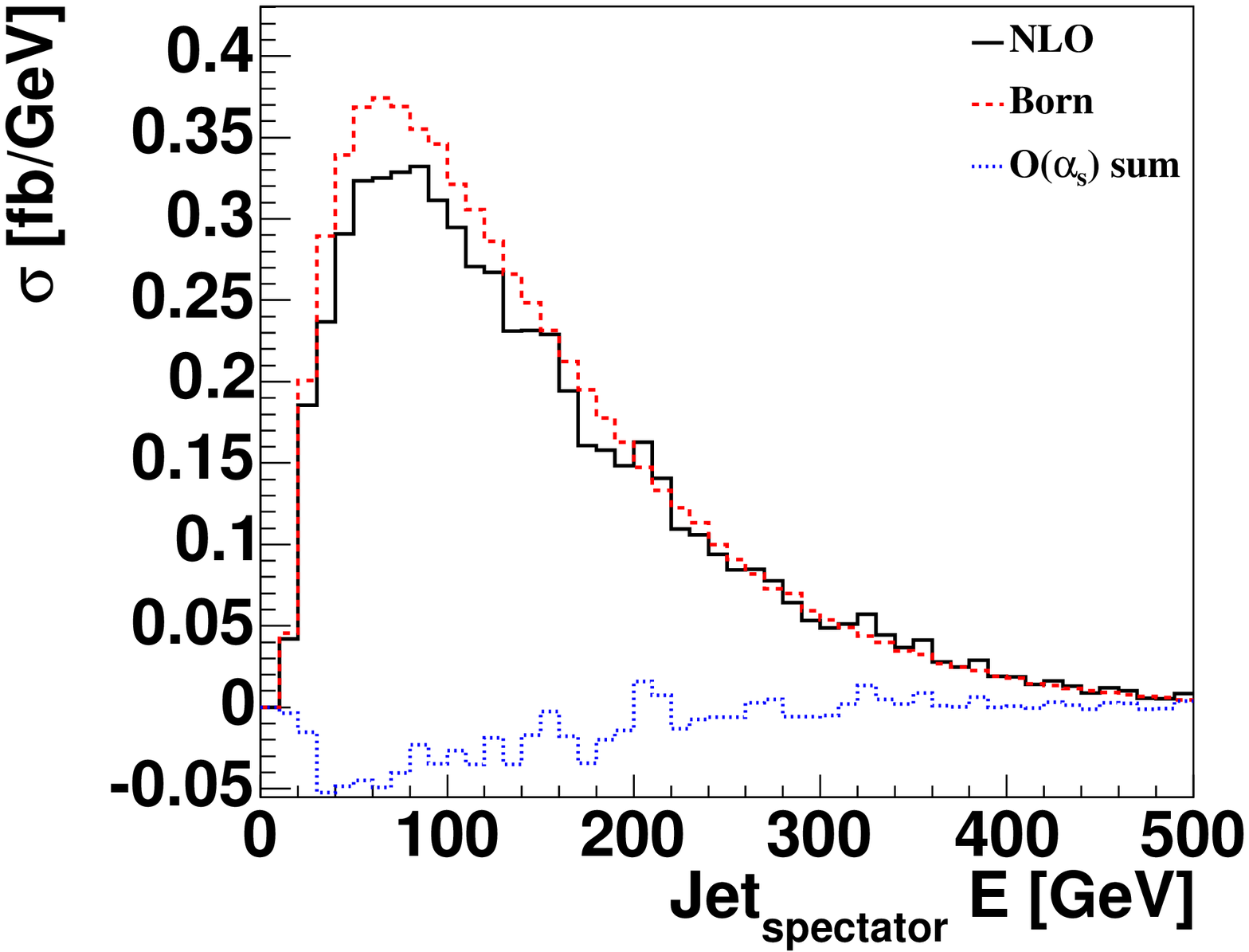}}\subfigure[]{\includegraphics[%
  width=0.35\linewidth,
  keepaspectratio]{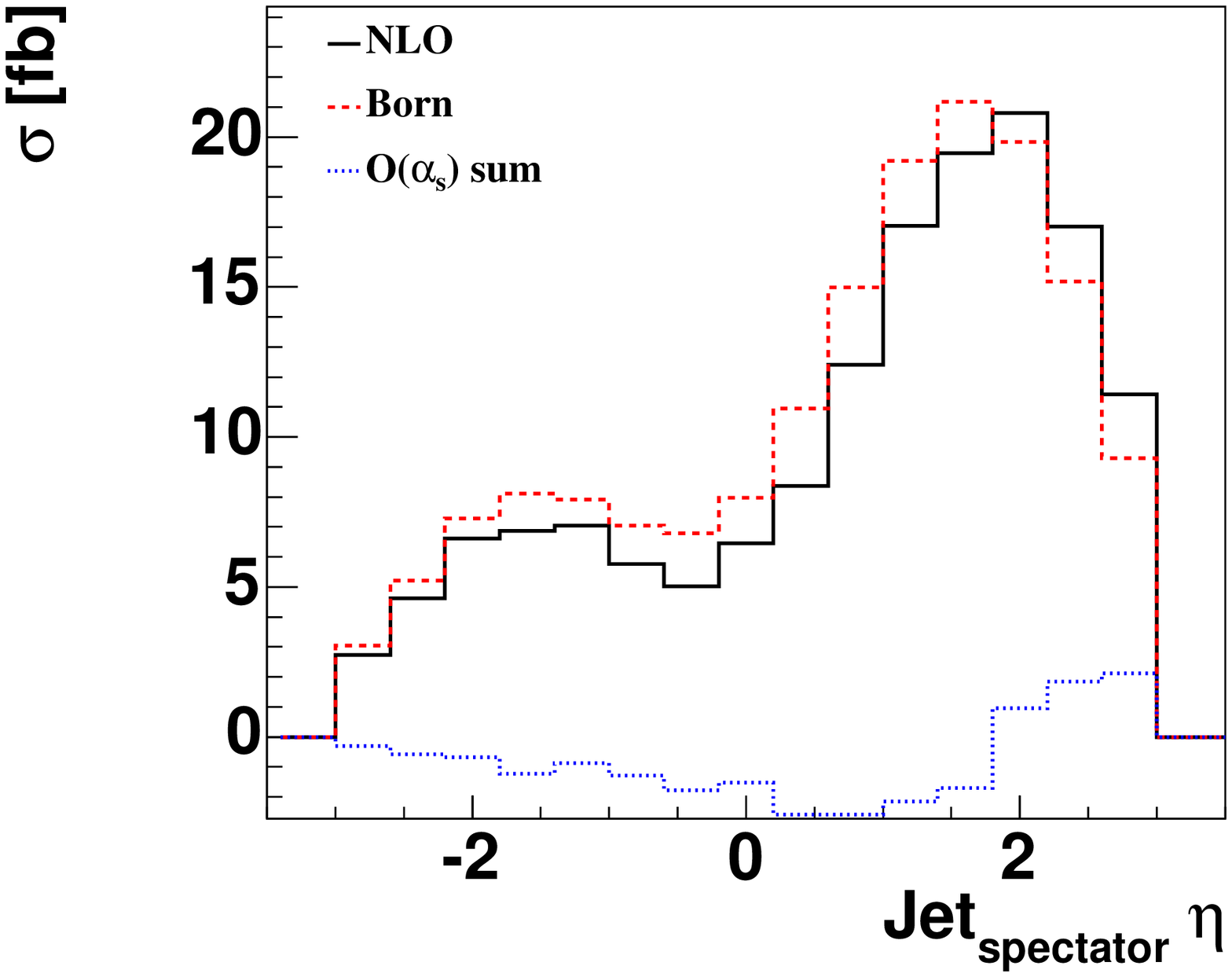}}

\caption{Transverse momentum $p_{T}$ (a), energy $E$ (b) and the pseudo-rapidity
$\eta$ (c) of the spectator jet after selection cuts, comparing Born-level
to $\oalphas$ corrections. \label{fig:spectator}}
\end{figure}

The differences between the $s$-channel and $t$-channel single top
processes at the Born-level are that the former has two $b$-taggable
jets in the final state (the $b$~quark and the $\bar{b}$~quark)
while the latter only has one $b$-taggable~jet and also has one
light quark jet. This untagged jet (spectator jet) is a unique feature
of the $t$-channel process which can be used to disentangle $t$-channel
single top quark events from the copious backgrounds. Therefore, its
kinematic distributions need to be well studied, especially the impact
of $O(\alpha_{s})$ corrections on the features which make this jet
a unique signature, such as the spectator jet pseudo-rapidity, cf.
Fig.~\ref{fig:spectator}.

\begin{figure}
\subfigure[]{\includegraphics[%
  width=0.40\linewidth,
  keepaspectratio]{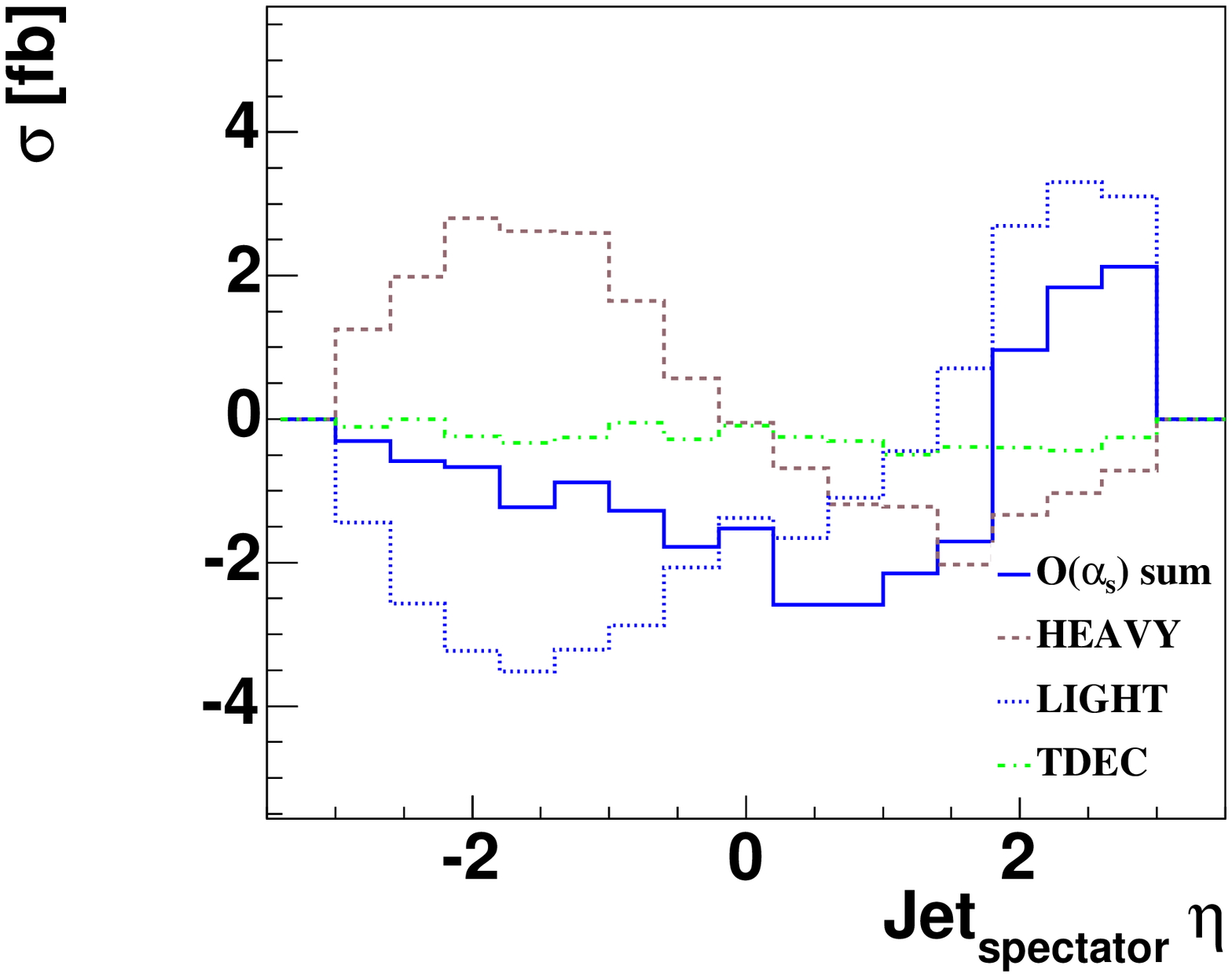}}\subfigure[]{\includegraphics[%
  width=0.40\linewidth,
  keepaspectratio]{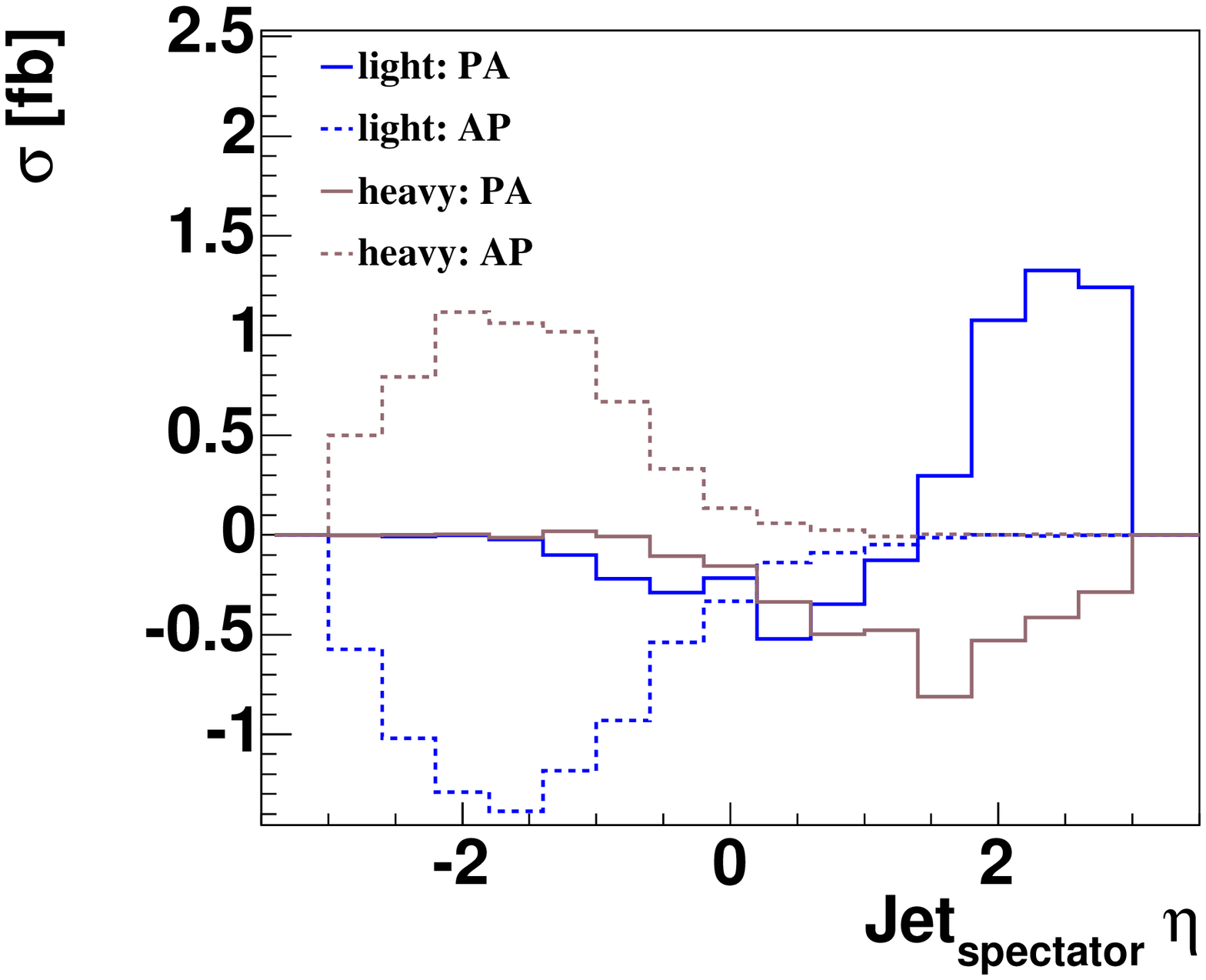}}

\caption{Each individual contribution of the $O(\alpha_{s})$ corrections
to the spectator jet pseudo-rapidity, summed (a), separately for the
case when the incoming up-type quark is from the proton and anti-proton
(b). Here, ``PA'' and ``AP'' denotes the initial state light quark
originating from proton and anti-proton, respectively.\label{fig:spectator_eta_nlo}}
\end{figure}

The pseudo-rapidity distribution of the spectator jet is asymmetric
because the Tevatron is a $p\bar{p}$ collider \cite{Yuan:1989tc}.
In order to produce a heavy top quark decaying to a positively charged
lepton, the valence quark from the proton is most important, implying
that the light quark will tend to move in the proton direction. We
define the positive $z$-direction to be the proton direction in the
laboratory frame, thus the pseudo-rapidity of the spectator jet will
tend be positive. Similarly, the spectator jet in an anti-top quark
event produced from the $t$-channel process will preferably be at
a negative pseudo-rapidity due to the large anti-up quark parton distribution
inside the antiproton. The $O(\alpha_{s})$ corrections shift the
spectator jet to even more forward pseudo-rapidities due to additional
gluon radiation. However, since the $O(\alpha_{s})$ corrections are
small compared to the Born-level contribution, the spectator jet pseudo-rapidity
distribution only shifts slightly. As Fig.~\ref{fig:spectator_eta_nlo}
shows, the LIGHT and HEAVY contributions have almost opposite behavior.
The former shifts the spectator jet to even higher pseudo-rapidities,
while the later shifts it more to the central rapidity region. This
behavior is due to two different effects, as illustrated in Fig.~\ref{fig:spectator_eta_nlo}
(b), in which ``PA'' denotes that the light quarks come from the proton
while the bottom quarks from the anti-proton and vice versa for ``AP''.
After separating the contributions by whether the light quark is from
the proton or the antiproton, it can be seen that the HEAVY corrections
shift the proton contribution down and the antiproton contribution
up due to the slight change in acceptance caused by the additional
jet. The LIGHT corrections show the opposite tendency. For the TDEC
contribution, all corrections have similar shapes and the sum of them
leaves the spectator jet pseudo-rapidity unchanged, as expected. After
summing the negative soft-plus-virtual corrections with the real emission
corrections, we obtain the result shown in Fig.~\ref{fig:spectator},
which shows that the $\oalphas$ correction shifts the spectator jet
to be in even more forward direction.

Besides its forward rapidity, the spectator jet also has large transverse
momentum. Since it comes from the initial state quark after emitting
the effective $W$~boson, the transverse momentum peaks around $\sim M_{W}/2$,
cf. Fig.~\ref{fig:spectator}. By comparison, the third jet is most
often much softer, we can thus use $p_{T}$ of the jet to identify
the the spectator jet when considering exclusive three-jet events.

\subsubsection{$b$ Jet}

\begin{figure}
\subfigure[]{\includegraphics[%
  width=0.40\linewidth,
  keepaspectratio]{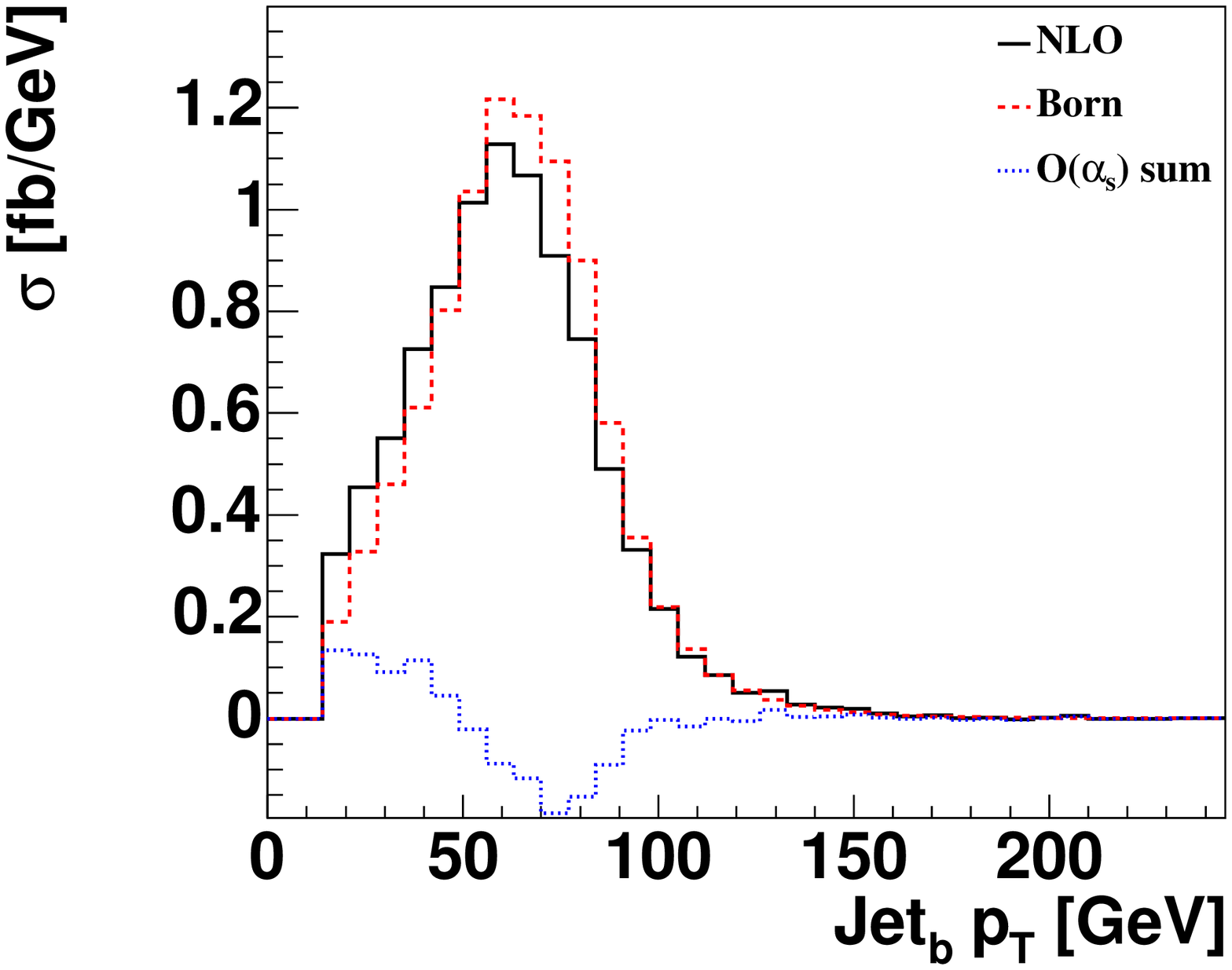}}\subfigure[]{\includegraphics[%
  width=0.40\linewidth,
  keepaspectratio]{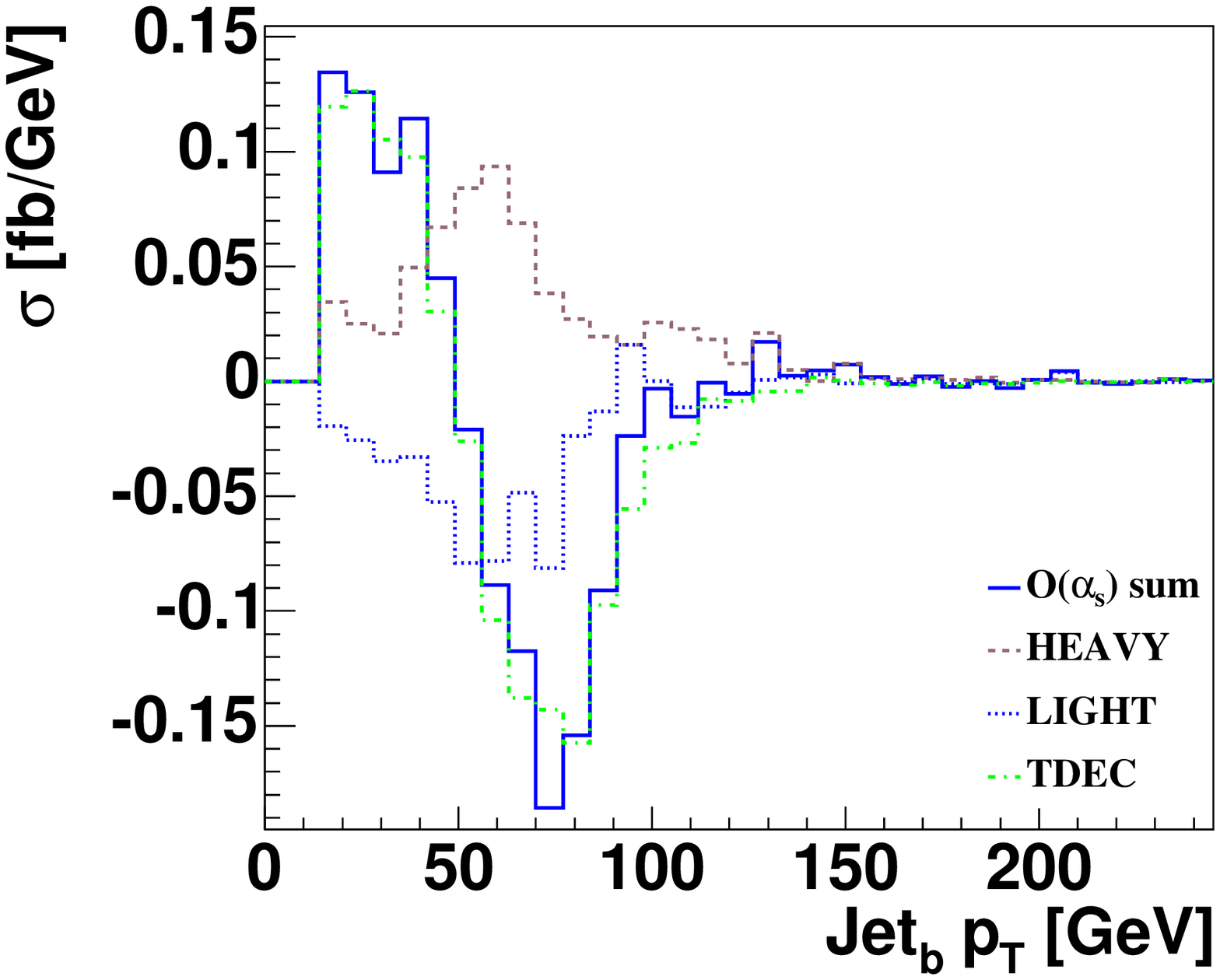}}

\caption{Transverse momentum $p_{T}$ of the $b$ jet after selection cuts,
comparing Born-level to $\oalphas$ corrections (a) and the individual
$\oalphas$ contributions (b).\label{fig:ptb}}
\end{figure}

Compared to the lepton and $\met$, the effects of the $\oalphas$
corrections on the reconstructed $b$~jet are more pronounced. Figure~\ref{fig:ptb}
shows a comparison of the $b$~jet $p_{T}$ distribution between
the Born-level and $\oalphas$ corrections. The $p_{T}$ distribution
of the $b$~jet is predominantly determined by the top quark mass
and therefore peaks at $\sim m_{t}/3$. The NLO QCD corrections broaden
the transverse momentum distribution and shift the peak position to
lower values. The location of the mean of the distributions depends
on the cuts that are applied because the different $\oalphas$ corrections
have different effects: in the case of the loose cuts it increases
from 64~GeV at the Born-level to 62~GeV at full NLO. The LIGHT corrections
shift the mean of the $b$~jet $p_{T}$ distribution up, the HEAVY
corrections leave it mostly unchanged, and the TDEC corrections tend
to shift it down. The $b$~jet $p_{T}$ distribution receives a large
contribution from the TDEC corrections, as expected. When a gluon
is radiated from the top quark decay, it tends to move along the $b$~jet
direction due to collinear enhancements and therefore shifts the $b$~jet
$p_{T}$ distribution to the small $p_{T}$ region, as shown in Fig.~\ref{fig:ptb}. 

\begin{figure}
\includegraphics[%
  width=0.40\linewidth,
  keepaspectratio]{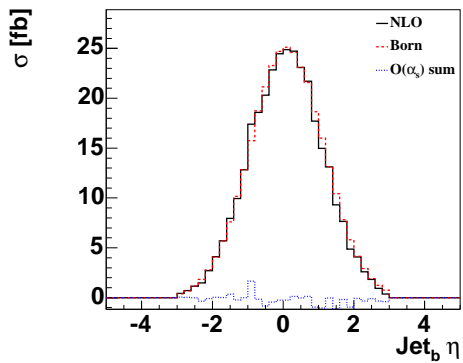}

\caption{Pseudo-rapidity $\eta$ of the $b$-jet after selection cuts, comparing
Born level to $\oalphas$ corrections. \label{fig:etab}}
\end{figure}

The $b$~jet pseudo-rapidity distribution is less affected by the
$\oalphas$ corrections, as can be seen in Fig.~\ref{fig:etab}.
The top quark is so heavy that it is mostly produced in the central
rapidity region and thus the $b$~jet from its decay also peaks around
a pseudo-rapidity of zero. The shape of the $b$~jet pseudo-rapidity
distribution remains almost unchanged compared to the Born-level because
it comes from the top quark decay. 

\begin{figure}
\subfigure[]{\includegraphics[%
  width=0.40\linewidth,
  keepaspectratio]{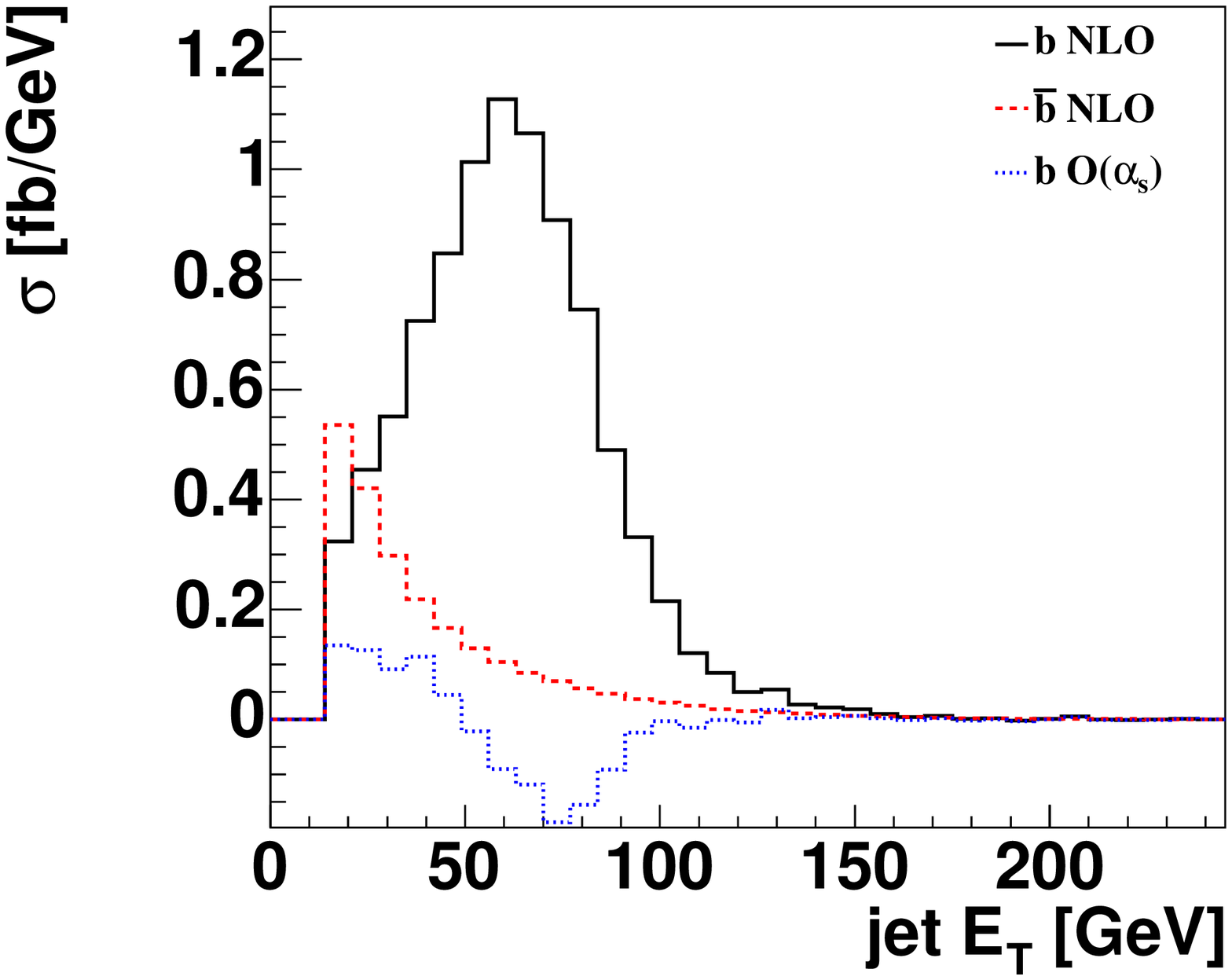}}\subfigure[]{\includegraphics[%
  width=0.40\linewidth,
  keepaspectratio]{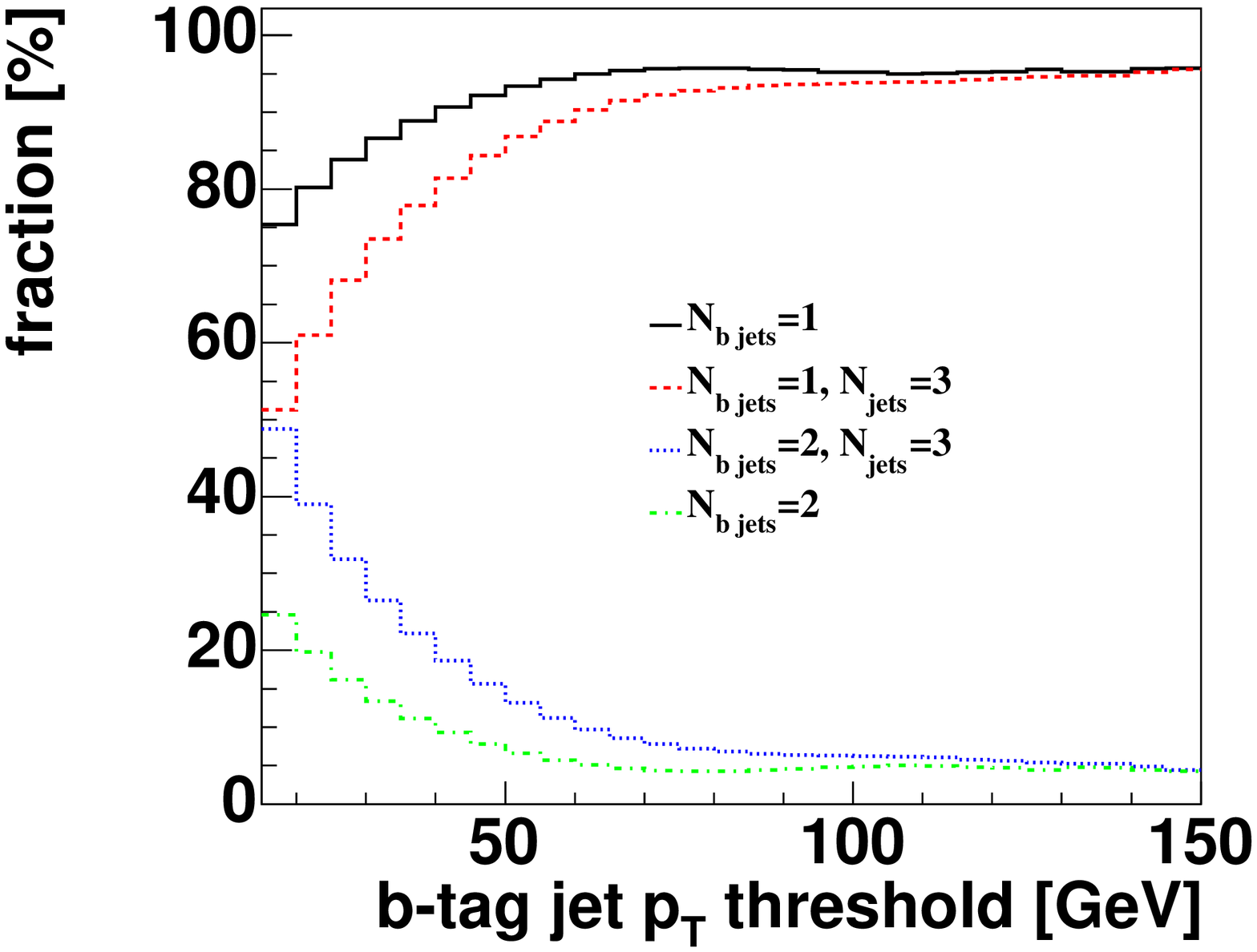}}

\caption{Momentum of the $b$ and $\bar{b}$ jets (a) and fraction of events
with one or two $b$-tagged jets as a function of the jet $p_{T}$
threshold, for both inclusive two-jet and exclusive three-jet events
(b).\label{fig:ptbbbar}}
\end{figure}

Figure~\ref{fig:ptbbbar} compares the momentum of the $b$~jet
and the $\bar{b}$~jet at NLO and examines the fraction of events
containing one or two $b$-tagged jets in the final state. Events
with both $b$~and $\bar{b}$~jets originate from the $W$-gluon
fusion subprocess, $qg\to q^{\prime}\bar{b}t(\to bW(\to\ell^{+}\nu))$.
While the fraction of events with two $b$-tagged jets is high for
low jet $p_{T}$, it drops quickly. For a jet $p_{T}$ above 50~GeV,
the fraction of events with both $b$~and $\bar{b}$~jets is less
than 5\%. Figure~\ref{fig:ptbbbar} also shows that for exclusive
three-jet events, the fraction of events with two $b$-tagged jets
is much higher. About half of the events contain two $b$-tagged jets
for the lowest threshold of 15~GeV, and the fraction only goes down
to 5\% when the jet $p_{T}$ reaches about 140~GeV.

\subsubsection{$H_{T}$ Distribution}

The impact that different $\oalphas$ corrections have on the $p_{T}$
of the jets is also reflected in event-wide energy variables such
as the total transverse energy in the event ($H_{T}$), defined as\begin{equation}
H_{T}=p_{T}^{lepton}+\met+\sum_{jets}p_{T}^{jet}.\label{eq:HT}\end{equation}
\begin{figure}
\subfigure[]{\includegraphics[%
  width=0.40\linewidth,
  keepaspectratio]{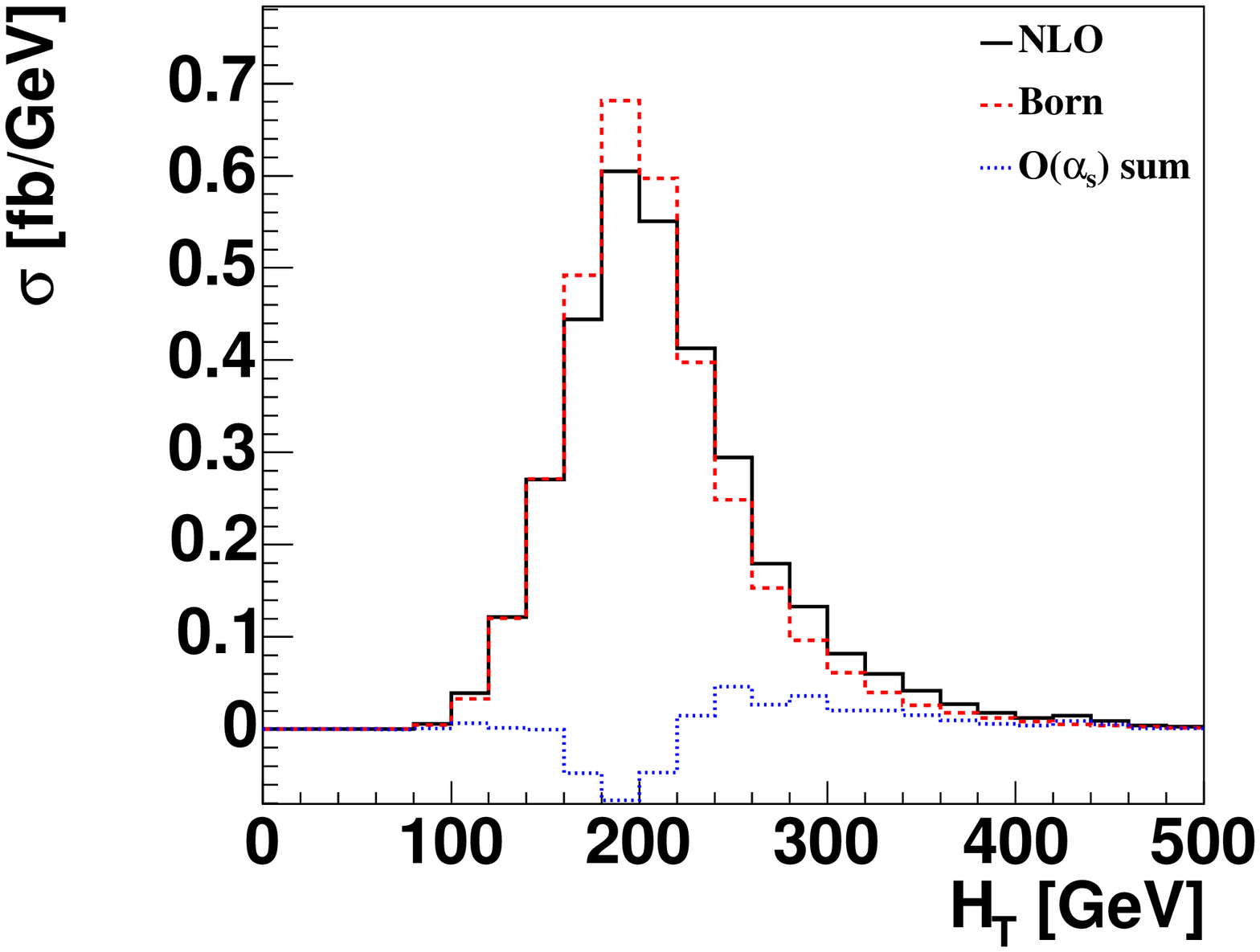}}\subfigure[]{\includegraphics[%
  width=0.40\linewidth,
  keepaspectratio]{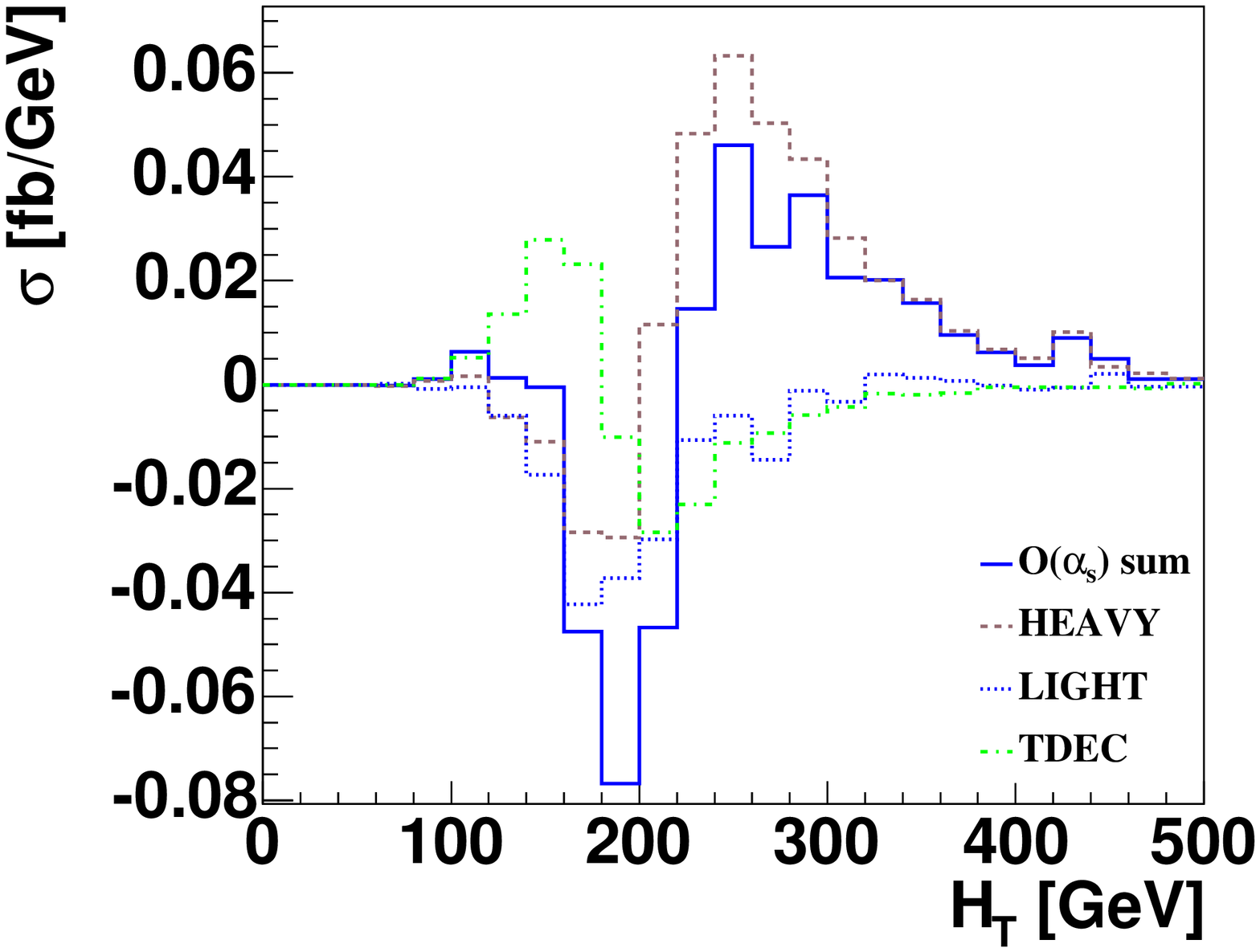}}

\caption{Total event transverse energy $H_{T}$ after selection cuts, comparing
Born level to $\oalphas$ corrections (a) and the individual $\oalphas$
contributions (b).\label{fig:HT}}
\end{figure}
The distribution of $H_{T}$ for $t$-channel single top quark events
is shown in Fig.~\ref{fig:HT}. The Born-level $H_{T}$ distribution
peaks around 200 GeV. Both the HEAVY and LIGHT contributions decrease
the height of the peak and shift it to higher values, and all three
$O(\alpha_{s})$ contributions broaden the distribution.

\subsection{Event Reconstruction\label{sub:Event-Reconstruction}}

When analyzing single top quark events we would like to take advantage
not only of simple single-object kinematics but also of correlations
between objects. In order to take full advantage of the correlations,
we need to reconstruct the event completely, not just the final state
jets but also intermediate particles, in particular the $W$-boson
and the top quark. 

The $W$~boson can be reconstructed from the final state electron
and the observed missing transverse energy, $\met$. The lack of information
about the beam-direction component of the neutrino momentum ($p_{z}^{\nu}$)
that would prevent this reconstruction is typically overcome by requiring
that the invariant mass of the electron-neutrino system be equal to
the mass of the $W$~boson. This additional constraint results in
two possible solutions for $p_{z}^{\nu}$. One usually follows the
prescription given in Ref.~\cite{Kane:1989vv} of choosing the solution
which has the smaller $\left|p_{z}^{\nu}\right|$. This picks the
correct $p_{z}^{\nu}$ in about $70\%$ of the events. The price paid
for this method is that the $W$~boson will blur the spin correlation
in the reconstructed single top quark event. We will see later that
we can improve on this by using a top mass constraint.

In order to reconstruct the top quark, the reconstructed $W$~boson
then needs to be combined with the $b$~jet from the top quark decay.
The challenge to overcome here is the proper identification of the
$b$~jet. In the $s$-channel single top process, we used the ``best-jet''
algorithm to find the correct $b$~jet among the two possible $b$-tagged
jets in the final state, making use of the known top quark mass ($m_{t}=178\,{\rm GeV}$)~\cite{Cao:2004ap}.
The effectiveness of the best-jet algorithm is mainly limited by the
efficiency of the $W$~boson identification method; if the $W$~boson
is not reconstructed properly, then identifying the $b$~jet from
the top quark decay becomes a random choice. 

\begin{figure}
\subfigure[]{\includegraphics[%
  width=0.50\linewidth,
  keepaspectratio]{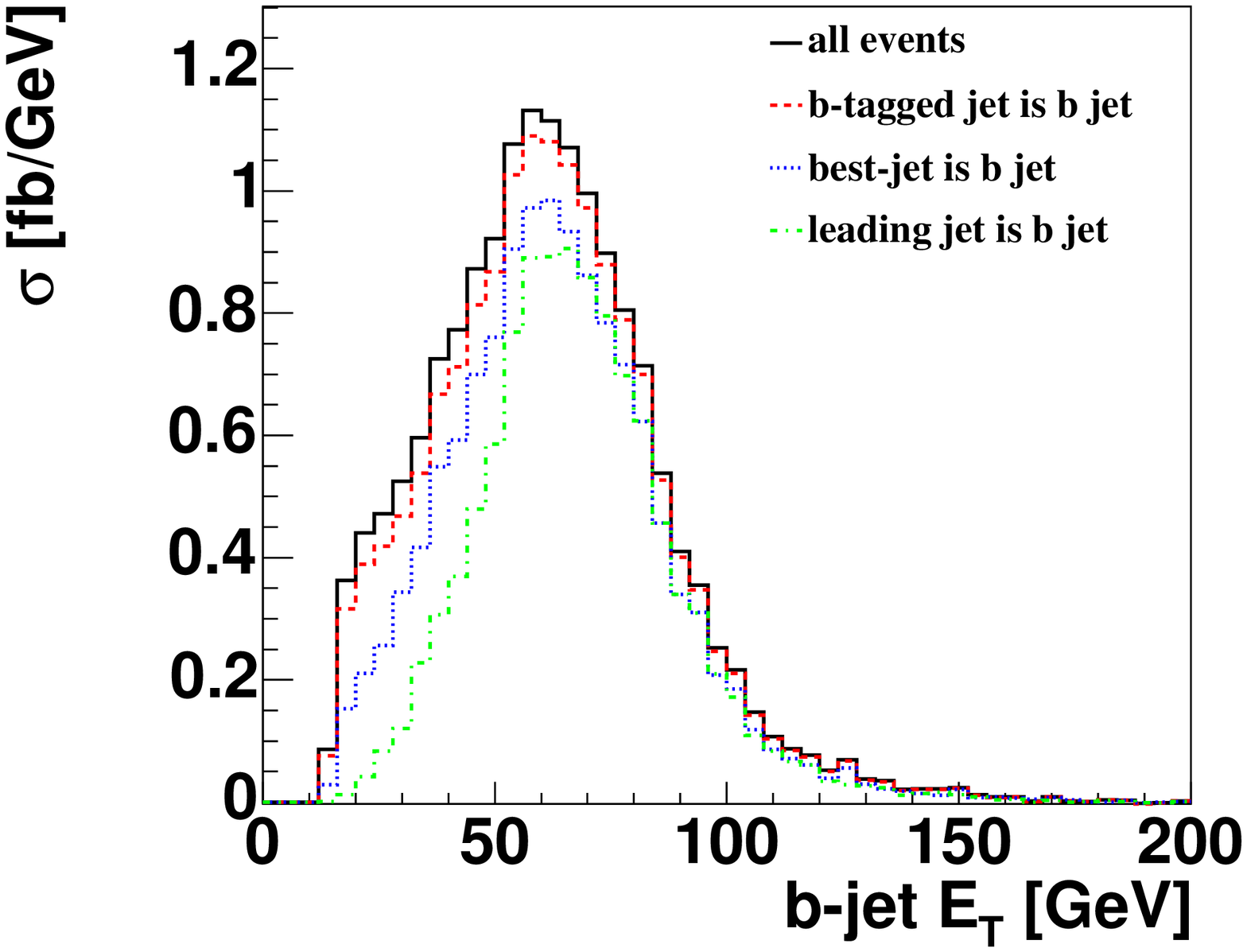}}\subfigure[]{\includegraphics[%
  width=0.50\linewidth,
  keepaspectratio]{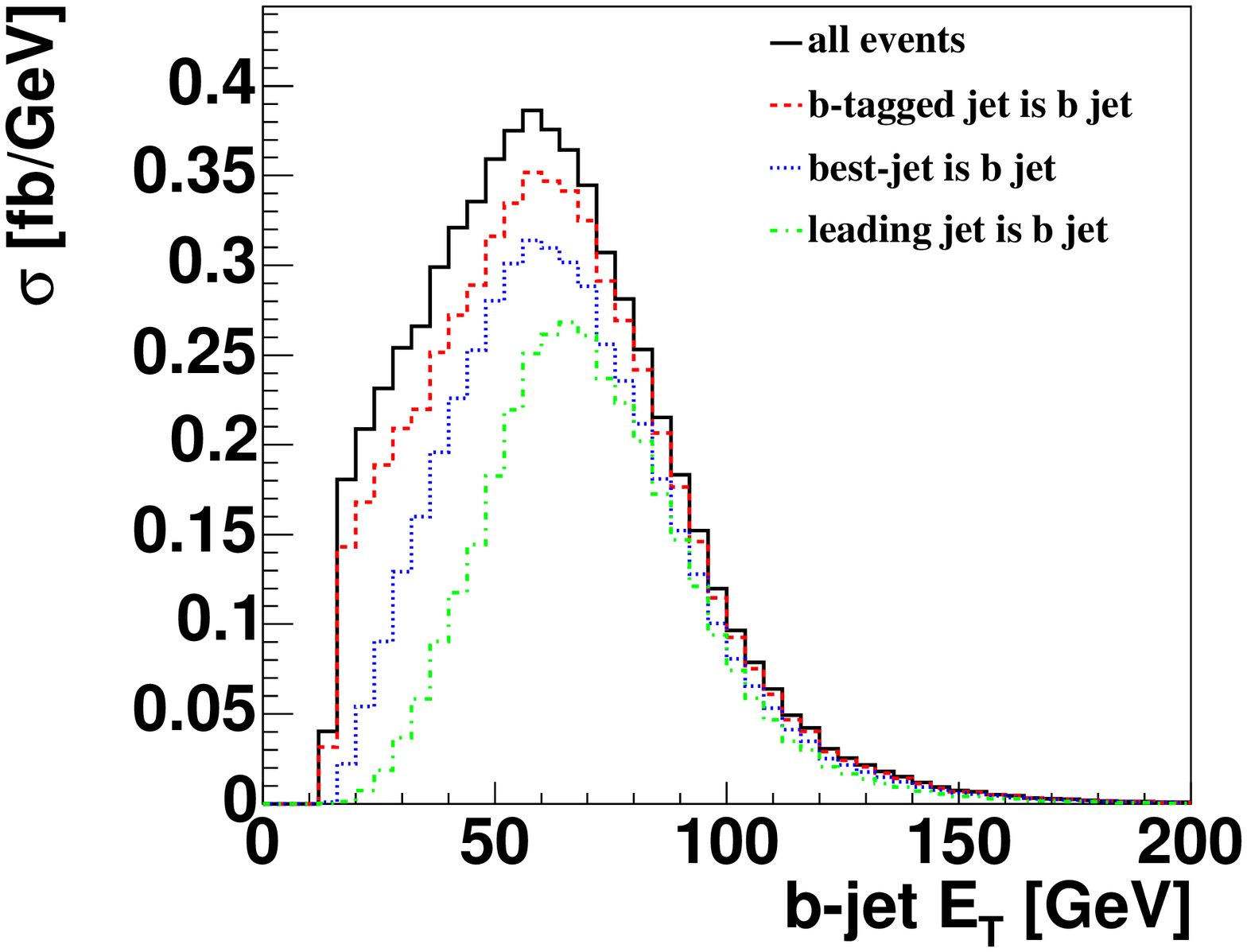}}

\caption{Transverse momentum of the $b$~quark jet, for all events (solid
histogram), only for those events for which the leading $b$-tagged
jet corresponds to the $b$~quark jet (dashed histogram), only for
those events in which the best jet corresponds to the $b$~quark
jet (dotted histogram), and only for those events in which the leading
jet corresponds to the $b$ quark jet (dash-dotted histogram). The
transverse momentum is shown for the inclusive two-jet sample (a)
and for the exclusive three-jet sample (b).\label{fig:b_jet_eff}}
\end{figure}

Unlike the $s$-channel processes, inclusive two-jet events of the
$t$-channel single top processes typically only contain one $b$-tagged
jet in the final state, corresponding to the $b$~quark from the
top quark decay. Figure~\ref{fig:b_jet_eff} shows a comparison of
three different algorithms to identify the $b$~jet from the top
quark decay: a) using the leading jet (highest $p_{T}$) in the event,
b) using the leading $b$-tagged jet in the event, and c) using the
best jet. The leading jet in this case could be either the $b$~jet
or the light quark jet, thus the efficiency for identifying the $b$-jet
from top quark decay with method a) is rather low. The leading $b$-tagged
jet is either the $b$~jet from the top quark decay or the $\bar{b}$~jet
originating from the $W$-gluon fusion subprocess. Figure~\ref{fig:b_jet_eff}
shows that the leading $b$-tagged jet corresponds to the $b$~quark
from the top decay most of the time; it identifies the $b$~quark
correctly in 95\% of the events in the inclusive sample after the
loose selection cuts. We expect the efficiency to drop in the three-jet
sample because of the additional $\bar{b}$~quark in the final state.
However, as shown in Fig.~\ref{fig:ptbbbar}, the $b$~jet from
the top quark decay is much harder than the $\bar{b}$~jet radiated
from the initial state gluon. As a result, the leading $b$-tagged
jet corresponds to the $b$~quark from the top quark decay in 90\%
of the exclusive three-jet events. Figure~\ref{fig:b_jet_eff} shows
that this fraction is slightly smaller for low $b$~jet $p_{T}$
values, and reaches 100\% above about 100~GeV. By comparison, the
best jet corresponds to the $b$~quark in only 80\% of the events
in the inclusive sample and about 72\% of the events in the exclusive
three-jet sample. We therefore choose the method of identifying the
leading $b$-tagged jet as the $b$~quark from the top decay. Having
identified the $b$~quark in this manner with no additional assumptions
allows us to use it to reconstruct the top quark and the $W$~boson
more accurately, in particular to determine the longitudinal momentum
of the neutrino ($p_{z}^{\nu}$). We reconstruct the $W$~boson as
above using a $W$~mass constraint. We then identify the proper neutrino
$p_{z}^{\nu}$ by choosing the solution for which the invariant mass
of the reconstructed $W$ and the leading $b$-tagged jet is closest
to the true top quark mass, i.e. $178\,{\rm GeV}$.

This method identifies the correct neutrino $p_{z}^{\nu}$ about 92\%
of the time at the Born-level and about 84\% of the time at NLO. These
efficiencies are slightly smaller than the $b$~quark identification
efficiencies shown in Fig.~\ref{fig:b_jet_eff} due to the finite
widths of the $W$~boson and the top quark.

\begin{figure}
\subfigure[]{\includegraphics[%
  width=0.50\linewidth,
  keepaspectratio]{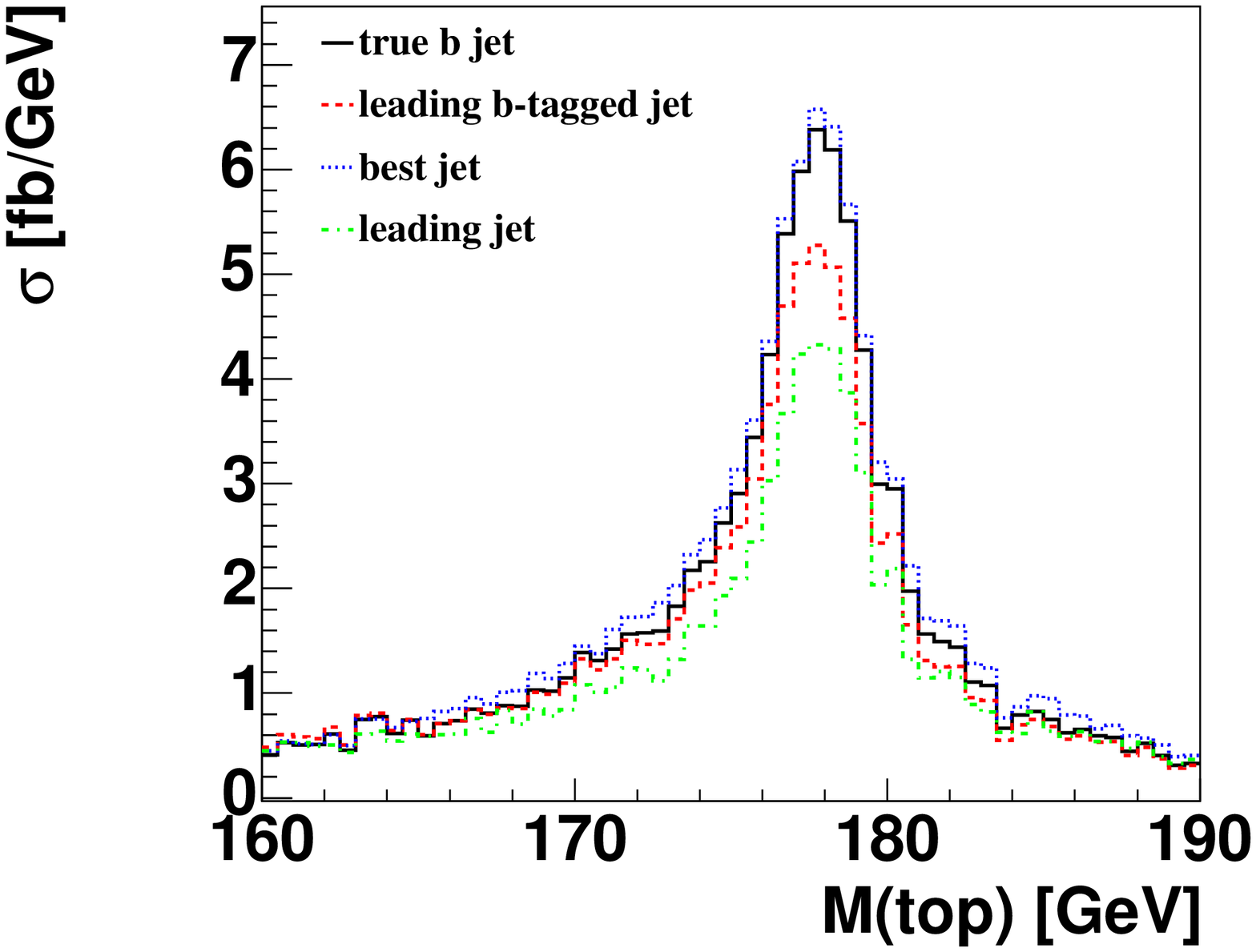}}\subfigure[]{\includegraphics[%
  width=0.50\linewidth,
  keepaspectratio]{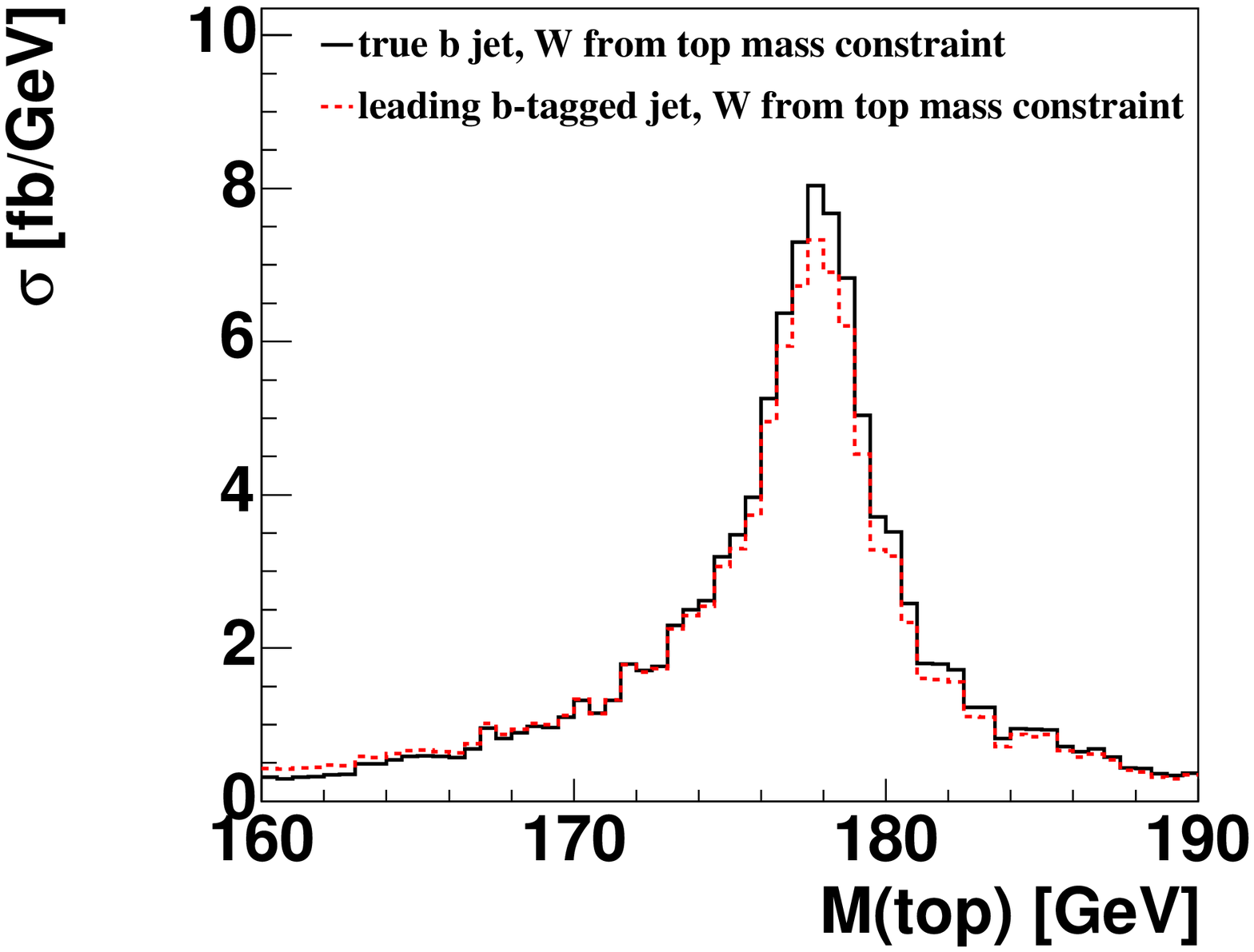}}

\caption{Invariant mass of the reconstructed $W$ and a jet object. This jet
object is: the jet containing the $b$ quark from the top decay, using
parton information (solid line), the leading $b$-tagged jet (dashed
line), the best jet (dotted line), or the leading jet (dot-dashed
line). The $W$ is reconstructed either using the standard neutrino
$p_{z}^{\nu}$ constraint (a) or a top mass constraint (b). \label{fig:TopM}}
\end{figure}

Figure~\ref{fig:TopM} shows the invariant mass of the reconstructed
top quark, comparing the different methods to identify the $b$~quark
and the $W$~boson from the top quark decay. The top quark width
is larger than it would be at parton level even though no kinematic
smearing was applied in this study to mimic the detector effect. This
is the result of using the reconstructed kinematics of the $W$~boson
(in particular the neutrino $z$-momentum) rather than parton level
information. Furthermore, since the same reconstructed $W$~boson
was used for all curves in Fig.~\ref{fig:TopM}, differences among
the individual curves are solely due to the jet choice used to reconstruct
the top quark. Finally, because parton level information is used for
the $b$~jet curve, it functions as a reference and upper limit for
the other methods. 

Figure~\ref{fig:TopM}(a) shows the top mass reconstructed with the
$W$~boson using the standard method of choosing the smaller $\left|p_{z}^{\nu}\right|$.
As expected, using the leading jet gives the worst performance, and
using the leading $b$-tagged is a better choice. Given this choice
for the neutrino, however, the best jet algorithm looks best because
for those cases where the neutrino has been misreconstructed, the
best jet algorithm sometimes chooses the wrong jet and thus it falsely
appears to give well reconstructed top mass because the algorithm
chooses an invariant mass close to 178~GeV. Figure~\ref{fig:TopM}(b)
shows the top mass reconstructed with the $W$~boson reconstructed
from a top mass constraint. The overall height of the mass peak is
higher in Fig.~\ref{fig:TopM}(b) than in Fig.~\ref{fig:TopM}(a),
indicating that this method is able to reconstruct the $W$~boson
and $b$~jet correctly more often. It can also be seen that for this
choice of $W$~boson reconstruction, we can properly reconstruct
the true top quark with the leading $b$-tagged jet with very high
efficiency. Using the leading $b$-tagged jet together with a top
mass constraint to reconstruct the $W$~boson gives the best reconstructed
top quark and is the most efficient event reconstruction algorithm
for $t$-channel single top quark events. We will henceforth focus
on this algorithm.

\begin{figure}
\subfigure[]{\includegraphics[%
  width=0.40\linewidth,
  keepaspectratio]{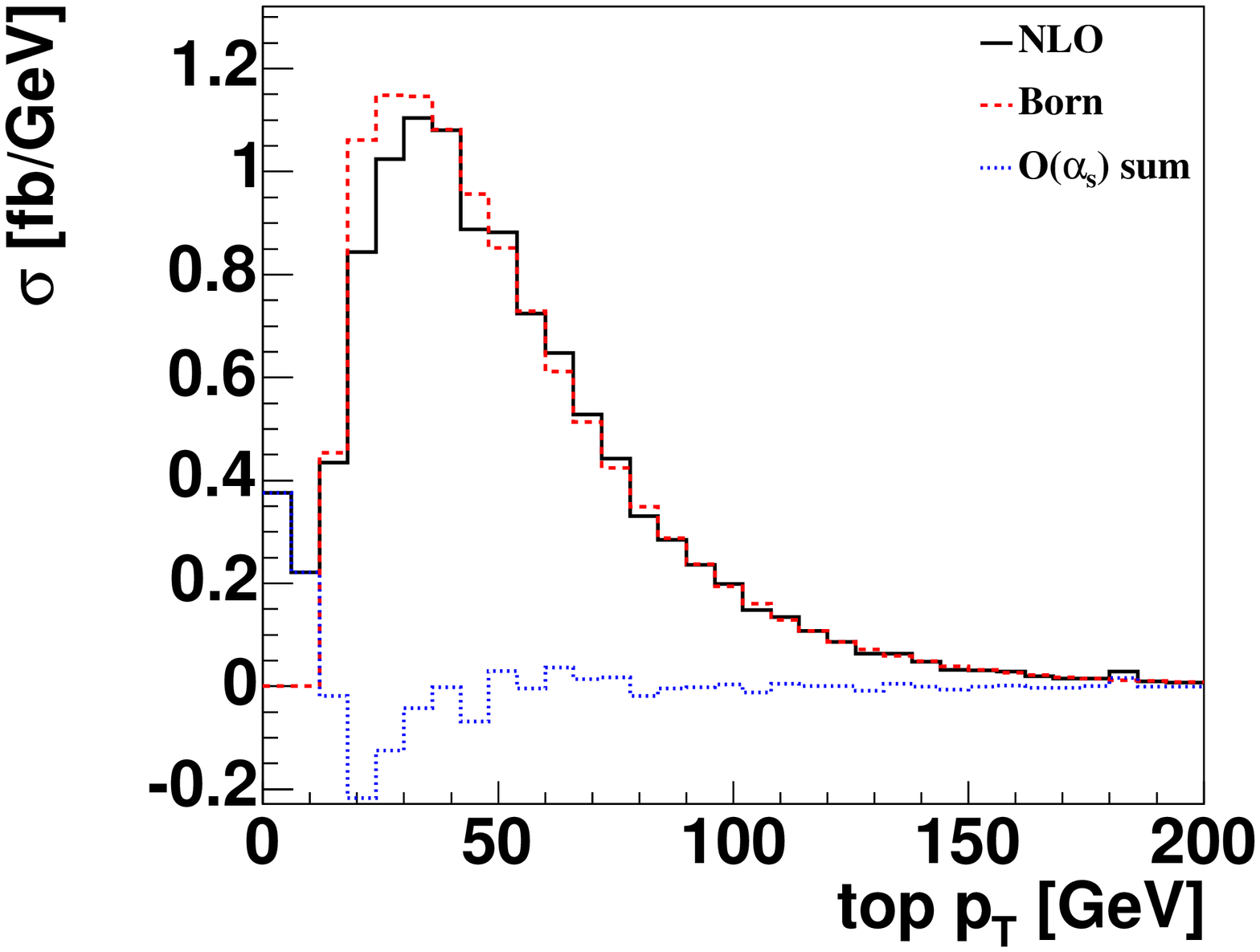}}\subfigure[]{\includegraphics[%
  width=0.40\linewidth,
  keepaspectratio]{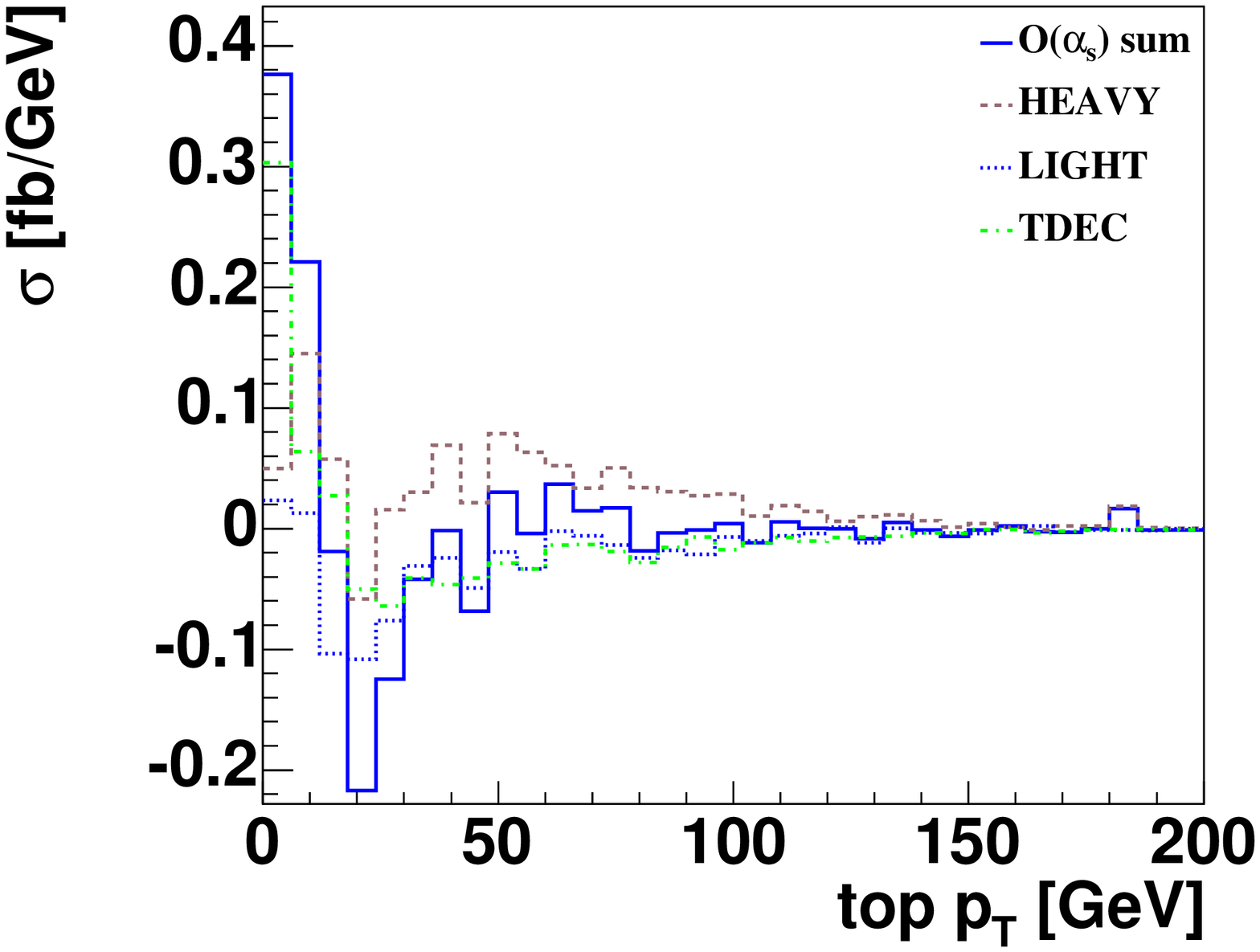}}

\caption{Transverse momentum of the reconstructed top quark, comparing Born-level
to $\oalphas$ corrections(a) and the individual $\oalphas$ contributions
(b).\label{fig:TopPt}}
\end{figure}

\begin{figure}
\subfigure[]{\includegraphics[%
  width=0.40\linewidth,
  keepaspectratio]{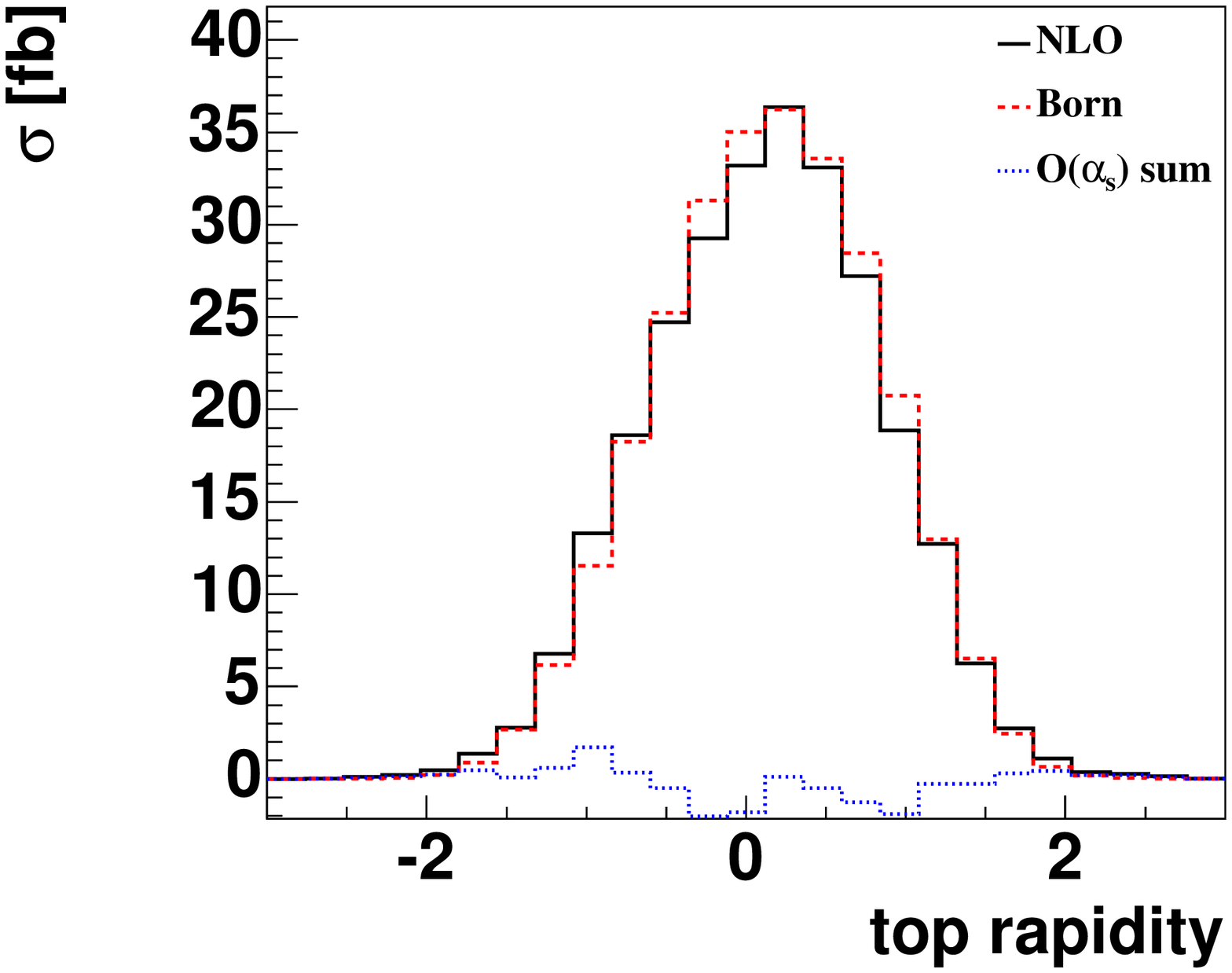}}\subfigure[]{\includegraphics[%
  width=0.40\linewidth,
  keepaspectratio]{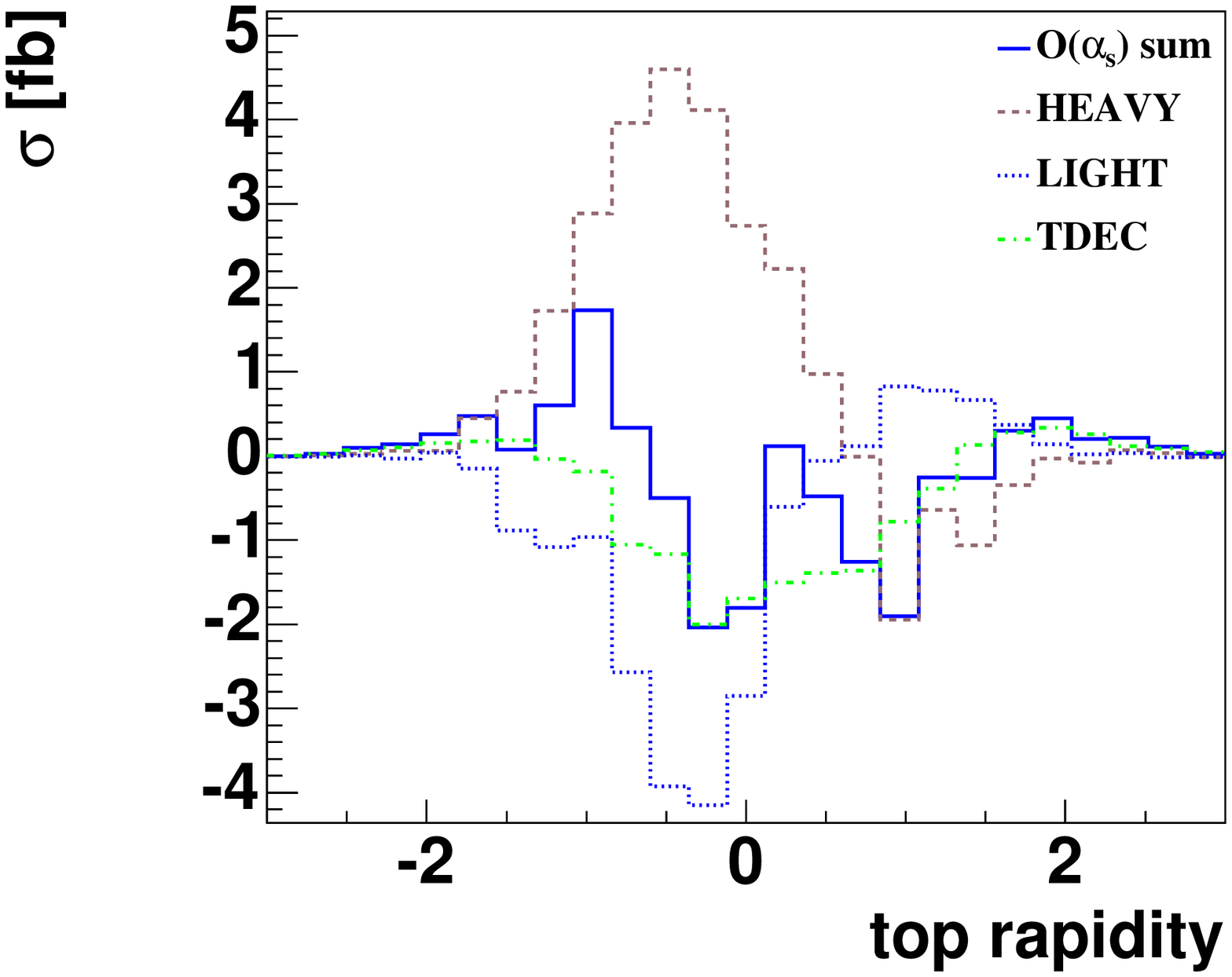}}

\caption{Rapidity of the reconstructed top quark, comparing Born-level to
$\oalphas$ corrections(a) and the individual $\oalphas$ contributions
(b).\label{fig:TopRapidity}}
\end{figure}
Figure~\ref{fig:TopPt} shows the transverse momentum of the top
quark reconstructed from the leading $b$-tagged jet together with
the top mass constraint to reconstruct the $W$~boson, hence, to
determine $p_{z}^{\nu}$. The LIGHT and HEAVY corrections tend to
shift the top quark $p_{T}$ to higher values, while the TDEC correction
lowers it as expected. The different $\oalphas$ corrections have
only a small effect on the top quark rapidity shown in Fig.~\ref{fig:TopRapidity},
similar to the $b$~jet above. The HEAVY correction tends to shift
the top quark to negative rapidities, while the LIGHT correction tends
to shift it to positive rapidities, resulting in a net effect of only
a small change in rapidity.

\subsection{Top Quark Polarization\label{sub:Object-Correlations}}

Having identified the $b$~jet from the top quark decay and the $W$~boson,
we can now study correlations expected from event kinematics. As noted
in the previous section, we shall focus on the study of top quark
polarization using the leading $b$-tagged jet and the top mass constraint
for the $W$~boson, because it reconstructs the final state correctly
with the highest efficiency.

In the SM, the top quark produced in single top quark events is highly
polarized, and this polarization can in principle be measured. The
top quark is by far the heaviest known fermion and the only known
fermion with a mass at the electroweak symmetry-breaking scale. Thus,
it is hoped that a detailed study of top quark coupling to other particles
will be of great utility in determining if the SM mechanism for electroweak
symmetry-breaking is the correct one, or if new physics is responsible~\cite{Chen:2005vr}.
Angular correlations among the decay products of polarized top quarks
provide a useful handle on these couplings. It has been pointed out
that among the decay products of the top quark, the charged lepton
is maximally correlated with the top quark spin~\cite{Mahlon:1995zn,Parke:1996pr}.
We can thus obtain the most distinctive distribution by plotting the
angle between the charged lepton and the spin axis of the top quark
in its rest frame.

\begin{figure}
\includegraphics[%
  width=1.0\linewidth,
  keepaspectratio]{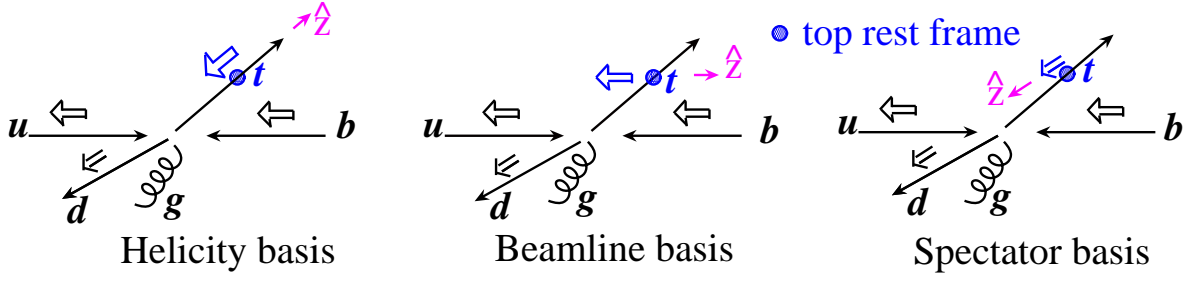}

\caption{Illustration of the three choices for the top quark spin basis. The
circle denotes the top quark rest frame and the blue arrows denote
the top quark spin direction.\label{fig:TopSpinBasis}}
\end{figure}

Although the top quark is produced via the left-handed charged current,
there is no reason to believe that the helicity basis will give the
best description of the top quark spin. Choosing an appropriate basis
could maximize spin correlation effects. Three polarization bases
have been studied in the literature for the $t$-channel process,
and they differ by the reference frame used to define the polarization:
the helicity basis, the beamline basis, and the so-called {}``spectator''
basis~\cite{Mahlon:1996pn}. All three work in the top quark rest
frame, but they have different reference axis for the top quark spin,
cf. Fig.~\ref{fig:TopSpinBasis}. In the more commonly used helicity
basis, the top quark spin is measured along the top quark direction
of motion in the center of mass frame which is chosen as the frame
of the (reconstructed top quark, spectator jet) system after event
reconstruction. In the beamline basis, the top quark spin is measured
along the incoming proton direction. In the spectator basis we can
maximize spin correlations by taking advantage of the fact that the
top quark produced through the $t$-channel single top processes is
almost $100\%$ polarized along the direction of the spectator quark.
In the discussion below, we will examine the polarization of single
top quark events in these three bases. 

In helicity basis, the c.m. frame needs to be reconstructed in order
to define the top quark momentum. Due to additional jet radiation,
the determination of the c.m. frame at NLO is more complicated than
at the Born-level. The additional radiation will also blur the spin
correlation, therefore, choosing the appropriate frame will reduce
this effect. In this study, two options for reconstructing the c.m.
frame are investigated:

\begin{enumerate}
\item $tq(j)$-frame: the c.m. frame of the incoming partons. This is the
rest frame of all the final state objects (reconstructed top quark
and all others jets). In exclusive two-jet events, this frame is the
same as the c.m. frame at the Born-level, i.e. reconstructed from
summing over momentum of the top quark and spectator jet. In exclusive
three-jet events, this frame is reconstructed by summing over the
4-momenta of top quark, spectator jet, and the third-jet from our
parton level calculation. 
\item $tq$-frame: the c.m. frame of the top quark and spectator jet. In
this case, even in exclusive three-jet events, the reference frame
is constructed by summing over only the 4-momenta of the top quark
and spectator jet. Note that this differs from the $tq(j)$-frame
only in exclusive three-jet events.
\end{enumerate}
As shown in Table~\ref{tab:toppol-2jet} and discussed below, the
degree of top polarization is larger in the $tq$-frame than in the
$tq(j)$-frame. Therefore, in the figures below we only display the
top quark polarization in the $tq$-frame. 

\begin{figure}
\subfigure[]{\includegraphics[%
  width=0.40\linewidth,
  keepaspectratio]{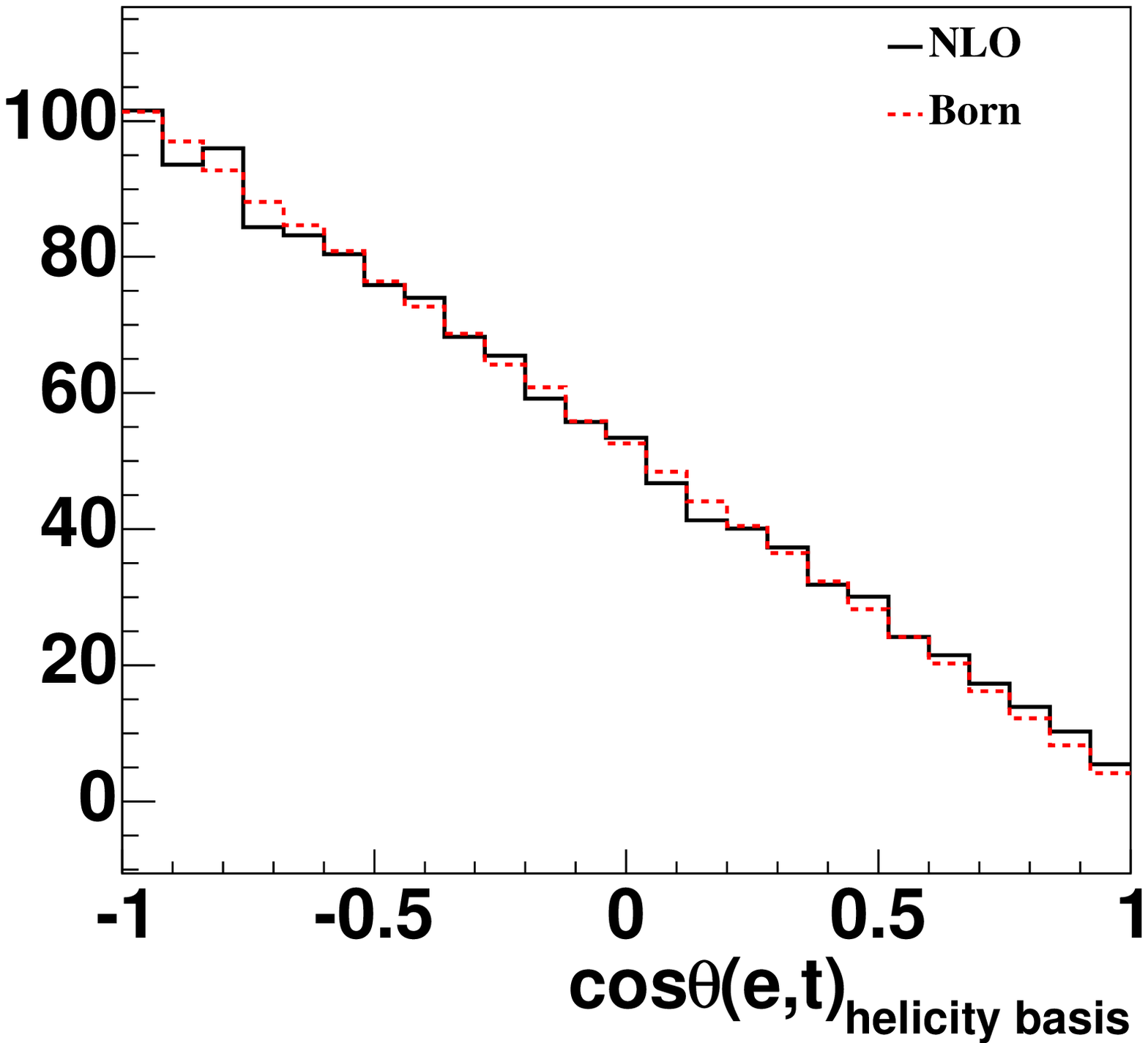}}\subfigure[]{\includegraphics[%
  width=0.40\linewidth,
  keepaspectratio]{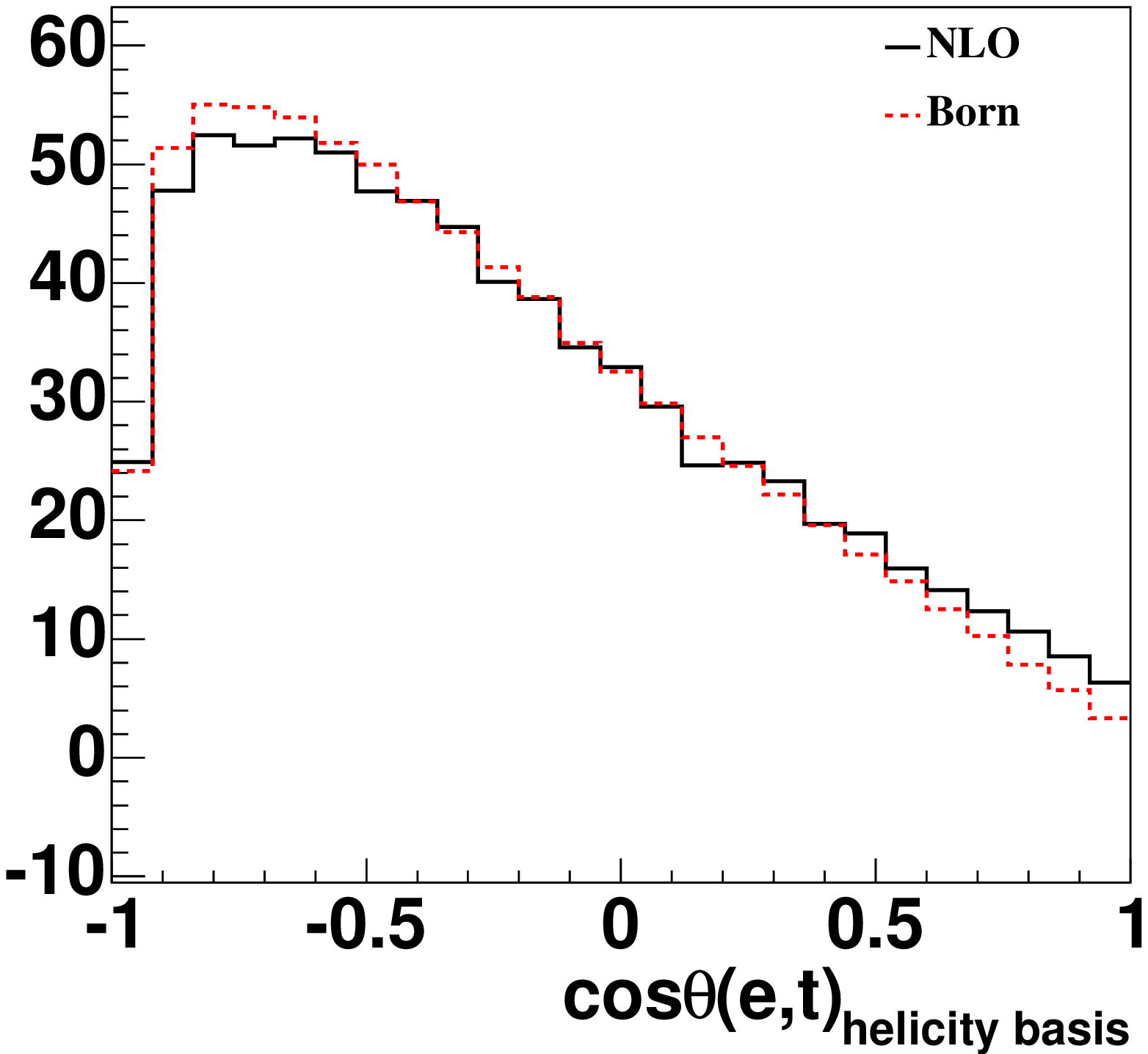}}

\caption{Top quark polarization in the helicity basis using the full parton
information (a) and after event reconstruction with selection cuts
(b), comparing Born-level to $\oalphas$ corrections. The Born-level
and NLO curves have been normalized to the same area.\label{fig:TopPolHel}}
\end{figure}

In the helicity basis, the polarization of the top quark is examined
as the angular distribution ($\cos\theta_{hel}$) of the lepton in
the c.m. frame relative to the moving direction of the top quark in
the same frame. The angular correlation in this frame is given by
\begin{eqnarray}
\cos\theta_{hel}=\frac{\vec{p}_{t}\cdot\vec{p}_{\ell}^{\;*}}{|\vec{p}_{t}||\vec{p}_{\ell}^{\;*}|},\end{eqnarray}
where $\vec{p}_{\ell}^{\;*}$ is the charged lepton three-momentum
defined in the rest frame of the top quark, whose three momentum is
denoted as $\vec{p_{t}}$, which is in turn defined in the c.m. frame.
For a left-handed top quark, the angular correlation of the lepton
$\ell^{+}$ is given by $(1-\cos\theta_{hel})/2$, and for a right-handed
top quark, it is $(1+\cos\theta_{hel})/2$. Figure~\ref{fig:TopPolHel}(a)
shows that this linear relationship for $\cos\theta_{hel}$ is indeed
a valid description for $t$-channel single top quark events at the
parton level. The figure also shows that the top quark is not completely
polarized in the helicity basis, and that this polarization is only
slightly weakened when including $\oalphas$ corrections. Figure~\ref{fig:TopPolHel}
(b) shows that this weakening is amplified after event reconstruction,
where the effect of the lepton-jet separation cut can also be seen,
as a drop-off of the $\cos\theta_{hel}$ distribution close to a value
of $-1$. 

\begin{figure}
\subfigure[]{\includegraphics[%
  width=0.40\linewidth,
  keepaspectratio]{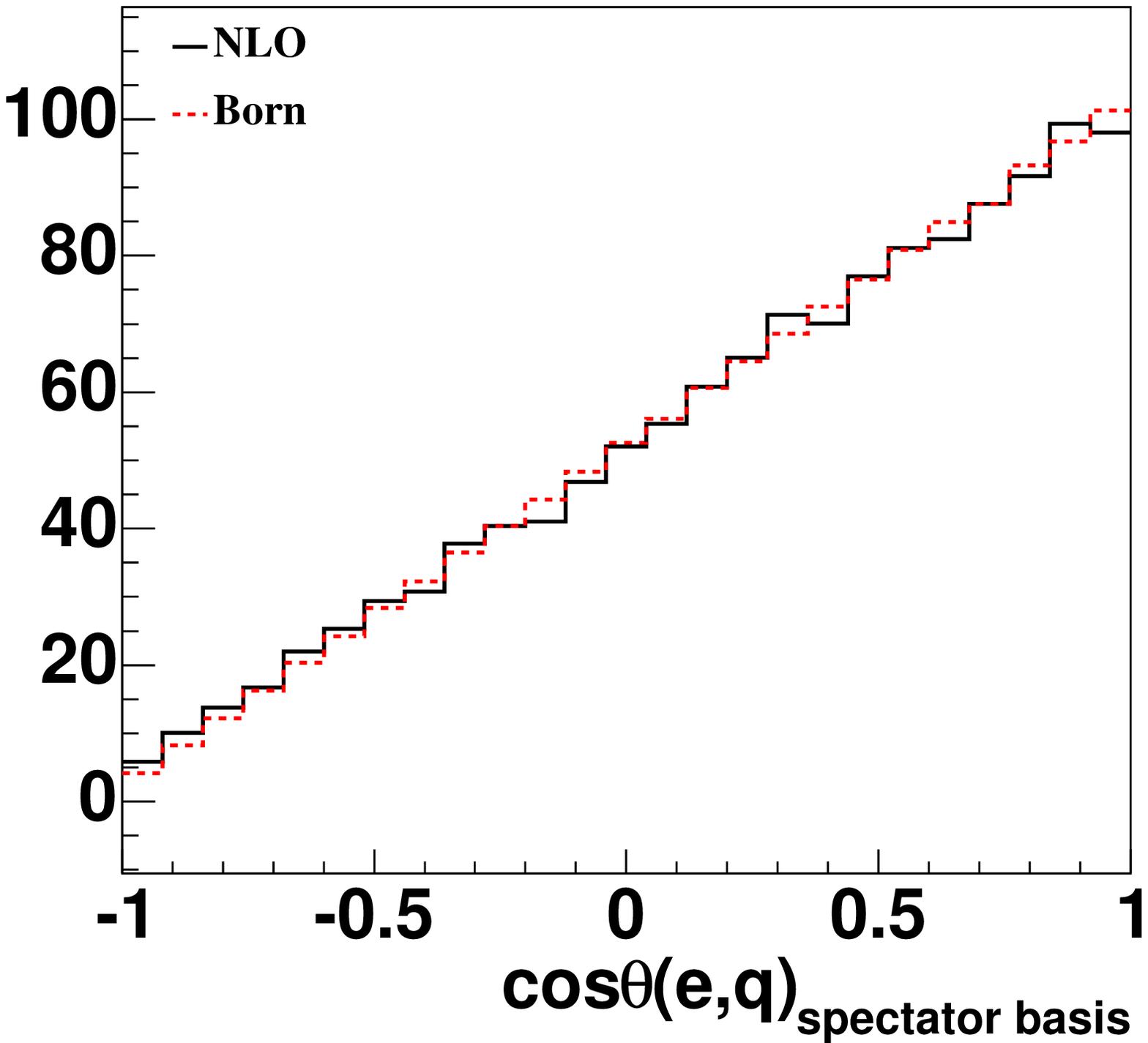}}\subfigure[]{\includegraphics[%
  width=0.40\linewidth,
  keepaspectratio]{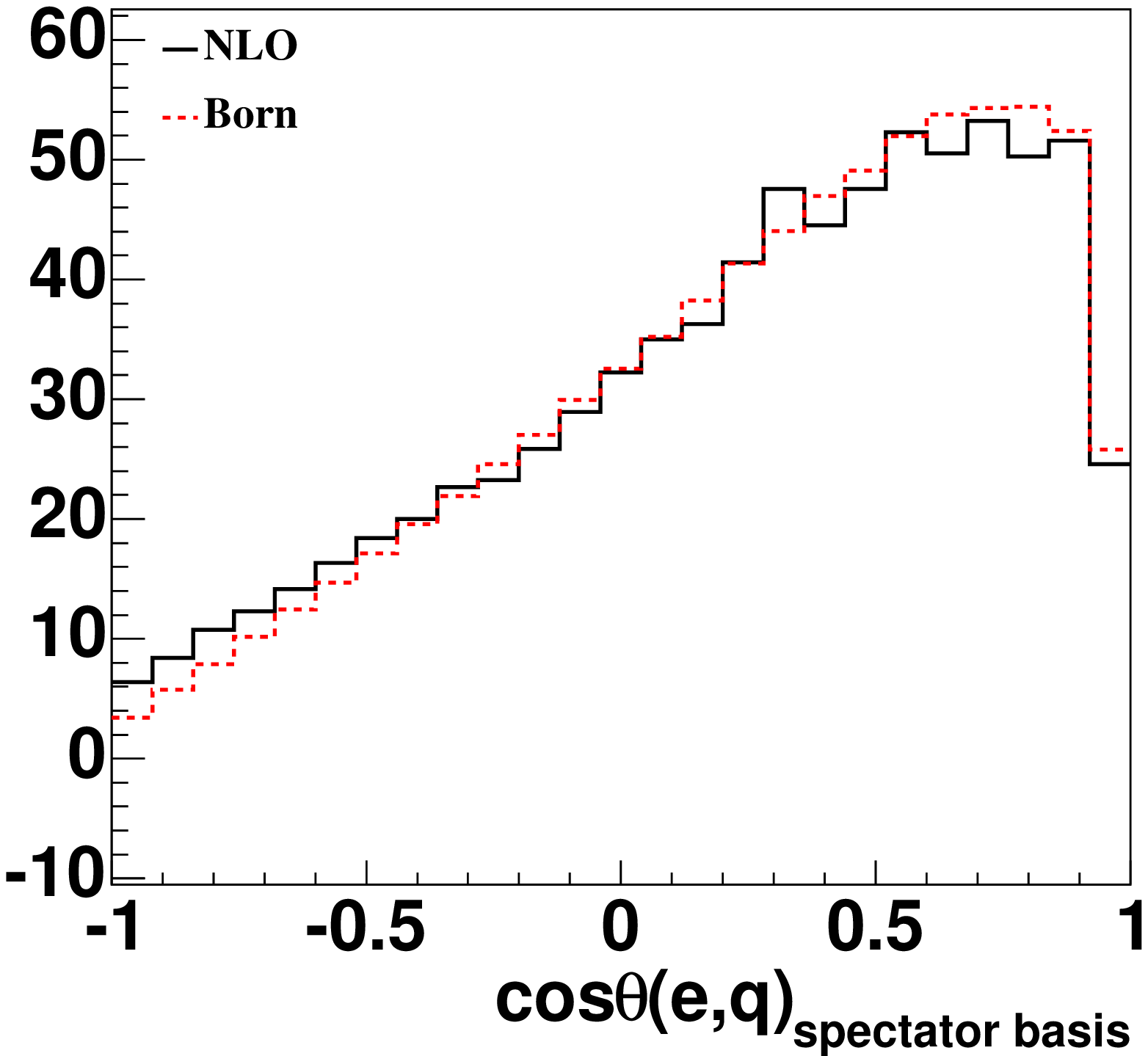}}

\caption{Top quark polarization in the spectator basis using the full parton
information (a) and after event reconstruction with selection cuts
(b), comparing Born-level to $\oalphas$ corrections. The Born-level
and NLO curves have been normalized to the same area.\label{fig:TopPolSpec}}
\end{figure}

In the {}``spectator'' basis, the relevant angular correlation for
the $t$-channel process is $\cos\theta_{spec}$, defined as \begin{eqnarray}
\cos\theta_{spec}=\frac{\vec{p}_{spec}^{\;*}\cdot\vec{p}_{\ell}^{\;*}}{|\vec{p}_{spec}^{\;*}||\vec{p}_{\ell}^{\;*}|}\,,\end{eqnarray}
 where $\vec{p}_{spec}^{\;*}$ is the spectator jet three-momentum
in the top quark rest frame and $\vec{p}_{\ell}^{\;*}$ is the lepton
three-momentum in the top quark rest frame. Although this basis picks
the wrong spin axis direction for the $\bar{d}b$ and $b\bar{d}$
initial states, it is correct most of the time at the Tevatron collider.
This is because the Tevatron is a $p\bar{p}$ collider which means
that the production rate of $p\bar{p}(ub,bu)\to dt$ is much larger
than $p\bar{p}(\bar{d}b)\to\bar{u}t$. The large slope found in Fig.~\ref{fig:TopPolSpec}
shows that the spectator basis indeed results in a large degree of
correlation for $t$-channel single top quark events at the parton
level. The slope shown in Fig.~\ref{fig:TopPolSpec}(a) is opposite
to that in Fig.~\ref{fig:TopPolHel}(a) due to the fact that the
spin quantization axis points in opposite directions for the two basis,
cf. Fig.~\ref{fig:TopSpinBasis}. The degree of polarization of the
top quark is larger than that in the helicity basis, and the $\oalphas$
corrections blur the spin correlation effects only slightly, both
at parton level and after event reconstruction. The reconstructed
$\cos\theta_{spec}$ distribution again shows a drop-off due to the
lepton-jet separation cut, in this case at high $\cos\theta_{spec}$.

\begin{figure}
\subfigure[]{\includegraphics[%
  width=0.40\linewidth,
  keepaspectratio]{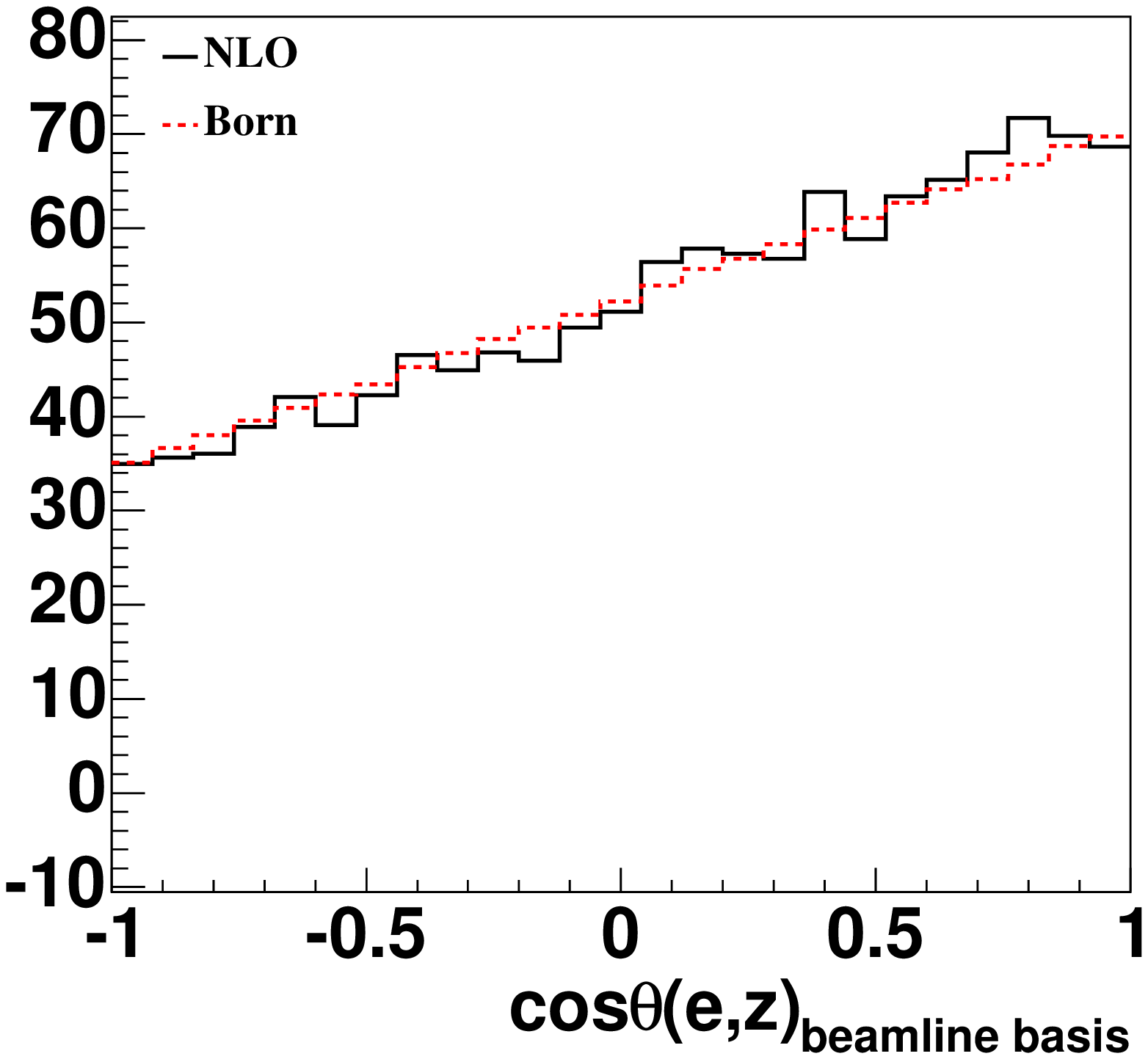}}\subfigure[]{\includegraphics[%
  width=0.40\linewidth,
  keepaspectratio]{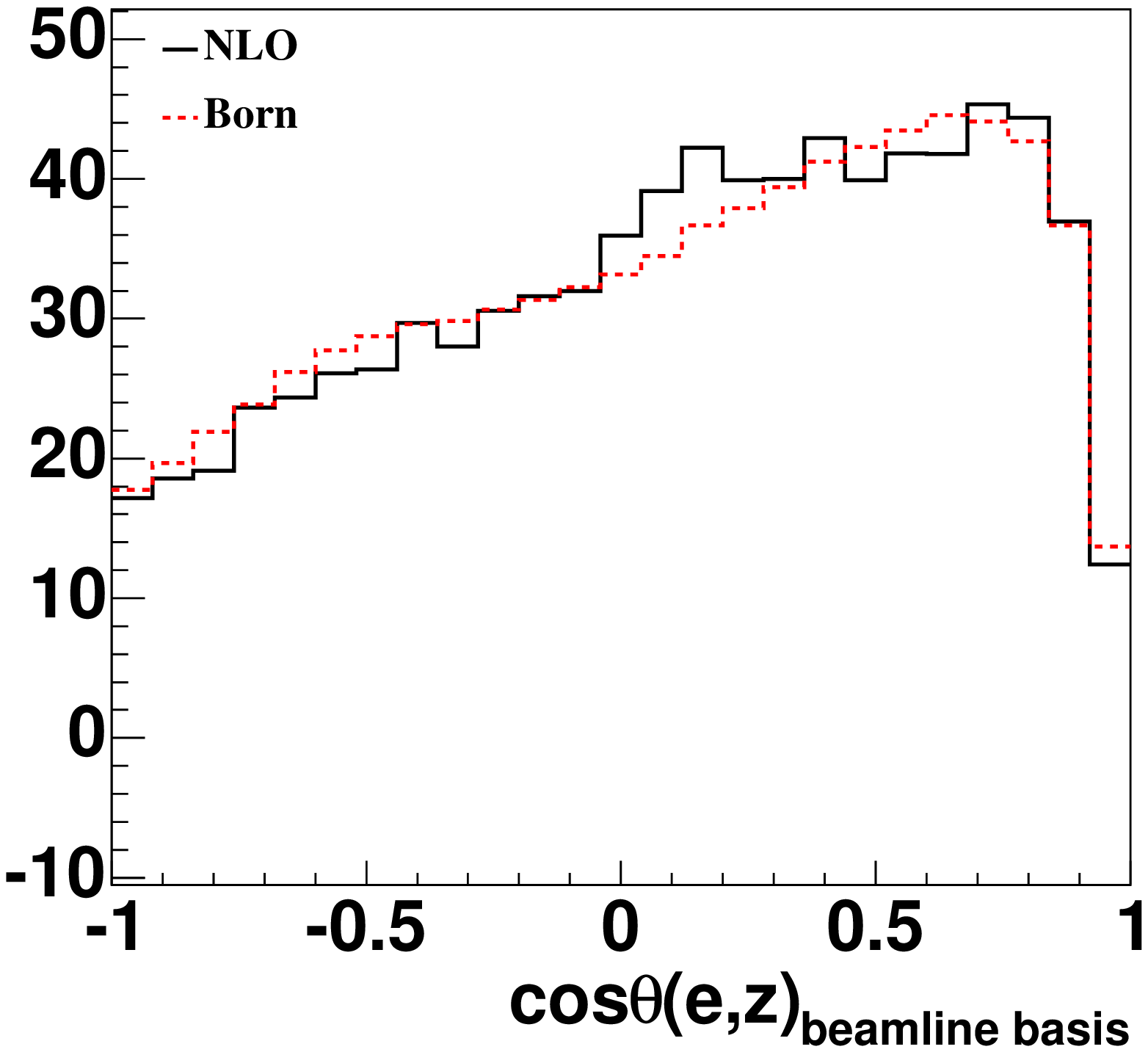}}

\caption{Top quark polarization in the beamline basis using the full parton
information (a) and after event reconstruction with selection cuts
(b), comparing Born-level to $\oalphas$ corrections. The Born-level
and NLO curves have been normalized to the same area.\label{fig:TopPolBeam}}
\end{figure}

In the {}``beamline'' basis, the relevant angular correlation for
the $t$-channel process is $\cos\theta_{beam}$, defined as \begin{eqnarray}
\cos\theta_{beam}=\frac{\vec{p}_{p}^{\;*}\cdot\vec{p}_{\ell}^{\;*}}{|\vec{p}_{p}^{\;*}||\vec{p}_{\ell}^{\;*}|}\,,\end{eqnarray}
 where $\vec{p}_{p}^{\;*}$ is the proton three-momentum in the top
quark rest frame and $\vec{p}_{\ell}^{\;*}$ is the lepton three-momentum
in the top quark rest frame. In this analysis, we orient the coordinate
system such that protons travel in the positive $z$ direction and
anti-protons travel in the negative $z$ direction. For a top quark
polarized along the proton moving direction, the angular distribution
of the lepton $\ell^{+}$ is $(1+\cos\theta_{beam})/2$, while for
a top quark polarized along the anti-proton moving direction it is
$(1-\cos\theta_{beam})/2$. Figure~\ref{fig:TopPolBeam} shows that
this linear relationship for $\cos\theta_{beam}$ is a valid description
for $t$-channel single top quark events at the parton level. However,
the top quark is less polarized in the beamline basis at parton level.
In this case, the $\oalphas$ corrections actually improve the spin
correlation effects at parton level. After event reconstruction the
situation is similar as before, the spin correlation is further reduced
and shows a drop-off close to 1.

To better quantify the change in polarization, it is useful to define
the degree of polarization $\mathcal{D}$ of the top quark. This is
given as the ratio\[
\mathcal{D}=\frac{N_{-}-N_{+}}{N_{-}+N_{+}},\]
 where $N_{-}$ ($N_{+}$) is the number of left-handed (right-handed)
polarized top quarks in the helicity basis. Similarly, in the spectator
(beamline) basis, $N_{-}$ ($N_{+}$) is the number of top quarks
with polarization against (along) the direction of the spectator jet
(proton) three momentum in the top quark rest frame. The angular distribution
is then given by~\cite{Mahlon:1998uv}\textbf{\begin{eqnarray*}
\frac{1}{\sigma}\frac{d\sigma}{d(\cos\theta)} & = & \frac{N_{-}}{N_{-}+N_{+}}\frac{1+\cos\theta}{2}+\frac{N_{+}}{N_{-}+N_{+}}\frac{1-\cos\theta}{2}\\
 & = & \frac{1}{2}\left(1+D\cos\theta_{i}\right).\end{eqnarray*}
}Simple algebra leads to the following identity: \begin{eqnarray}
\mathcal{D} & = & -3\int_{-1}^{1}x\,{\frac{d\sigma}{\sigma dx}}\, dx\,,\label{dpola}\end{eqnarray}
 where ${\displaystyle \frac{d\sigma}{\sigma dx}}$ is the normalized
differential cross section as a function of the polar angle $x$.
Here, $x$ denotes $\cos\theta_{hel}$ in the helicity basis, etc.
Based on the degree of polarization $\mathcal{D}$, we can easily
get the spin fractions $\mathcal{F}_{\pm}$ as:\begin{eqnarray*}
\mathcal{F}_{-} & = & \frac{N_{-}}{N_{-}+N_{+}}=\frac{1+\mathcal{D}}{2},\\
\mathcal{F}_{+} & = & \frac{N_{+}}{N_{-}+N_{+}}=\frac{1-\mathcal{D}}{2}.\end{eqnarray*}
Note that $\mathcal{F}_{-}$($\mathcal{F}_{+}$) is the fraction of
left-handed (right-handed) polarized top quarks in the helicity basis,
etc. 

We can also define the asymmetry ${\mathcal{A}}$ of the distribution
as \begin{eqnarray}
{\mathcal{A}} & = & \frac{\int_{-1}^{0}d\sigma(\cos\theta)-\int_{0}^{1}d\sigma(\cos\theta)}{{\int_{-1}^{0}d\sigma(\cos\theta)+\int_{0}^{1}d\sigma(\cos\theta)}}.\end{eqnarray}
It is easy to check that without imposing any kinematic cuts, $D=2{\mathcal{A}}$.
Furthermore, the ratio of top quarks with spin along the basis direction
will be $r_{\uparrow}=0.5-{\mathcal{A}}$ when no cuts are applied.
However, when cuts are imposed, these two relations break down. Table~\ref{tab:toppol-2jet}
shows that the relationship $\mathcal{D}=2{\mathcal{A}}$ indeed holds
at parton level (within rounding errors) and is still approximately
true at $O(\alpha_{s})$.

\begin{table}
\begin{center}\begin{tabular}{cc|c|c|c|c|c|c}
\hline 
&
&
\multicolumn{2}{c|}{$\mathcal{D}$ }&
\multicolumn{2}{c|}{\textbf{$\mathcal{F}$}}&
\multicolumn{2}{c}{$\mathcal{A}$}\tabularnewline
\cline{3-4} \cline{5-6} \cline{7-8} 
&
&
LO&
NLO&
LO&
NLO&
LO&
NLO\tabularnewline
\hline
Helicity basis: &
Parton level ($tq(j)$-frame)&
0.96&
0.74&
0.98&
0.87&
0.48&
0.37\tabularnewline
&
Parton level ($tq$-frame)&
0.96&
0.94&
0.98&
0.97&
0.48&
0.47\tabularnewline
&
Reconstructed events ($tq(j)$-frame)&
0.84&
0.73&
0.92&
0.86&
0.46&
0.41\tabularnewline
&
Reconstructed events ($tq$-frame)&
0.84&
0.75&
0.92&
0.88&
0.46&
0.42\tabularnewline
Spectator basis: &
Parton level&
-0.96&
-0.94&
0.98&
0.98&
-0.48&
-0.47\tabularnewline
&
Reconstructed events&
-0.85&
-0.77&
0.93&
0.89&
-0.46&
-0.42\tabularnewline
Beamline basis:&
Parton level&
-0.34&
-0.38&
0.67&
0.69&
-0.17&
-0.19\tabularnewline
&
Reconstructed events&
-0.30&
-0.32&
0.65&
0.66&
-0.17&
-0.20\tabularnewline
\hline
\end{tabular}\end{center}

\caption{Degree of polarization $\mathcal{D}$, polarization fraction $\mathcal{F}$,
and asymmetry $\mathcal{A}$ for inclusive two-jet single top quark
events, at the parton level before cuts and after selection cuts and
event reconstruction, in the $t$-channel process. Here, $\mathcal{F}$
corresponds to $\mathcal{F}_{-}$ in the helicity basis for left-handed
top quarks and to $\mathcal{F}_{+}$ in the spectator and beamline
bases for top quarks with polarization along the direction of the
spectator-jet and proton three momentum, respectively. Also, the $tq(j)$-frame
in the helicity basis denotes the c.m. frame of the incoming partons,
while the $tq$-frame denotes the rest frame of the top quark and
spectator jet. \label{tab:toppol-2jet}}
\end{table}

In Table~\ref{tab:toppol-2jet}, we present our results for inclusive
two-jet events at the parton level before selection cuts and after
the loose set of cuts and event reconstruction (cf. Sec.~\ref{sub:Event-Reconstruction}).
The result for exclusive three-jet events is shown in Table~\ref{tab:toppol-3jet}.
For completeness, we also include the same study for the $s$-channel
single top process in the Appendix. 

\begin{itemize}
\item We note that in the helicity basis, the degree of top quark polarization
is larger in the $tq$-frame than in the $tq(j)$-frame (the usual
c.m. frame of the incoming partons) at the parton level. The degree
of top quark polarization is $94\%$ in the $tq$-frame and only $74\%$
in $tq(j)$-frame. This is due to the fact the degree of top quark
polarization in inclusive two-jet events is a mixture of contributions
from both exclusive two-jet events and exclusive three-jet events.
Table~\ref{tab:toppol-3jet} shows that for exclusive three-jet events,
the degree of top quark polarization is larger in the $tq$-frame
than in the $tq(j)$-frame. This reduction in the degree of polarization
for the $tq(j)$-frame is due to events in which the additional jet
is produced before the exchange of the virtual $W$~boson. After
event reconstruction, the two frames give almost the same degree of
top quark polarization. 
\item We find that at the parton level stage, there is very little difference
between the helicity basis (using the $tq$-frame) and the spectator
basis, and that both of them give significantly better polarization
than the beamline basis both at Born-level and NLO. The top quark
is almost completely polarized in the helicity and spectator bases,
and the $O(\alpha_{s})$ corrections only degrade that picture slightly.
The similarity between these two bases is due to the fact that the
degree of polarization of the top quark is dominated by exclusive
two-jet events for which these two bases are equivalent, cf. the Born-level
results in Table~\ref{tab:toppol-2jet}. After event reconstruction
for the inclusive two-jet sample, the degree of polarization is reduced
as expected in both the helicity and spectator bases. 
\item In the beamline basis, the polarization actually increases when going
from Born-level to NLO, but it is still much lower than in the other
two bases. After event reconstruction, the degree of polarization
is also reduced. 
\item In the exclusive three-jet sample, the degree of polarization is further
reduced because the third-jet affects the kinematics of either spectator
jet or the top quark. The helicity basis with $tq$-frame gives almost
the same degree of polarization as the spectator basis. 
\end{itemize}
Our study shows that the helicity basis (using the $tq$-frame) and
the spectator basis are equally good to study the top quark polarization.
Unlike the $s$-channel process in which the $W$-boson is not perfectly
reconstructed in the best-jet algorithm and thus the polarization
measurement was significantly degraded after event reconstruction,
using the leading $b$-tagged jet and the top mass constraint gives
excellent final state reconstruction in the $t$-channel process,
and the degree of top quark polarization is only somewhat degraded
after event reconstruction. 

\begin{table}
\begin{center}\begin{tabular}{cc|c|c|c}
\hline 
&
&
\multicolumn{1}{c|}{$\mathcal{D}$ }&
\multicolumn{1}{c|}{\textbf{$\mathcal{F}$}}&
\multicolumn{1}{c}{$\mathcal{A}$}\tabularnewline
\hline
Helicity basis: &
Parton level ($tq(j)$-frame)&
0.65&
0.83&
0.35\tabularnewline
&
Parton level ($tq$-frame)&
0.78&
0.89&
0.43\tabularnewline
&
Reconstructed events ($tq(j)$-frame)&
0.63&
0.81&
0.34\tabularnewline
&
Reconstructed events ($tq$-frame)&
0.70&
0.85&
0.38\tabularnewline
\hline
Spectator basis: &
Parton level&
-0.78&
0.89&
-0.42\tabularnewline
&
Reconstructed events&
-0.70&
0.85&
-0.38\tabularnewline
\hline
Beamline basis:&
Parton level&
-0.26&
0.63&
-0.16\tabularnewline
&
Reconstructed events&
-0.27&
0.63&
-0.16\tabularnewline
\hline
\end{tabular}\end{center}

\caption{Degree of polarization $\mathcal{D}$, polarization fraction $\mathcal{F}$,
and asymmetry $\mathcal{A}$ for exclusive three-jet single top quark
events, at parton level and after event reconstruction, in the $t$-channel
process. Here, $\mathcal{F}$ corresponds to $\mathcal{F}_{-}$ in
the helicity basis for left-handed top quarks and to $\mathcal{F}_{+}$
in the spectator and beamline bases for top quarks with polarization
along the direction of the spectator-jet and proton three momentum,
respectively. The $tq(j)$ frame in the helicity basis denotes the
c.m. frame of the incoming partons, while the $tq$ frame denotes
the rest frame of the reconstructed top quark and light quark. \label{tab:toppol-3jet}}
\end{table}

\subsection{Distributions for Three-jet Events\label{sub:Distributions-for-Three-jet}}

As shown in Fig.~\ref{fig:njets_jet_pt}, a large fraction of the
events passing the loose selection cuts contain three jets. In this
section we focus on the properties of these three-jet events and the
additional jet. 

\textcolor{black}{In events containing two untagged jets, we can use
the $p_{T}$ of the untagged jet to pick up the spectator jet. When
the $\bar{b}$~jet from the initial state gluon splitting in the
$W$-gluon fusion process is mistagged, it will also contribute an
untagged jet to the event. For simplicity, we will assume fully efficient,
perfect $b$-tagging here and consider events with single $b$-tags
and double $b$-tags separately.}

\begin{figure}
\subfigure[]{\includegraphics[%
  width=0.40\linewidth,
  keepaspectratio]{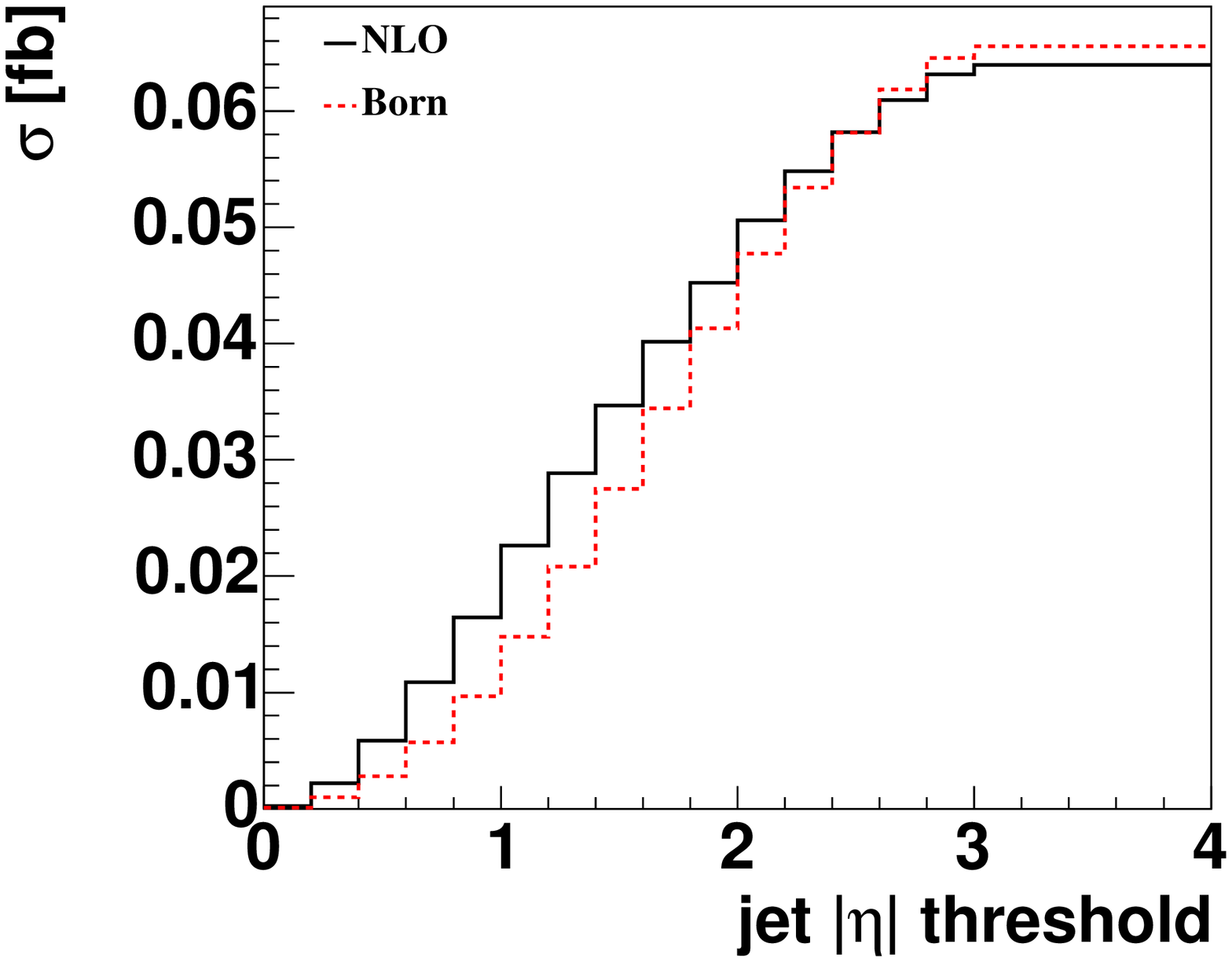}}\subfigure[]{\includegraphics[%
  width=0.40\linewidth,
  keepaspectratio]{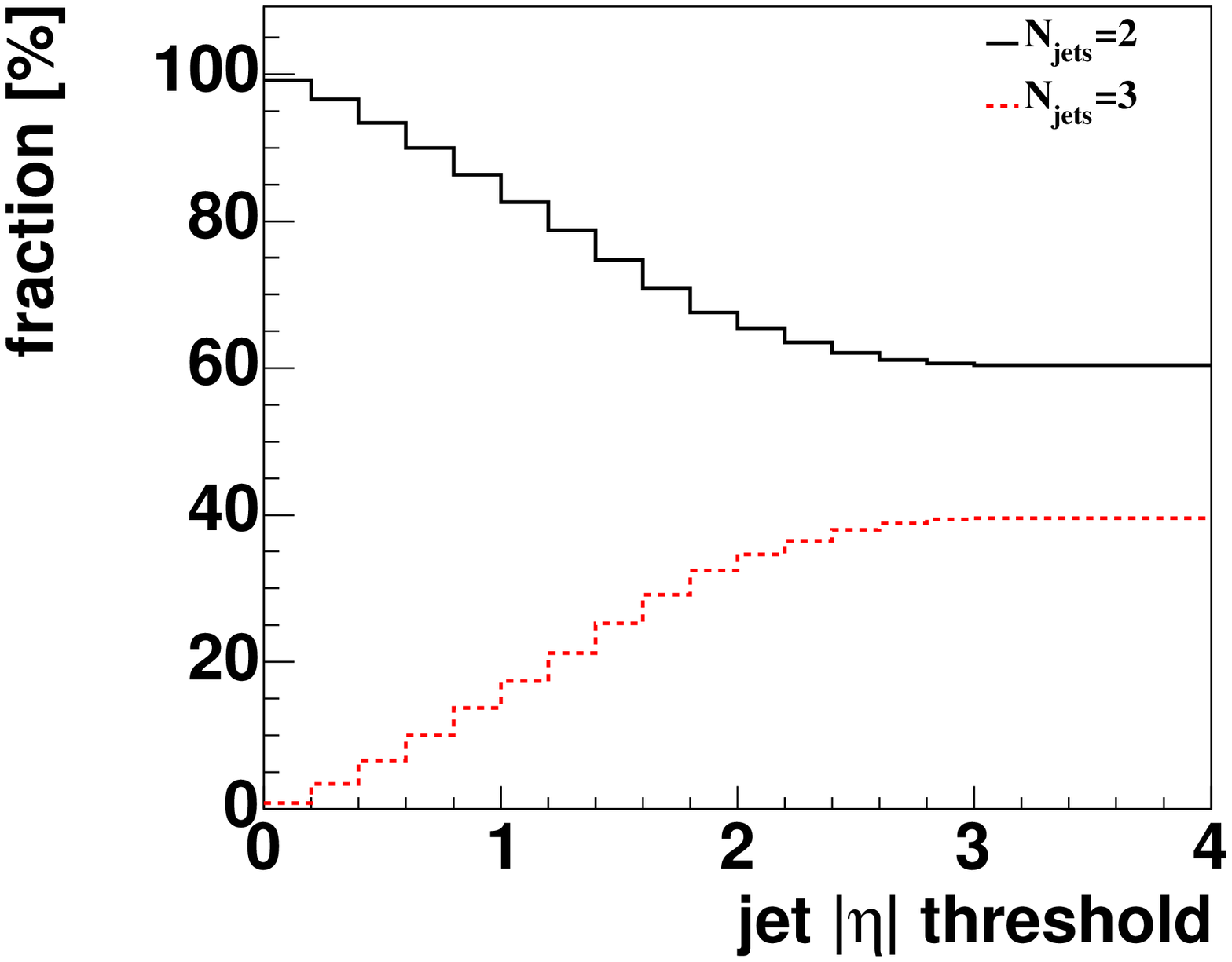}}

\caption{Inclusive cross section and fraction of three-jet events at NLO for
varying jet pseudo-rapidity cuts, after the loose selection cuts.
Shown is the total cross section as a function of the jet pseudo-rapidity
cut (a) and the fraction of two-jet and three-jet events as a function
of the jet pseudo-rapidity cut (b). \label{fig:njets_jet_eta}}
\end{figure}

From Fig.~\ref{fig:njets_jet_pt} it is clear that the jet multiplicity
at NLO depends strongly on the jet $p_{T}$ cut. Figure~\ref{fig:njets_jet_eta}
shows that it also depends strongly on the jet pseudo-rapidity cut.
The dependence of the total cross section on the jet pseudo-rapidity
cut is different between the Born-level and NLO, mostly as a result
of the third jet. Figure~\ref{fig:njets_jet_pt} also shows that
not only the cross section but also the jet multiplicity depends strongly
on the jet pseudo-rapidity cut. Only for jet pseudo-rapidity cuts
above 3 are cross section and jet multiplicities stable.

\begin{figure}
\subfigure[]{\includegraphics[%
  width=0.40\linewidth,
  keepaspectratio]{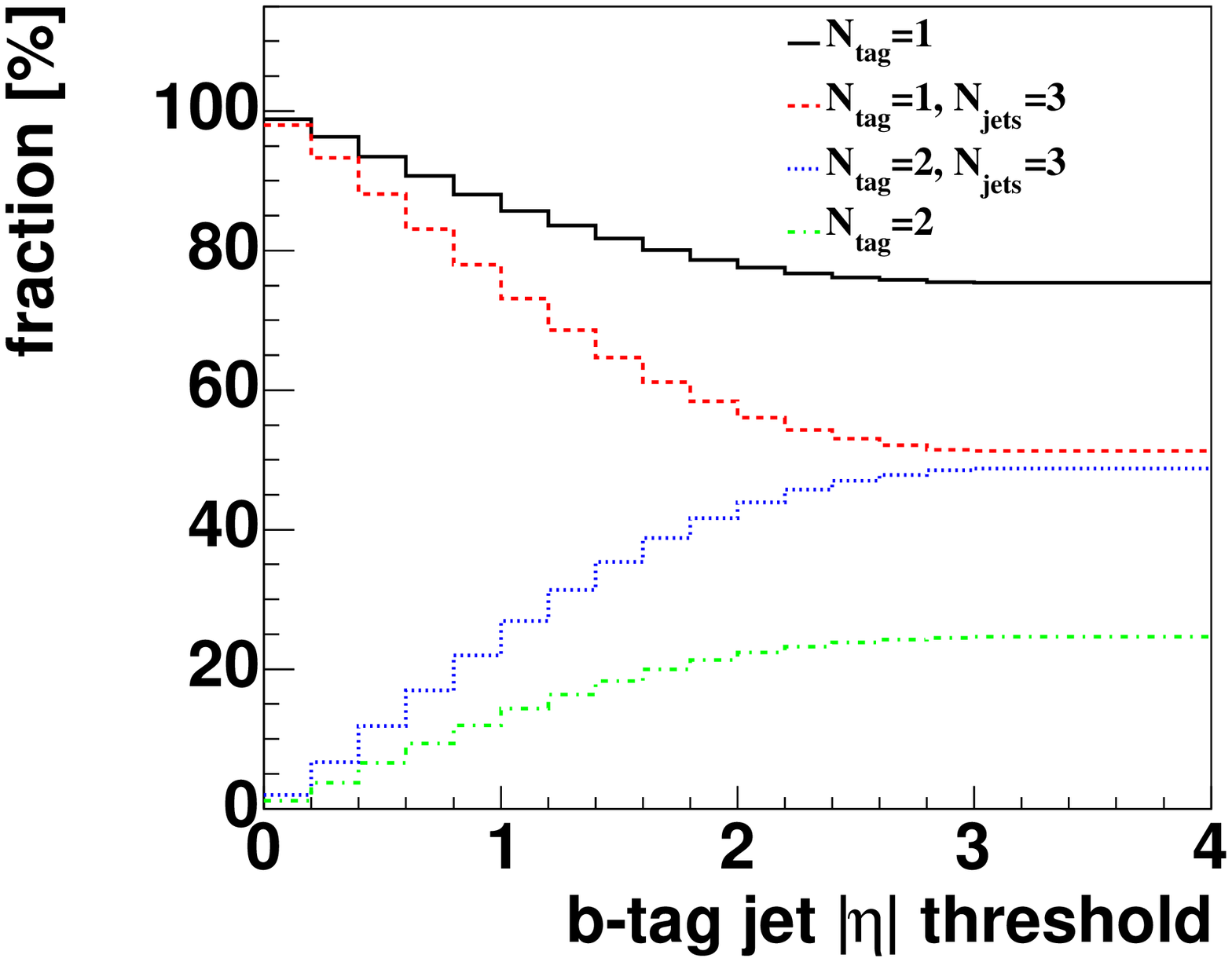}}\subfigure[]{\includegraphics[%
  width=0.40\linewidth,
  keepaspectratio]{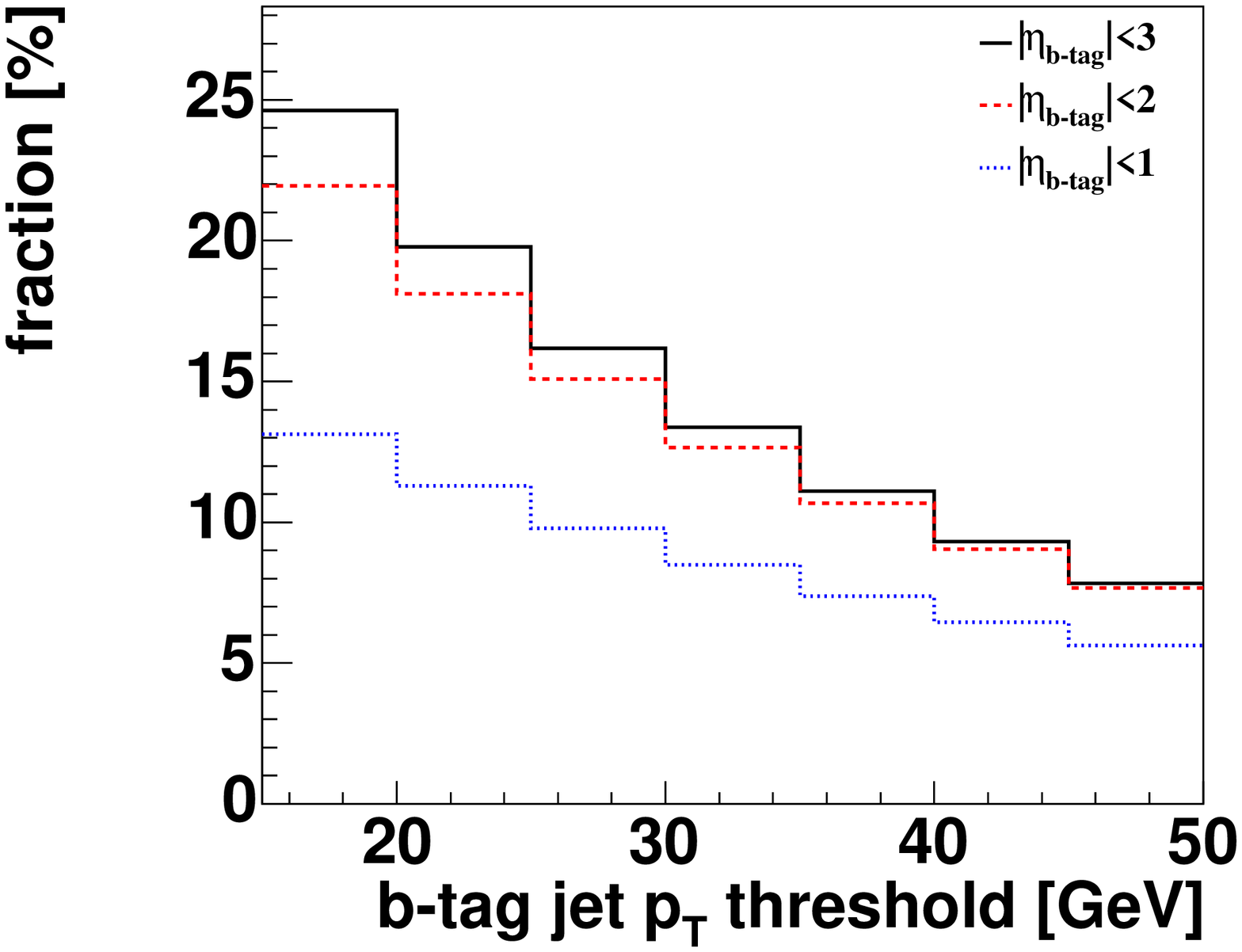}}

\caption{Fraction of events with one or two $b$-tagged jets for the inclusive
sample and the exclusive three-jet sample as a function of the pseudo-rapidity
cut on the $b$-tagged jets (a), and fraction of events with two tagged
jets as a function of the $p_{T}$ threshold on the $b$-tagged jets
for three different pseudo-rapidity cuts (b), after the loose selection
cuts. \label{fig:nbjets_jet_pteta}}
\end{figure}

The fraction of $b$-tagged jets also depends strongly on the jet
$p_{T}$ cut, as shown in Fig.~\ref{fig:ptbbbar}. There is a large
fraction of events in which the third jet comes from the $\bar{b}$~quark,
especially for low $b$-tag $p_{T}$ thresholds. This can also be
seen in Fig.~\ref{fig:nbjets_jet_pteta}, which shows the dependence
of the fraction of events with 1 $b$-tag and 2 $b$-tags as a function
of the cut on the $b$-tagged jet $p_{T}$ and $\eta$. As before,
the fraction of events with 2 $b$-tagged jets is stable only for
pseudo-rapidity thresholds above 3. At that point, about one quarter
of the inclusive events contain two $b$-tagged jets, and half of
the exclusive three-jet events contain two $b$-tagged jets. In the
following analysis, we shall require at least one $b$-tagged jet
in the event and do not distinguish the identity of the third jet,
unless specified otherwise.

\subsubsection{Kinematic Distribution of the Extra Jet}

Initial- and final-state emission of additional gluons occurs before
the top quark goes on shell and can thus be considered as {}``production-stage
emission'', while decay-stage emission occurs only after the top
quark goes on shell. In principle, an event with an extra jet can
thus be classified as production-stage or decay-stage by looking at
the invariant mass of the decay products. In production-stage emission
events, the $W$~boson and $b$~quark momenta will combine to give
the top quark momentum. In decay-stage emission events, the gluon
momentum must also be included to reconstruct the top quark momentum.
This interpretation is exact at the NLO parton level in the narrow
width approximation. Finite top width effects can blur the above classification
due to interference between production- and decay-stage emission.
This classification is nevertheless still useful in our case because
the top width of 1.5~GeV is small compared to the hard gluon $E_{T}$
cut imposed in the MC calculations. It should be kept in mind that
in an actual experiment, the production-decay distinction is further
blurred by the experimental jet energy resolution and ambiguities
associated with properly assigning partons to jets. 

\begin{figure}
\subfigure[]{\includegraphics[%
  width=0.40\linewidth,
  keepaspectratio]{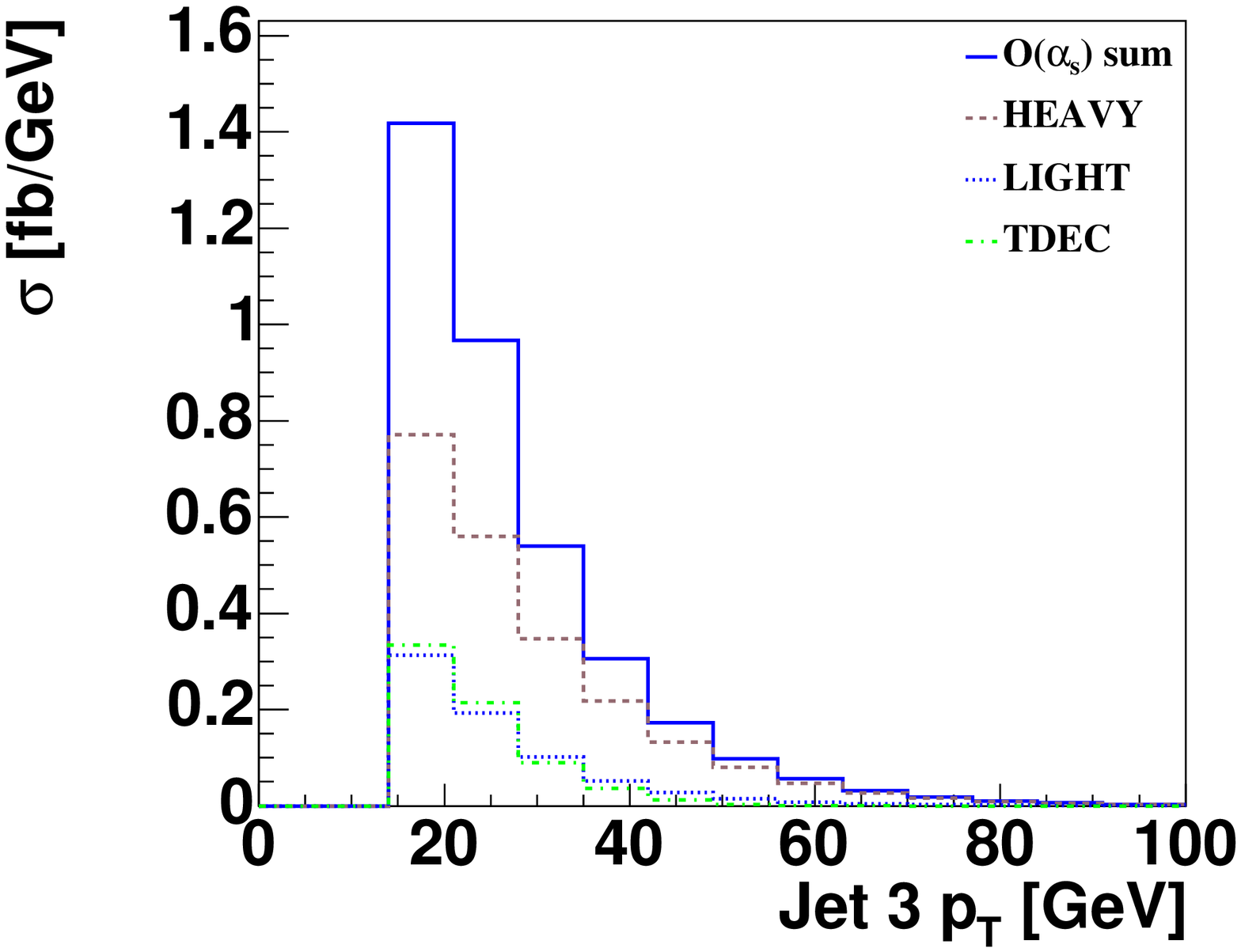}}\subfigure[]{\includegraphics[%
  width=0.40\linewidth,
  keepaspectratio]{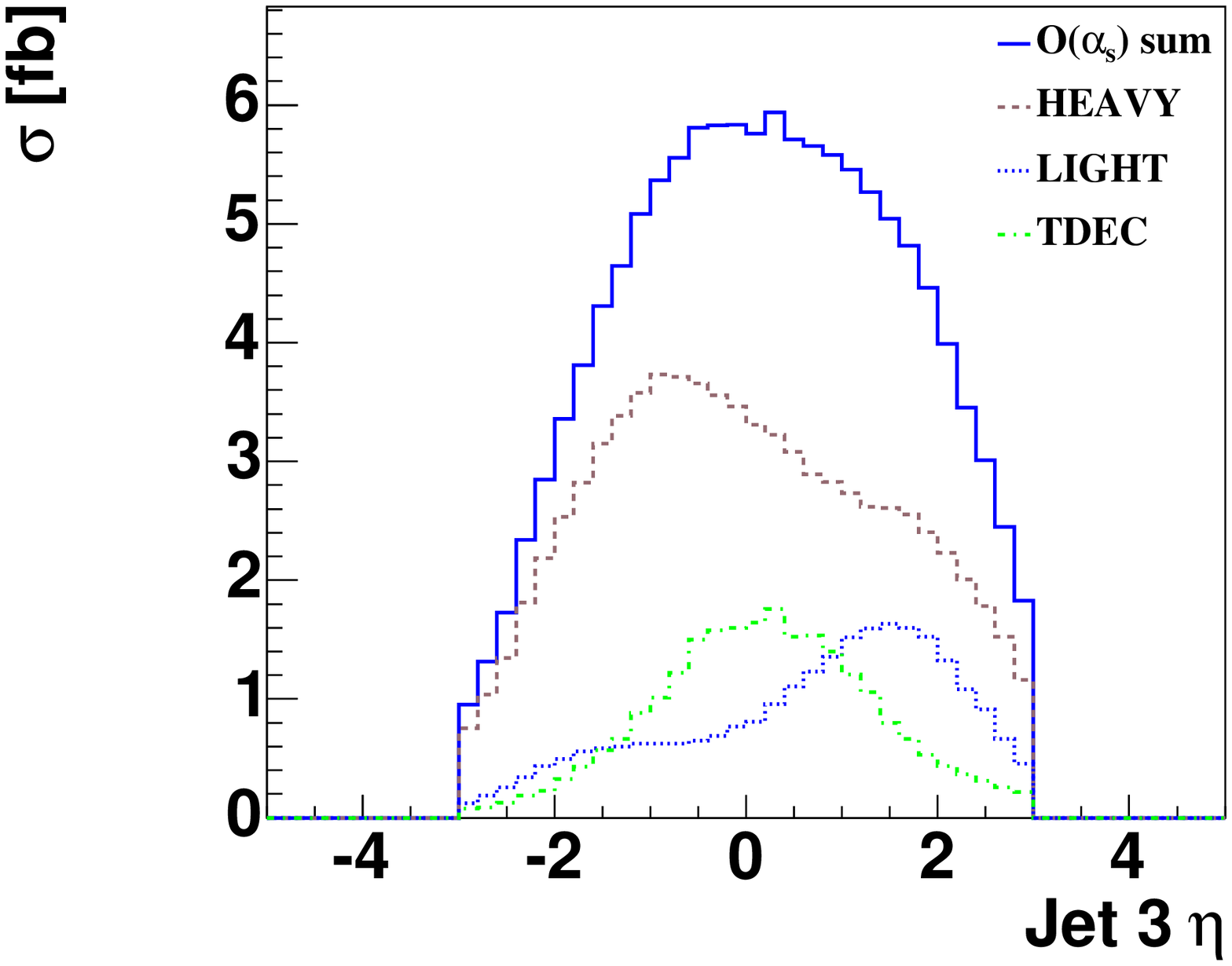}}

\caption{Transverse momentum (a) and pseudo-rapidity (b) of the third jet
after selection cuts for the various $\oalphas$ contributions.\label{fig:ptetaJet3}}
\end{figure}

Figure~\ref{fig:ptetaJet3} shows the transverse momentum distribution
as well as the pseudo-rapidity distribution of the third jet in three-jet
events. This jet corresponds to the gluon in about 70\% of the events
after the loose set of cuts. The transverse momentum distribution
of the third jet for those events where that jet corresponds to the
$\bar{b}$~quark (from $W$-gluon fusion subprocess) can be seen
in Fig.~\ref{fig:ptbbbar}. It comprises about $80\%$ of the HEAVY
correction after imposing the loose selection cuts, which dominates
over LIGHT and TDEC radiative corrections. 

As expected, the $E_{T}$ distribution is steeply falling for all
contributions, but it extends to much higher $p_{T}$ values for HEAVY
emission. The smaller values of $E_{T}$ to which TDEC emission is
constrained are again a consequence of the available phase space from
the top quark decay. Recall that the top quarks are produced with
relatively modest transverse momentum (cf. Fig.~\ref{fig:TopPt}),
so that gluons from the decay do not receive much of a boost. Note
also that an increase in the $p_{T}$ cut on the jet would result
in a further reduction in relative size of the decay contribution
compared to production. Figure~\ref{fig:ptetaJet3} also shows the
distribution in pseudo-rapidity of the extra jet. The third jet radiated
from the LIGHT and HEAVY quark lines has a relative larger magnitude
in its (non-symmetric) rapidity, as compared to the more central TDEC
emission. This is consistent with our intuition that decay-stage radiation,
dominated by the gluon radiated from the bottom quark which tends
to appear in the central pseudo-rapidity region, is also likely to
be produced centrally. However, this TDEC contribution is small and
the HEAVY radiation dominates even in the central region. 

\begin{figure}
\subfigure[]{\includegraphics[%
  width=0.40\linewidth,
  keepaspectratio]{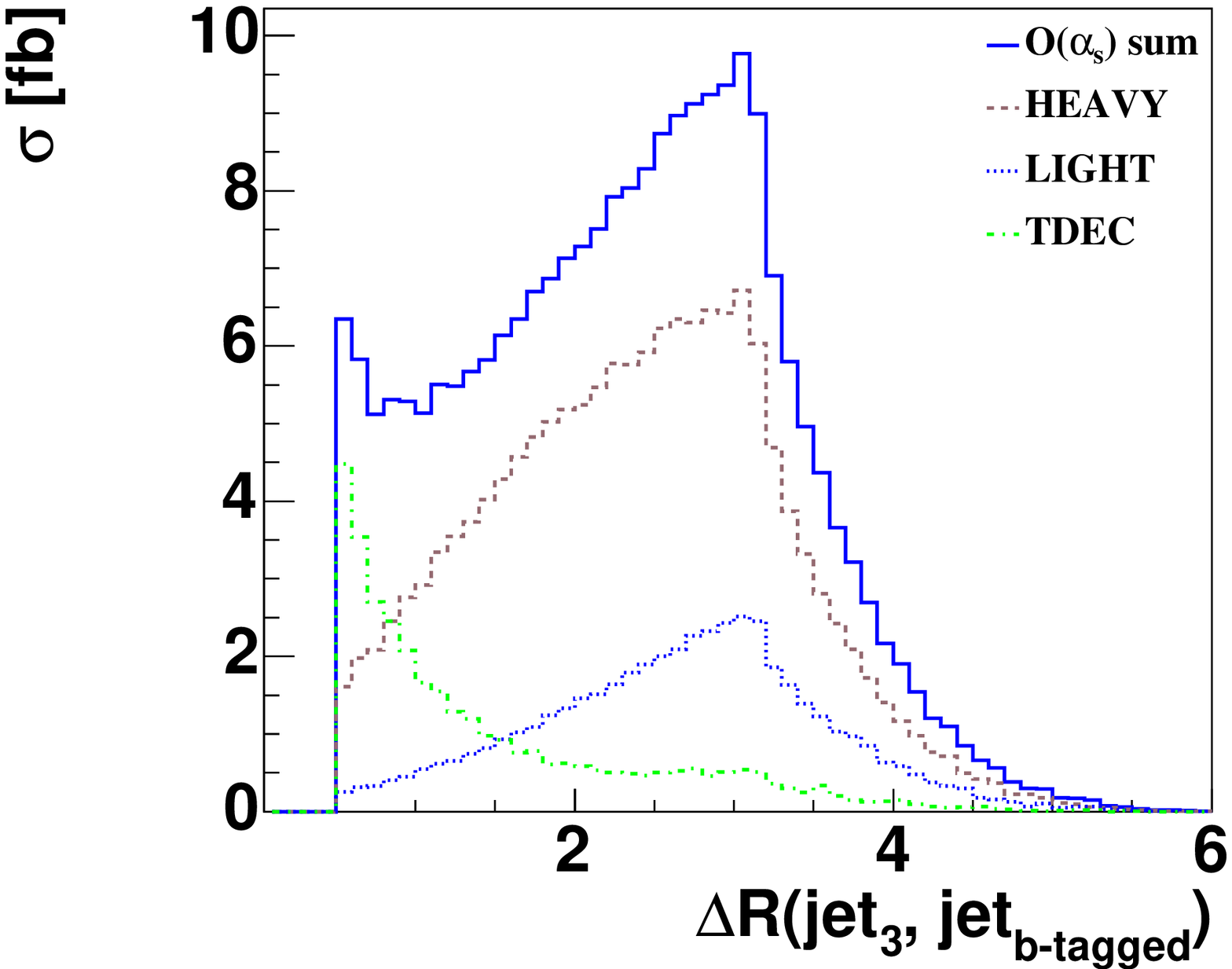}}\subfigure[]{\includegraphics[%
  width=0.40\linewidth,
  keepaspectratio]{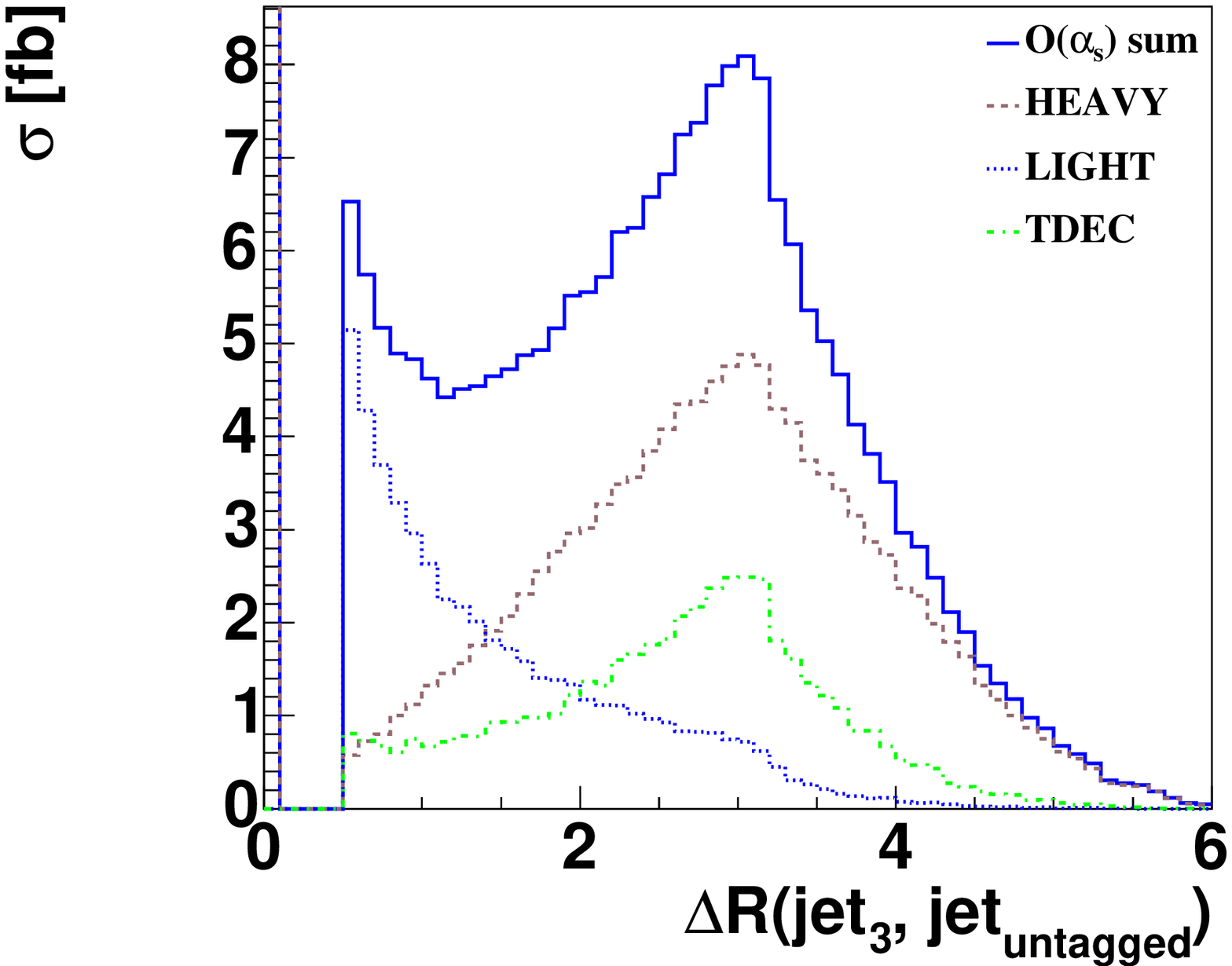}}

\caption{Separation between the third jet and the tagged jet (a) and between
the third jet and the untagged jet (b) after selection cuts for the
various $\oalphas$ corrections. \label{fig:dRJet3bJet}}
\end{figure}

This tendency of decay-stage radiation to be associated with the final-state
$b$~quark might lead one to expect that if the extra jet is {}``near''
the $b$~jet it should be included in the top quark reconstruction,
and if it is not then it should be excluded. Figure~\ref{fig:dRJet3bJet},
which shows the angular separation $\Delta R$ between the extra jet
and the leading $b$-tagged-jet as well as the leading untagged jet,
confirms that the decay-stage radiation peaks close to the leading
$b$-tagged jet, and production-stage radiation peaks farther away.
Figure~\ref{fig:dRJet3bJet}(a) clearly shows that the decay contribution
dominates in the low $\Delta R$ region. This is different from the
$s$-channel single top process, in which the production contribution
dominates over decay emission even in the small $\Delta R$ region,
cf. Fig.~17 of Ref.~\cite{Cao:2004ap}. A higher $p_{T}$ cut on
the jet would make this situation worse because it would increase
the relative size of the production emission. 

Nevertheless, the figure suggests that it might be possible to further
improve the top quark reconstruction for exclusive three-jet events.
In this paper, we have been using the leading $b$-tagged jet to identify
the $b$~quark from the top quark decay, and constraints on $m_{t}$
and $M_{W}$ to obtain the correct $p_{z}^{\nu}$. With this procedure
the final state can be reconstructed accurately about 84~\% of the
time in inclusive two-jet events, cf. Fig.~\ref{fig:b_jet_eff}.
Part of the efficiency loss is due to the gluon radiated in TDEC emission,
and Fig.~\ref{fig:dRJet3bJet} indicates that some of this loss could
be reclaimed by including the third jet in the top quark reconstruction
if it is close to the $b$-tagged jet. The actual cut on $\Delta R$
would need to be tuned to maximize the top quark reconstruction efficiency.
However, tuning a prescription for dealing with the extra jet in $t$-channel
single top quark events is complicated because the effects of multiple
emission, hadronization, and detector resolutions will affect the
result. For simplicity, we thus chose not to include the third jet
in the top quark reconstruction algorithm used in this paper. 

In Fig.~\ref{fig:dRJet3bJet}(b), the equivalent distribution in
$\Delta R$ between the extra jet and the untagged jet is also shown,
where the LIGHT quark line radiation peaks close to the untagged jet
as expected from gluon radiation in the final state. Obviously, the
radiation of the HEAVY quark line is dominated by initial state radiation,
therefore, its contribution to $\Delta R(j_{3},j_{{\rm untagged}})$
is small in the region of small $\Delta R$. The peak at zero in this
distribution corresponds to events containing two $b$-tagged jets
in the the $W$-gluon fusion subprocess.

\subsubsection{Angular Correlation Between the Extra Jet and the Best Jet }

As discussed above, the $\Delta R$ separation between the third jet
and the leading $b$-tagged jet can be used to distinguish decay-stage
emission from production-stage emission. It was shown in Ref.~\cite{Cao:2004ap}
that the best-jet algorithm is effective in reconstructing the $s$-channel
single top quark event, while retaining the top quark polarization
information, but can also distinguish the decay-stage gluon radiation
from the production-stage radiation by studying which jets are chosen
as the best jets. Similarly, we can also study the correlations between
the extra jet and the best jet in $t$-channel single top quark events
using the best jet algorithm.

\begin{figure}
\subfigure[]{\includegraphics[%
  width=0.40\linewidth,
  keepaspectratio]{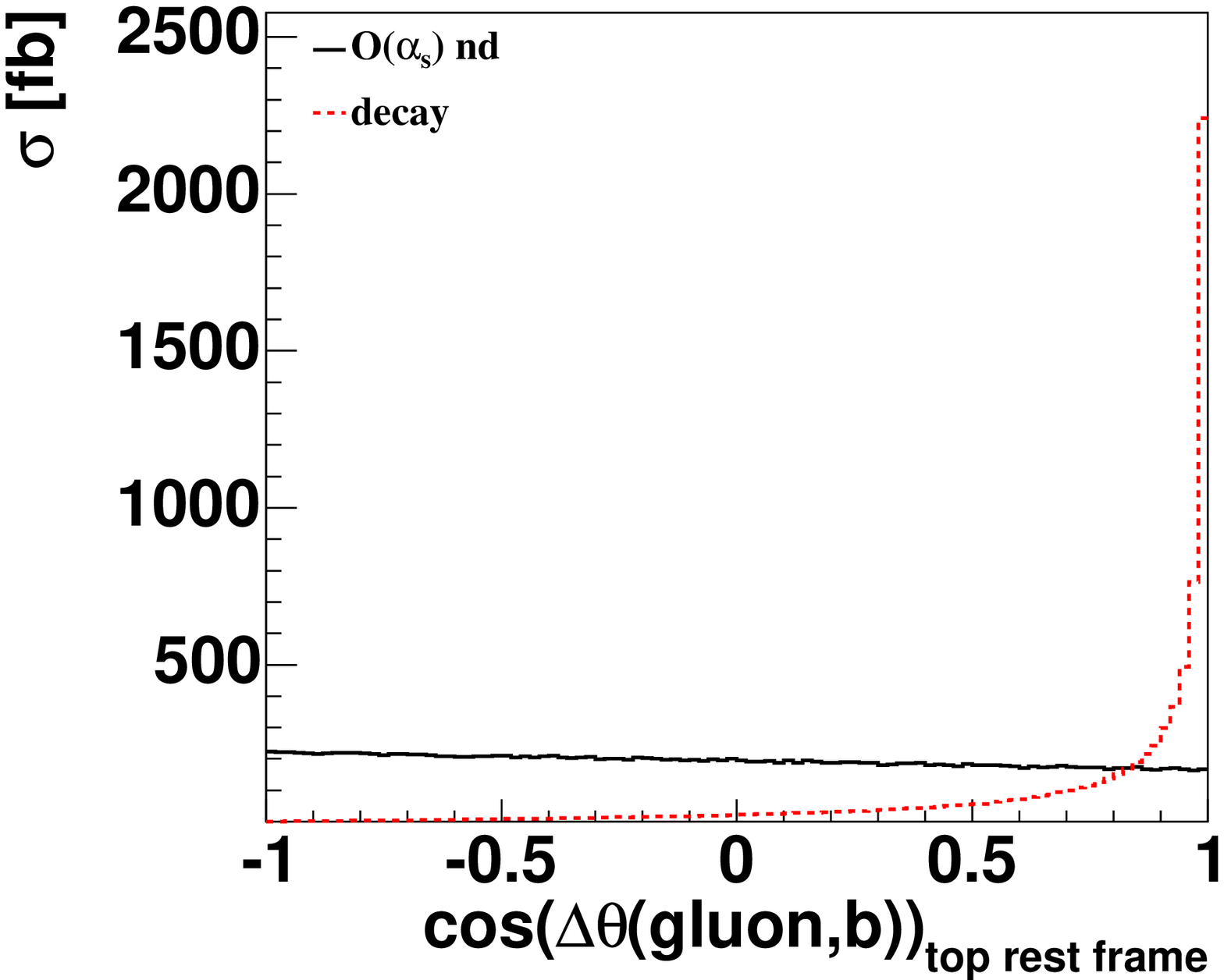}}\subfigure[]{\includegraphics[%
  width=0.40\linewidth,
  keepaspectratio]{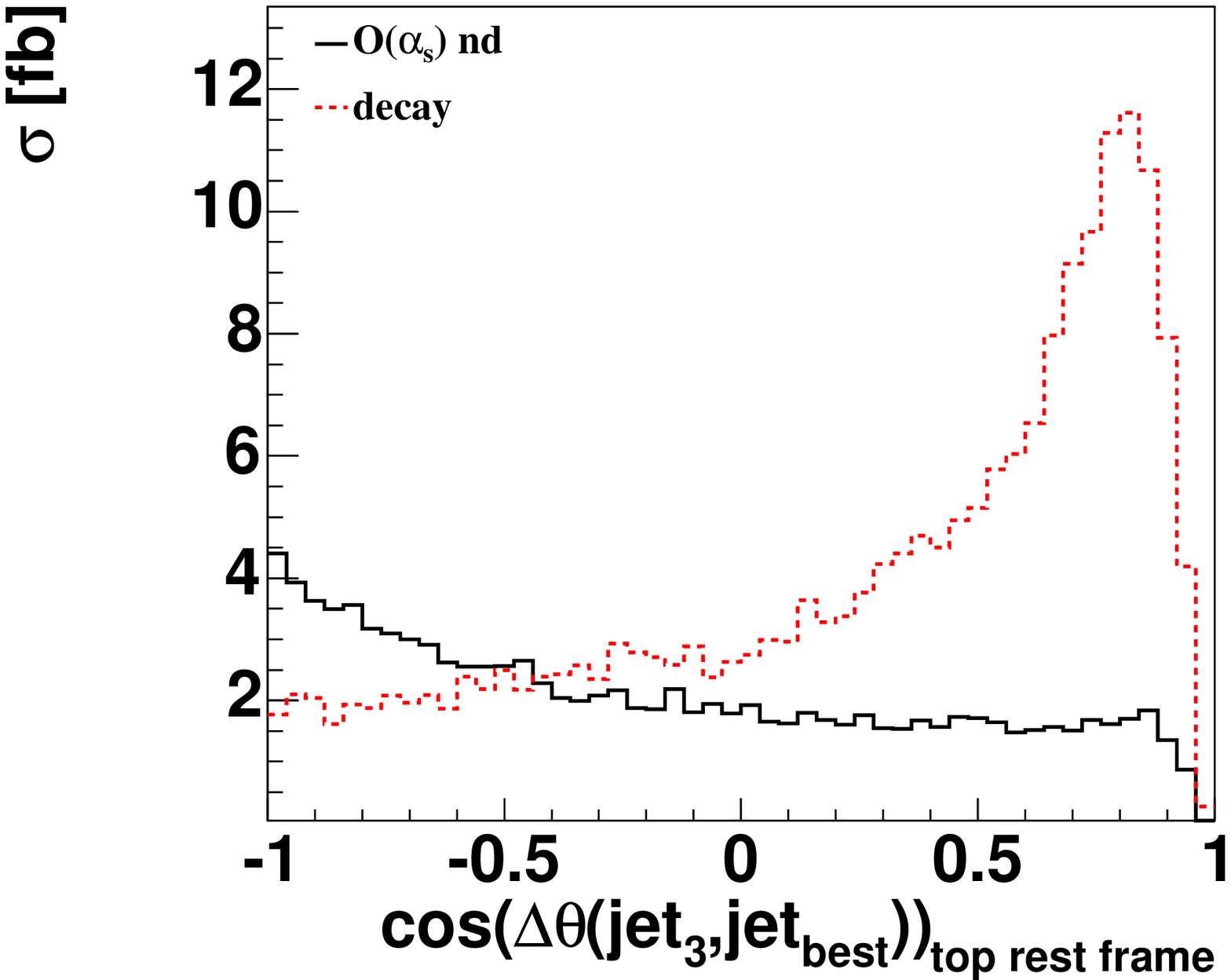}}

\caption{Angular correlation $\cos\theta$ between the gluon and the b quark
before any selection cuts using the full parton information (a) and
between the third jet and the best jet after selection cuts (b). The
solid line shows all $\oalphas$ contributions except for the decay
part, while the dashed line shows only the $\oalphas$ decay contribution.\label{fig:CosThetaJet3BgJet}}
\end{figure}

Figure~\ref{fig:CosThetaJet3BgJet} (a) shows the angular correlation
$\cos\theta$ between the radiated parton (gluon or $\bar{b}$) and
the $b$~quark at parton level before cuts. Figure~\ref{fig:CosThetaJet3BgJet}
(b) shows the same correlation after event reconstruction between
the third jet and the best jet. Only events for which the best jet
algorithm chooses a two-jet system are included in the figure. In
this case there is a clear separation between production-stage and
decay-stage emission, and the best jet algorithm can be used to separate
the two. This approach thus allows a detailed experimental study of
the radiation pattern in top quark decays in the $t$-channel single
top quark events.

\section{Conclusions\label{sec:Conclusions}}

We have presented a next-to-leading order study of $t$-channel single
top quark events at the Tevatron, including $\oalphas$ QCD corrections
to both the production and decay of the top quark. To obtain an accurate
prediction of the inclusive rate of $t$-channel single top quark
production, a modified narrow width approximation has been adopted
to link the production of the top quark with its decays (thus preserving
top quark spin information) instead of the usual narrow width approximation.
The impact of kinematical cuts on the acceptances has been studied
for several different sets of cuts. We found that the acceptances
are sensitive to the $\Delta R_{{\rm cut}}$ we imposed on the jet
cone size and the lepton isolation. With the choice of $\Delta R_{{\rm cut}}=0.5$,
we found that the difference between the Born-level and NLO acceptances
is about $10\%$ for a loose cut set (a) and $3\sim6\%$ for a tight
cut set (c). The above difference becomes significantly larger when
changing $\Delta R_{{\rm cut}}$ from 0.5 to 1.0. 

We categorize the $\oalphas$ contributions to the $t$-channel single
top process into three gauge invariant sets: the light quark line
corrections, the heavy quark line corrections and the top quark decay
corrections. Keeping track of the different categories facilitates
the comparison between event generators and exact NLO predictions. 

The $\oalphas$ corrections are small in size and contribute about
5.5\% of the inclusive cross section at NLO. They do however modify
the event kinematics and in particular result in a large fraction
of events containing three reconstructed jets in the final state for
the loose set of kinematic cuts. The acceptance for $t$-channel single
top quark events and this fraction of 3-jet events depend strongly
on the pseudo-rapidity cut on the jets. Although the radiative corrections
to the inclusive $t$-channel single top production rate are small,
they affect the shape of some of the important kinematic distributions
that can be used experimentally to separate the $t$-channel single
top signal from the various backgrounds, such as the pseudo-rapidity
distribution of the spectator jet. We find that the $\oalphas$ LIGHT
and HEAVY corrections have almost opposite contributions to various
pseudo-rapidity distributions, due to the difference in the parton
distribution functions between the valence quarks and sea quarks.
The former shifts the spectator jet to even higher pseudo-rapidities,
while the later shifts it to more central pseudo-rapidity regions.
The summed contributions cause the spectator jet to be even more forward
which will change the prediction of the acceptance for $t$-channel
single top quark events. Also, a large fraction of three jet events
contain two $b$-jets due to the collinear enhancement in the $W+g$
fusion process. This implies that higher order corrections need to
be calculated in order to correctly predict the behavior of the $b$-jet
in the small $p_{T}$ region.

In order to study top quark properties such as the top quark polarization,
induced from the effective $t$-$b$-$W$ couplings, we need to reconstruct
the top quark by combining the reconstructed $W$ boson with the $b$-tagged
jet. Most of the $t$-channel single top quark events contain only
one $b$-jet in the final state. Thus, we can use the leading $b$-tagged
jet algorithm to identify the $b$-jet in the final state, rather
than the best-jet algorithm which is more appropriate for the $s$-channel.
We found that the leading $b$-tagged jet algorithm effectively picks
up the correct $b$-jet in the event. Since this algorithm makes no
assumptions about the other particles in the event, we can use it
to also determine the longitudinal momentum of the neutrino ($p_{z}^{\nu}$)
accurately through a top quark mass constraint. Information about
the reconstructed final state can be used to explore correlations
between different objects in the event. After reconstruction of the
top quark event, we study spin correlations in the top quark decay
in three different bases: the helicity basis, the spectator basis,
and the beamline basis. We find that the degree of top polarization
is very large, especially in the helicity and spectator bases. This
is true even after event reconstruction because we are able to reconstruct
the top quark final state almost perfectly. As one expects, the degree
of top polarization is reduced slightly by the $\oalphas$ corrections,
both at parton level and after event reconstruction with the loose
selection cuts. We also note that using the $tq$-frame (the rest
frame of the reconstructed top quark and spectator jet) in the helicity
basis gives almost the same degree of polarization as in the spectator
basis. 

Finally, we point out that the above conclusion does not strongly
depend on the choice of jet algorithm. We have checked that at the
parton level using the Durham $k_{T}$ algorithm~\cite{Catani:1992zp,Ellis:1993tq}
leads to similar conclusion on the relative importance of the production-
and decay-stage gluon emission.

\begin{acknowledgments}
We thank Hong-Yi Zhou for collaboration in the early stage of this
project . This work was supported in part by NSF grants PHY-0244919
and PHY-0140106. 
\end{acknowledgments}
\appendix

\section{Top quark polarization in the s-channel process}

Above we presented the top quark spin correlations in the helicity
basis in the $t$-channel using two different definitions for the
c.m. frame. In our previous paper on the $s$-channel, the degree
of top quark polarization in the helicity basis was calculated only
in the $t\bar{b}$ frame (see Sec.~IV of Ref.~\cite{Cao:2004ap})
and given only for inclusive 2-jet events. That frame corresponds
to the c.m. frame of the incoming partons strictly speaking only for
events without initial or final state gluon radiation. For completeness,
we also present here the results for the top quark polarization in
the $s$-channel single top process for two different c.m. frame definitions.
Similar to the $t$-channel single top process, we define two reference
frames to study the degree of top quark polarization at the parton
level in the helicity basis for $s$-channel single top quark events.

\begin{enumerate}
\item $t\bar{b}(j)$-frame: the c.m. frame of the incoming partons. This
is the rest frame of all the final state objects (reconstructed top
quark and all others jets). In exclusive two-jet events, this frame
is the same as that at the Born-level, i.e. reconstructed from summing
over momentum of the top quark and non-best-jet. In exclusive three-jet
events, this frame is reconstructed by summing over the 4-momenta
of top quark, non-best-jet, and the third-jet from the parton level
calculation. 
\item $t\bar{b}$-frame: the c.m. frame of the top quark and non-best-jet.
In this case, even in the exclusive three-jet events, the reference
frame is constructed by summing over only the 4-momenta of the top
quark and non-best-jet. Note that this differs from the $t\bar{b}(j)$-frame
only in exclusive three-jet events.
\end{enumerate}
\begin{table}
\begin{center}\begin{tabular}{cc|c|c|c|c|c|c}
\hline 
&
&
\multicolumn{2}{c|}{$\mathcal{D}$ }&
\multicolumn{2}{c|}{\textbf{$\mathcal{F}$}}&
\multicolumn{2}{c}{$\mathcal{A}$}\tabularnewline
\cline{3-4} \cline{5-6} \cline{7-8} 
&
&
LO&
NLO&
LO&
NLO&
LO&
NLO\tabularnewline
\hline
Helicity basis: &
Parton level ($t\bar{b}(j)$-frame)&
0.63&
0.54&
0.82&
0.77&
0.32&
0.27\tabularnewline
&
Parton level ($t\bar{b}$-frame)&
0.63&
0.58&
0.82&
0.79&
0.32&
0.29\tabularnewline
&
Reconstructed events ($t\bar{b}(j)$-frame)&
0.46&
0.37&
0.73&
0.68&
0.21&
0.21\tabularnewline
&
Reconstructed events ($t\bar{b}$-frame)&
0.46&
0.37&
0.73&
0.68&
0.26&
0.21\tabularnewline
Beamline basis: &
Parton level&
-0.96&
-0.92&
0.98&
0.96&
-0.48&
-0.46\tabularnewline
&
Reconstructed events&
-0.48&
-0.42&
0.74&
0.71&
-0.24&
-0.21\tabularnewline
\hline
\end{tabular}\end{center}

\caption{Degree of polarization $\mathcal{D}$, polarization fraction $\mathcal{F}$,
and asymmetry $\mathcal{A}$ for inclusive two-jet single top quark
events, at the parton level and after event reconstruction, in the
$s$-channel process. Here, $\mathcal{F}$ corresponds to $\mathcal{F}_{-}$
in the helicity basis for left-handed top quarks and to $\mathcal{F}_{+}$
in the beamline basis for top quarks with polarization along the direction
of anti-proton three momentum, respectively. The $t\bar{b}g$ frame
in the helicity basis denotes the c.m. frame of the incoming partons
while $t\bar{b}$ frame denotes the rest frame of the reconstructed
top quark and $\bar{b}$ quark. \label{tab:toppol-schan-2jet}}
\end{table}
\begin{table}
\begin{center}\begin{tabular}{cc|c|c|c}
\hline 
&
&
\multicolumn{1}{c|}{$\mathcal{D}$ }&
\multicolumn{1}{c|}{\textbf{$\mathcal{F}$}}&
\multicolumn{1}{c}{$\mathcal{A}$}\tabularnewline
\hline
Helicity basis: &
Parton level ($t\bar{b}g$ frame)&
0.45&
0.72&
0.25\tabularnewline
&
Parton level ($t\bar{b}$ frame)&
0.49&
0.74&
0.27\tabularnewline
&
Reconstructed events ($t\bar{b}g$ frame)&
0.36&
0.68&
0.21\tabularnewline
&
Reconstructed events ($t\bar{b}$ frame)&
0.37&
0.68&
0.21\tabularnewline
\hline
Beamline basis: &
Parton level&
-0.81&
0.91&
-0.45\tabularnewline
&
Reconstructed events&
-0.38&
0.69&
-0.19\tabularnewline
\hline
\end{tabular}\end{center}

\caption{Degree of polarization $\mathcal{D}$, polarization fraction $\mathcal{F}$,
and asymmetry $\mathcal{A}$ for exclusive three-jet single top quark
events, at parton level and after event reconstruction, in the $s$-channel
process. Here, $\mathcal{F}$ corresponds to $\mathcal{F}_{-}$ in
the helicity basis for left-handed top quarks and to $\mathcal{F}_{+}$
in the beamline bases for top quarks with polarization along the direction
of anti-proton three momentum, respectively. The $t\bar{b}g$ frame
in the helicity basis denotes the c.m. frame of the incoming partons
while the $t\bar{b}$ frame denotes the rest frame of the reconstructed
top quark and $\bar{b}$ quark. \label{tab:toppol-schan-3jet}}
\end{table}

We present the resulting top quark polarization in the helicity and
optimal bases in Table~\ref{tab:toppol-schan-2jet} for inclusive
2-jet events and in Table~\ref{tab:toppol-schan-3jet} for exclusive
3-jet events. Similar to the $t$-channel, the degree of polarization
is larger in the helicity basis in the $t\bar{b}$ frame. The polarization
is reduced when including the possible third jet to reconstruct the
c.m. frame in particular for those events where that jet is produced
in the initial state, before the virtual $W$~boson is created. Table~\ref{tab:toppol-schan-3jet}
also shows that the degree of polarization is generally smaller for
3-jet events, which is again similar to the $t$-channel.

\bibliographystyle{apsrev}
\clearpage\addcontentsline{toc}{chapter}{\bibname}\bibliography{reference}

\end{document}